\documentclass[numberedappendix, appendixfloats]{emulateapj}
\usepackage{epstopdf} %% for using pdf figure
\usepackage{graphicx} %% for using pdf figure
\usepackage{natbib}
\usepackage{color}
\usepackage{comment}

\shortauthors{T. Hosokawa et al.}
\shorttitle{Massive Primordial Stars}

\newcommand{\msun}{M_{\odot}}

\newcommand{\msunyr}{M_\odot~{\rm yr}^{-1}}
\newcommand{\mdot}{\dot{M}_*}

\newcommand{\hii}{H{\sc ii} }
\newcommand{\cmc}{{\rm cm}^{-3}}
\newcommand{\vvec}{\mbox{\boldmath$v$}}

\newcommand{\evecr}{\mbox{\boldmath$e_r$}}
\newcommand{\press}{{\rm dynes~cm}^{-2}}

\begin{document}

%\begin{comment}
%\end{comment}

\title{Formation of Massive Primordial Stars: \\
Intermittent UV Feedback with Episodic Mass Accretion}
\author{Takashi Hosokawa\altaffilmark{1,2},
        Shingo Hirano\altaffilmark{2},
        Rolf Kuiper\altaffilmark{3},
        Harold W. Yorke\altaffilmark{4}, \\
        Kazuyuki Omukai\altaffilmark{5},
        Naoki Yoshida\altaffilmark{2,6}}

\altaffiltext{1}{Research Center for the Early Universe,
the University of Tokyo, Tokyo 113-0033, Japan;
takashi.hosokawa@phys.s.u-tokyo.ac.jp, hosokwtk@gmail.com}
\altaffiltext{2}{Department of Physics, School of Science, 
the University of Tokyo, Tokyo 113-0033, Japan}
\altaffiltext{3}{University of T\"ubingen, Institute of Astronomy 
and Astrophysics, Auf der Morgenstelle 10, 
D-72076 T\"ubingen, Germany}
%
%\altaffiltext{3}{Max Planck Institute for Astronomy,
%K\"onigstuhl 17, D-69117 Heidelberg, Germany}
%
\altaffiltext{4}{Jet Propulsion Laboratory, California Institute
of Technology, Pasadena CA 91109, USA}
\altaffiltext{5}{Astronomical Institute, Tohoku University, Sendai
980-8578, Japan}
\altaffiltext{6}{Kavli Institute for the Physics and Mathematics
of the Universe (WPI), Todai Institutes for Advanced Study,
the University of Tokyo, Chiba 277-8583, Japan}

\begin{abstract}
We present coupled stellar evolution (SE) and 3D radiation-hydrodynamic 
(RHD) simulations of the evolution of primordial protostars, their immediate 
environment, and the dynamic accretion history
under the influence of stellar ionizing and dissociating
UV feedback. 
Our coupled SE-RHD calculations result in a wide diversity of final stellar
masses covering $10~\msun \lesssim M_* \lesssim 10^3~\msun$. 
The formation of very massive ($\gtrsim 250~\msun$) 
stars is possible under weak UV feedback, whereas
ordinary massive (a few $\times 10~\msun$) stars form when
UV feedback can efficiently halt the accretion. 
This may explain the peculiar abundance pattern of a Galactic 
metal-poor star recently reported by \citet{Aoki14}, possibly the 
observational signature of very massive precursor primordial stars.
Weak UV feedback occurs in cases of variable accretion, 
in particular when repeated short accretion bursts temporarily exceed
$0.01~\msunyr$, causing the protostar to inflate.
In the bloated state, the protostar has low surface temperature and
UV feedback is suppressed until the star eventually contracts, on a
thermal adjustment timescale,
to create an \hii region. 
If the delay time between successive accretion bursts is
sufficiently short, the protostar remains bloated for extended periods,
initiating at most only short periods of UV feedback. 
Disk fragmentation does not necessarily reduce the final stellar mass.
Quite the contrary, we find that disk fragmentation enhances
episodic accretion as many fragments migrate inward and are accreted
onto the star, thus allowing continued stellar mass growth under conditions of
intermittent UV feedback. 
This trend becomes more prominent as we improve the resolution
of our simulations. We argue that simulations with significantly higher
resolution than reported previously are needed to derive accurate
gas mass accretion rates onto primordial protostars.
\end{abstract}


\hspace{-96mm} \copyright 2016. All rights reserved.

%%%%%%%%%%%%%%%%%%%%%%%
\section{Introduction}
\label{sec:intro}
%%%%%%%%%%%%%%%%%%%%%%%
\setcounter{footnote}{0}

Modern cosmology predicts that the early universe is
dominated by massive ($M_* \gtrsim 10~\msun$) primordial stars 
\citep[e.g.,][]{Bromm09,Greif15Rv}. 
The formation of such massive primordial stars is
thought to begin in mini-haloes at 
$z \simeq 20 - 30$ \citep[e.g.,][]{ABN02,Y03}, forming 
tiny embryo protostars ($M_* \sim 10^{-2}~\msun$)  
after the gravitational collapse of natal clouds 
\citep[e.g.,][]{ON98,Y08}.
These protostars grow in mass by accreting surrounding gas 
\citep[e.g.,][]{OP03,Tan04}.
Their evolution in the accretion stage 
and how they interact with their environment determine
the characteristics of the emerging massive stars.

%------------------------ UV feedback (in 2D) ---------------------------%

Recent theoretical studies suggest that 
the typical final stellar mass will be somewhat smaller
than that of the cloud mass, $\sim$$10^3~\msun$
\citep[e.g.,][]{Bromm02}.
Stellar radiative feedback is one of the crucial
processes which regulate the mass accretion
\citep[e.g.,][]{OI02,MT08}.
\citet[][hereafter HS11 and HS12]{HOYY11, HYOY12} 
show that, with coupled stellar evolution (SE) and 
2D axisymmetric radiation-hydrodynamic 
(RHD) simulations, UV feedback eventually
halts the accretion of material onto the protostar.
As the stellar UV emissivity rises with increasing stellar
mass, a bipolar \hii region forms and dynamically expands,
first slowing, then reversing the inward flow of gas in the accretion envelope.
The circumstellar accretion disk also loses gas via photoevaporation,
being exposed to the stellar UV radiation.
\citet[][hereafter HR14]{Hirano14} have greatly extended the above work
to follow the evolution in more than one hundred
primordial clouds taken from 3D cosmological simulations.
Although UV feedback eventually shuts off mass accretion
in general, the resulting final stellar masses have 
great diversity, spanning $10~\msun \lesssim M_* \lesssim 10^3~\msun$ 
\citep[see also][hereafter HR15]{Susa14,Hirano15}.

%-------------------- 3D effects (fragmentation) ------------------------%

Whereas the 2D simulations have revealed the dynamic nature
of stellar UV feedback, 3D simulations
show an additional aspect of mass accretion through
self-gravitating circumstellar disks
\citep[e.g.,][]{Stacy10,Cl11,Greif11,Greif12,Vorobyov13,MD14}.
In particular, these studies show that disks readily fragment
via gravitational instability. 
The fate of the fragments seems to have a great diversity;
some will survive, being expelled from the system via gravitational 
ejection, and others will migrate inward through the disk to merge with
each other or fall onto the central star causing accretion bursts.
Through such recurrent merger and ejection events, binary
or multiple stellar systems will finally appear
\citep[e.g.,][]{Stacy13}. 

%--------------- 3D effects (fragmentation+UV feedback) ------------------%
%------ very massive star needed? Aoki+14, direct-collapse ------%

The final stellar mass should ultimately be determined by the complex interplay
between UV feedback and the various 3D effects described above. 
Performing SE-RHD simulations including all of these effects is still
challenging, but nevertheless remains the ultimate goal for computational studies.
Whereas the first 3D simulations mostly focused on the early
evolution for $10 - 10^3$~years after the birth of the protostar,
\citet{Stacy12} follow for longer periods ($\simeq 5 \times 10^3$~years), 
namely until UV feedback begins to operate.
They show that a bipolar \hii region grows from a binary system
formed via disk fragmentation.
\citet{Susa13} and \citet{Susa14} follow for even longer periods
($\simeq 10^5$~years) of evolution, including the effect of photodissociating 
(FUV) feedback with their 3D simulations, and show that mass accretion onto
the star is ultimately halted by the feedback effects.

The above studies suggest that both UV feedback and disk
fragmentation might co-operatively reduce
the final stellar masses. This is because 
the accreting gas is divided among multiple protostars
and the accretion rates onto each protostar is correspondingly reduced
\citep[e.g.,][]{Peters10}. In fact, the mass spectrum
derived by \citet{Susa14} covers a somewhat lower mass range
than the 2D results in HR14.

%--------------------------------------------------------------------------%

In this paper, we cast a new light on a different mode of interplay
between UV feedback and mass accretion in 3D.
As described above, highly variable time-dependent 
mass accretion histories are commonly seen in 3D 
numerical simulations. 
Even if the disk undergoes fragmentation, a large
fraction of the fragments do not actually survive.
For instance, \citet{Greif12} show that about $2/3$ of the
emerging fragments eventually merge with other fragments or existing stars. 
This is driven by the inward migration of the fragments
due to strong gravitational torques caused by the asymmetric
disk structure (e.g., spiral arms).
As a result, accretion histories normally become highly variable, 
resulting in short accretion bursts followed by relatively
long quiescent phases \citep[e.g.,][]{Rowan12,Vorobyov13,DeSouza15}.

%------------------------------------------------------------------%

Stellar evolution calculations which include the effects of mass
accretion show that 
rapid accretion with $\mdot \gtrsim 10^{-2}~\msunyr$ 
dramatically changes the protostellar structure, 
causing the abrupt expansion of the star
\citep[e.g.,][]{HOY12,Hosokawa13,Schleicher13}.
Since a star's UV emissivity strongly varies with its
effective temperature or radius, the resulting UV 
feedback expected for accretion bursts potentially show 
a new feature during the accretion phase.

%------------------------ in this paper --------------------------%

Here, we present the results of coupled SE-RHD numerical simulations 
(SE in 1D; RHD in 3D) to follow the stellar growth and evolution
with variable mass accretion, including the effects of stellar UV feedback.
As in our previous studies HS11 and HR14, we employ a stellar
evolution code to numerically solve the interior structure 
under the assumption of spherical symmetry and mass accretion. 
We follow the long-term ($\sim$$10^5$~years) evolution 
of the surrounding gas
including the effects of both the hydrogen-ionizing (EUV: $h \nu \geq 13.6~{\rm eV}$) 
and hydrogen-photodissociating (FUV: $11.2~{\rm eV} \leq h \nu \leq 13.6~{\rm
eV}$) feedback. As shown below, our current calculations 
suggest a wide range of final stellar masses, analogous to
the results of previous studies \citep[e.g., HR14, HR15;][]{Susa14}. 
We show that, as in previous studies, 
UV feedback can efficiently halt mass 
accretion for ordinary massive ($M_* \lesssim 100~\msun$) stars.
On the other hand, it is also possible to form 
very massive ($\gtrsim 250~\msun$) stars, whose 
observational signature might have been discovered in 
the Galactic metal-poor star SDSS J001820.5-093939.
\citep{Aoki14}.
In direct contrast to popular belief, 
disk fragmentation does not necessarily reduce the final
stellar mass; episodic accretion and/or
successive mergers can actually enhance stellar mass growth.
For these cases, the rapid and sometimes extreme evolution of the stellar 
radius with variable mass accretion causes 
the recurrent extinction and formation of \hii regions.
Such intermittent UV feedback does not efficiently halt 
gas accretion onto the protostar until very massive 
stars finally form. 
We also stress that performing the stellar evolution calculations 
together with 3D RHD simulations in realtime enables us to follow 
the evolution consistently. This is an essential feature of our SE-RHD
code.

%---------------------------------------------------------------%

The rest of the paper is organized as follows. 
We first describe the numerical methods and the cases 
considered in Sections~\ref{sec:method} 
and \ref{sec:cases}.
The main results are presented in Section~\ref{sec:results}. 
We discuss the implications of our results in 
Section~\ref{sec:discussion} and summarize our
conclusions in Section \ref{sec:conclusions}.

%%%%%%%%%%%%%%%%%%%%%%%%%%%%
\section{Numerical Method}
\label{sec:method}
%%%%%%%%%%%%%%%%%%%%%%%%%%%%

%---------------------- overviewing ------------------------%

The evolution in the protostellar accretion stage is followed 
with a hybrid SE-RHD code by solving the 3D RHD of the 
accretion flow in the vicinity of a protostar simultaneously
with the evolution of the protostar inside a sink cell of
the RHD grid. This approach is basically the same 
as that developed by HS11. 
The dynamics of the accretion flow are followed by performing 
a 3D RHD simulation, whereby for each time step
the characteristics of the protostar (mass, radius, 
intrinsic luminosity, accretion luminosity, FUV flux, EUV flux) 
provide boundary conditions for the RHD 
code at the inner edge of the sink cell, necessary for the 
radiation transfer, heating/cooling and chemistry modules of the RHD code.
The RHD simulation in turn provides the time-dependent 
mass accretion rate into the sink cell.
The protostar's evolution is calculated
by numerically solving the interior structure with the
additional feature of mass accretion onto the stellar surface
\citep[e.g.,][]{KY13,VB15}.
We describe the methodologies for the RHD simulations
and for stellar evolution calculations in
Sections~\ref{ssec:rhd} and \ref{ssec:stellar} respectively.

%+++++++++++++++++++++++++++++++++++++++++++++++++++++%
\subsection{Radiation Hydrodynamics of Accretion Flow}
\label{ssec:rhd}
%+++++++++++++++++++++++++++++++++++++++++++++++++++++%

%------------------------------------------------------------------------------%
\begin{table*}[t]
\label{tb:cmd}
\begin{center}
Table 1. Cosmological Cases Considered \\[3mm]
{\scriptsize 
\begin{tabular}{lcccccc}
\hline
\hline
Cases & $N_R \cdot N_\theta \cdot N_\phi$
      & $r_{\rm max}$ ($10^4$~AU) & $M_{\rm g, tot}$ ($\msun$) 
      & $\Delta t$ ($10^4$~yr) & $M_{\rm *, 3D}$ ($\msun$)  
      & $M_{\rm *, 2D}$ ($\msun$) \\
\hline
A     &  $96 \cdot 32 \cdot 64$  & 150 & $1.0 \times 10^4$ & 7
		     & 462.4$^\ast$ & 751.3\\
B     &  $96 \cdot 32 \cdot 64$  & 150 & $1.3 \times 10^4$ & 7
		     & 598.8$^\ast$ & 283.9 \\
B-NF$^\dagger$  &   $96 \cdot 32 \cdot 64$ & 150 &  & 7 & \\
B-NF-HR2$^\ddag$-m0$^\star$ &  $128 \cdot 64 \cdot 128$ & 4 & $1.1
	     \times 10^3$ & 0.3 & \\
B-NF-HR2-m40    &  $128 \cdot 64 \cdot 128$ & 4 & & 0.3 & \\
B-NF-HR2-m70    &  $128 \cdot 64 \cdot 128$ & 4 & & 0.3 & \\
B-NF-HR2-m120   &  $128 \cdot 64 \cdot 128$ & 4 & & 0.3 & \\
B-NF-HR4$^\ddag$-m20  &  $192 \cdot 128 \cdot 256$ & 0.67 & & 0.1 & \\
B-NF-HR4-m20-fc$^\diamond$  & $192 \cdot 128 \cdot 256$ & 0.67 
                            & & 0.1 & \\
B-NF-HR4-m20-lsk$^\bullet$  & $192 \cdot 128 \cdot 256$ & 0.67 
                            & & 0.1 & \\
C     & $64 \cdot 32 \cdot 64$ & 4 & $1.0 \times 10^3$ & 12
      & 286.0 & 160.3 \\
C-HR2-m0      &  $128 \cdot 64 \cdot 128$ & 4 &   & 5.5 & \\ 
D     & $64 \cdot 32 \cdot 64$ & 4 & 490 & 8 & 67.4 & 55.5 \\
D-NF  & $64 \cdot 32 \cdot 64$ & 4 &  & 8 & \\
D-DF$^\dagger$  & $64 \cdot 32 \cdot 64$ & 4 &  & 8 & \\
E     & $96 \cdot 32 \cdot 64$ & 150  & $2.7 \times 10^3$ & 6
		     & 14.3 & 24.4 \\
\hline
\end{tabular}
}
\noindent
\end{center}
Col.\ 2: cell numbers, Col.\ 3: position of the 
radial outer boundary, Col.\ 4: total mass contained in the
 computational domain, Col.\ 5: time duration followed 
Col.\ 6: final stellar masses in the current work,
Col.\ 7: final stellar masses obtained in 2D SE-RHD simulations \citep{Hirano14} 
\\
\\
$^\ast$: Stellar masses at the end of the simulations, 
        $t = 7 \times 10^4$ years. \\ 
$^\dagger$: Suffix ``NF'' represent casess with no UV 
              feedback and ``DF'' with only dissociation (FUV) feedback. \\  
$^\ddag$: Suffix ``HR2'' represents cases with
            doubled and ``HR4'' with quadrupled spatial resolutions. \\
$^\star$: Suffix ``mX'' represents the points for the restart 
            after doubling the resolution when $M_* \simeq {\rm X}~\msun$. \\
$^\diamond$: Stringent limit for the cooling with $f_{\rm limit} = 24$ 
             (see Section~\ref{ssec:flimit}). \\
$^\bullet$: Larger sink cell with $r_{\rm min} \simeq 52$~AU
             ($r_{\rm min} = 30$~AU for all other cases).
\end{table*}
%---------------------------------------------------------------------------------%

%--------------------------------------------------------------%
\subsubsection{Simulation Overview and Numerical Configuration}
\label{sssec:numconfig}
%--------------------------------------------------------------%

%----------------------- PLUTO code --------------------------%

We use a modified version of the public 
multi-dimensional magneto-hydrodynamics
code {\tt PLUTO} 3.0 \citep[][]{Mignone07}. 
The original code had been modified to study
present-day high-mass star formation by implementing
the sink cell philosophy with the embedded stellar evolution
code outlined above, as well as self-gravity and radiation transfer modules
\citep[e.g.,][]{Kuiper10aa,Kuiper10,Kuiper11,KK13}.
For this investigation we added several physics modules necessary 
to study primordial star formation.

%----------------------------- basec eqs. -------------------------------%

The governing equations of hydrodynamics are 
\begin{equation}
\frac{\partial \rho}{\partial t} + \nabla \cdot (\rho \vvec) = 0 ,
\end{equation}
\begin{equation}
\frac{\partial (\rho \vvec)}{\partial t} 
+ \nabla \cdot ( \rho \vvec \otimes \vvec ) =
- \rho \nabla \Phi - \nabla p ,
\label{eq:eom}
\end{equation}
\begin{equation}
\frac{\partial e}{\partial t} + \nabla \cdot (e \vvec)
= - p \nabla \cdot \vvec + \Gamma - \Lambda ,
\end{equation}
\begin{equation}
p = (\gamma -1) e,
\end{equation}
where $\rho$ and $p$ are the gas mass density and pressure,
$\vvec$ the 3D velocity vector, 
$\Phi$ the gravitational potential, $e$ the gas thermal energy density,
$\Gamma$ and $\Lambda$ the heating and cooling rates per volume respectively.
We allow the adiabatic exponent $\gamma$ to vary depending on
the chemical abundances and gas temperature \citep[e.g.,][]{ON98}.  
We calculate the energy source term $\Gamma - \Lambda$
with the same thermal processes as considered in HS11
(Table S1 in HS11, but see Section~\ref{sssec:rad} 
below for some differences). 
Note that equation (\ref{eq:eom}) does not include the viscosity terms 
which enabled angular momentum transport under 2D axial symmetry
\citep[e.g., $\alpha$-viscosity,][]{SS73}.
Instead, gravitational torques induced by the non-axisymmetric disk structure
(e.g., spiral arms, clumps) operate in 3D.

%---------------------------- grid configuration --------------------------------%

As in \citet{Kuiper11}, we solve the above equations 
in spherical coordinates.
The spatial grids for the polar ($\theta$) and azimuthal ($\phi$)
directions are homogeneously distributed over $0 \leq \theta \leq \pi$
and $0 \leq \phi \leq 2 \pi$. 
The angular resolutions are thus $\Delta \theta = \pi / N_\theta$
and $\Delta \phi = 2 \pi / N_\phi$, where $N_\theta$ and $N_\phi$ 
are the grid numbers. We set $\Delta \theta = \Delta \phi$ 
by taking $N_\phi = 2 N_\theta$ (Table 1).
The radial grid size $\Delta r$ logarithmically increases 
with increasing $r$ as
\begin{equation}
\Delta r (r) = r (10^f - 1),
\end{equation}
where $f \equiv \log(r_{\rm max}/r_{\rm min})/N_r$, $N_r$ is the number of radial grid cells,
and $r_{\rm min}$ ($r_{\rm max}$) is
the radial inner (outer) boundary of the computational domain.

%----------------------------- central sink ---------------------------------%

In order to avoid very short timesteps required for the dense gas
near and within the star, we do not resolve the central structure.
We instead assume a fixed central spherical sink cell 
of radius $r_{\rm min}$, within which a single star grows. 
The sink cell is assumed to be semi-permeable, meaning
accretion flow into the sink is allowed, but there is no outflow. 
We evaluate mass accretion rates onto the star by measuring
the gas inflow rates into the sink
(also see Section~\ref{ssec:stellar}).
We set the sink radius $r_{\rm min} = 30$~AU for our fiducial cases. 
Although the radius of an accreting protostar is generally
much smaller than this sink size when 
$\mdot \lesssim 10^{-2}~\msunyr$ \citep[e.g.,][]{OP03},  
an accreting star expands rapidly when
$\mdot \gtrsim 10^{-2}~\msunyr$, and inflates
up to red supergiant radii. \citet{HOY12} derive the mass-radius relationship 
for such a ``supergiant protostar'' accreting material above the critical rate,
\begin{equation}
R_* (M_*) \simeq 11.14~{\rm AU} \left( \frac{M_*}{100~\msun}
				\right)^{1/2},
\label{eq:mrsgp}
\end{equation}
which is comparable to our sink cell size. 

%------------------------------------------------------------------%

In the following, we will see that accreting protostars enter the supergiant 
stage for many of the examined cases.
Even with the sink size $r_{\rm min} = 30$~AU, our spherical coordinate
system offers relatively high spatial resolution in the vicinity of the sink, 
$(\Delta r \times r \Delta \theta)_{\rm min} 
\simeq 3.57~{\rm AU} \times 3.12~{\rm AU}$ for fiducial cases.
This is even better than the finest resolution used in HS11 $\sim$$10$~AU, 
so that the disk scale height is always resolved with several 
grid cells even in the innermost regions. 

%----------------------------- gravity ----------------------------------%

With this sink-cell technique, 
the gravitational potential $\Phi$ has two components:
one arising from the central star $\Phi_*$ and one from
the gas' self-gravity $\Phi_g$,
\begin{equation}
\Phi = \Phi_* + \Phi_g . 
\end{equation}
The specific gravitational force due to the central star is 
\begin{equation}
%\avec_{\rm grav} = 
- \nabla \Phi_* = - \frac{G M_*}{r^2} \evecr ,
\end{equation}
where $\evecr$ is the radial unit vector. The potential for the
gas' self-gravity $\Phi_g$ is obtained by solving Poisson's equation
\begin{equation}
\Delta \Phi_g = 4 \pi G \rho .
\end{equation}
We make use of the Poisson solver developed in \citet{Kuiper10}, 
which has been well tested and used with the {\tt PLUTO} code.

%----------------------------------------------------------------%
\subsubsection{Spatially Resolving the Jeans Length}
\label{ssec:flimit}
%----------------------------------------------------------------%

\citet{Truelove97} have suggested that, in order to avoid 
artificial fragmentation in numerical simulations, the Jeans length
has to be spatially resolved by at least four cells.  
Recent studies argue that even higher resolutions are needed 
to correctly capture the gravity-driven turbulence
that generally appears in the star formation process
\citep[e.g.,][]{Federrath11, Turk12, Meece14},
suggesting that the Jeans
length should be resolved with 32-64 cells. 
Unfortunately, it is computationally prohibitive
to consider the long-term ($\sim$$10^5$ years) evolution 
while fully satisfying this condition. 
We adopt the following ansatz instead:

%---------------------------------------------------------------------%

For each cell and each timestep, we calculate the local Jeans length 
$\lambda_{\rm J}$ and monitor the ratio of $\lambda_{\rm J}$
to the greater of $\Delta r$ and $r \Delta \theta$.
When this ratio falls below a certain fixed number $f_{\rm limit}$, 
we artificially squelch the cooling function
in the energy equation, i.e.,
\begin{equation}
\Lambda \to 0 \quad {\rm for} \quad 
\frac{\lambda_{\rm J}}
{\max(\Delta r, r \Delta \theta)} < f_{\rm limit}.
\label{eq:flimit}
\end{equation}
We normally take $f_{\rm limit} = 12$ as our fiducial value.
In practice, for a smooth transition, we evaluate the ratio 
$\xi \equiv f_{\rm limit} \max(\Delta r, r \Delta \theta)/\lambda_{\rm J}$
and multiply the cooling rate by a suppressing factor 
when $\xi$ exceeds unity, 
\begin{equation}
C_{\rm limit} = \exp 
\left[  - \left( \frac{\xi - 1}{0.1} \right)^2 \right], 
\end{equation}
which falls below $10^{-4}$ for $\xi \gtrsim 1.3$.

%------------------------------------------------------------------%

With this procedure, the limit of radiative 
cooling varies across the grid due to differing spatial resolutions.
Choosing a finer grid resolution $\max(\Delta r, r \Delta \theta)$
overall means that the condition in equation (\ref{eq:flimit}) is
satisfied for fewer cells and cooling operates without suppression
for a larger number of cells. 
Because the disk becomes more unstable with
more efficient cooling, disk fragmentation
occurs more readily in simulations at higher grid resolution, 
which is also seen in other studies \citep[e.g.,][]{MD14}.
We study effects of varying the spatial resolution 
in Section~\ref{ssec:sres} and
show that, contrary to the predictions of previous studies, 
more fragmentation does not necessarily reduce the
final stellar masses, but rather facilitates the formation of massive
stars by inducing intermittent UV feedback more often.

%----------------------------------%
\subsubsection{Radiation Transfer}
\label{sssec:rad}
%----------------------------------%

\paragraph{Stellar UV radiation}

We consider different kinds of radiation which play different roles 
in our simulations. The stellar UV flux is the primary source of UV
feedback which potentially halts the mass accretion onto 
stars and fixes the final stellar masses. 
We solve the transfer of ionizing (EUV) photons 
emitted by a protostar with a frequency-averaged approximation, 
\begin{equation}
\frac{1}{r^2} \frac{\partial (r^2 F_{\rm EUV})}{\partial r}
= -n (1-x) \sigma_{\rm EUV} F_{\rm EUV} ,
\label{eq:euvtr}
\end{equation}
where $F_{\rm EUV}$ is the photon number flux above the Lyman limit, 
$n$ the particle density,
$x$ the degree of ionization, and $\sigma_{\rm EUV}$ the photoionization 
cross section per particle, which is a function of the 
stellar effective temperature $T_{\rm eff}$.
Although the frequency-dependent UV transport can somewhat broaden
primordial ionization fronts, which can enhance the thin-shell 
instabilities in the fronts \citep[][]{WN08}, 
we do not expect that such a effect largely modifies the dynamics
especially with dense media in the vicinity of the star.
Equation (\ref{eq:euvtr}) is easily integrated
\begin{equation}
F_{\rm EUV} = \frac{S_{\rm EUV}}{4 \pi r^2} \exp(- \tau_{\rm EUV}) ,
\label{eq:feuv}
\end{equation}
where $S_{\rm EUV}$ is the stellar EUV emissivity, and
$\tau_{\rm EUV}$ the optical depth given by
\begin{equation}
\tau_{\rm EUV} = \int_{r_{\rm min}}^r n (1-x) \sigma_{\rm EUV}~dr' 
\label{eq:taueuv}
\end{equation} 
(We implicitly assume no UV absorption within the sink cell).

%------------------------------------------------------------------%

We solve equations (\ref{eq:feuv}) and (\ref{eq:taueuv})
making use of the adopted spherical grid 
(Section~\ref{sssec:numconfig}), i.e., along the radial 
cells having the same angular coordinate $(\theta_j, \phi_k)$.  
The obtained EUV flux $F_{\rm EUV}$ is used to provide the
photo\-ionization rate in the chemical rate equations and the
associated heating rate in the energy equation. We consider
the ionization of hydrogen only, treating helium as ``hydrogenic'',
i.e., singly ionized within an \hii region.
For the purposes of our investigation, this approximation is 
not critical, but certainly could influence the precise numerical
values of our results.

%---------------------- diffuse component ------------------------%

Within an \hii region there is a component of
diffuse EUV radiation emitted by the recombining gas,
namely recombination of hydrogen directly into the ground state
and recombinations of helium into the lowest states.
HS11 have separately solved the transfer of diffuse
hydrogen recombination photons using the flux-limited diffusion (FLD) 
approximation \citep{LP81,RY97}. 
However, UV feedback on the accreting material is mostly caused
by the direct EUV stellar flux \citep[also see][]{TKei13}. Thus,
we only solve for the transfer of the direct stellar EUV component
and use the ``on-the-spot'' approximation (each recombination of
hydrogen directly into the ground state causes the immediate ionization
of a hydrogen atom in the vicinity). In praxis this is done by
using a hydrogen recombination coefficient that
excludes recombinations directly into the ground state.

%----------------------------- FUV ---------------------------------%

As for the FUV radiation transfer, we only consider the
stellar direct component, by exactly the same method
using the self-shielding function \citep[][]{DB96} as in HS11. 
We recognize that by excluding the diffuse components of EUV and FUV, 
one will encounter "shadowing" effects caused by UV-opaque
clumps and the accretion disk itself. Our previous comparison 
studies with the 2D SE-RHD code have shown that this simplification 
does not significantly change the accretion history or final 
masses of the stars formed, which is the primary goal of this
investigation.

\paragraph{Molecular hydrogen line emission}

We follow the method used in HS11 for calculating the
cooling rate via H$_2$ line emission; we sum up the contributions
from all ``allowed'' quadrupole
transitions between the rotational and vibrational
levels of $J = 0-20$ and $v=0-2$. 
To estimate the trapping effect for each transition, we 
approximate its optical depth $\tau$ by the local
Jeans length $\lambda_{\rm J}$, i.e., $\tau = \alpha \lambda_{\rm J}$,
where $\alpha$ is the transition's absorption coefficient.
This is a crude approximation compared 
to recently developed novel 
(and much more computationally expensive) methods 
\citep[e.g.,][]{Clark12,Hirano13,Greif14,Hartwig15}.
In our current simulations, however, it is not reasonable to
model the detailed thermal disk structure, because of 
our limitations in following the exact thermal state of the
densest gas (Section~\ref{ssec:flimit}). 
For a given spatial resolution, our treatment underestimates the 
cooling and thus leads to less gravitationally unstable disks.
It therefore affects the quantitative behavior of disk fragmentation,
such as the exact number and sizes of the emerging fragments. 
To assess the impact of the details of disk fragmentation on our
final results,
we vary our spatial resolution (see Section~\ref{ssec:sres}).

\paragraph{Continuum emission}

As described in HS11, continuum cooling via H$^-$ free-bound
emission is important for the gas infalling onto the disk surface.
HS11 have implemented this cooling process by solving continuum 
radiation transfer using the FLD approximation.
In their 2D simulations, however, the H$^-$ continuum emission is
always optically thin throughout the evolution. 
This knowledge allows us to use a simpler approximation for the
current 3D simulations;
we simply subtract the H$^-$ bounding energy 0.755~eV for each formation
reaction H + e $\to$ H$^-$ + $\gamma$ (R10 in Table~2) from the
gas' internal energy.
We have also implemented this method in our 2D SE-RHD 
code and confirmed the validity of this approximation.

%------------------------- H2 continuum emission ----------------------------%

We do not include the cooling via H$_2$ collision-induced 
continuum emission \citep[CIE,][]{ON98}. 
This is because CIE only becomes important for the
dense molecular gas with $n \gtrsim 10^{13}~\cmc$, 
which only occasionally occurs in our simulations. 
Such dense gas normally lies close to the central star, 
primarily within the sink cell. 
Only with the enhanced spatial resolution considered in 
Section~\ref{ssec:sres} below, however, does the density 
attain $\sim$$10^{14}~\cmc$ in the
densest parts of disks. 
Note that we do not pursue the exact thermal state of such
tiny dense parts anyway, because of our inability to follow the 
long-term evolution at sufficiently high spatial resolution
(see discussion in Section~\ref{ssec:flimit}).

%-------------------------------------------------%
\subsubsection{Chemical Network for Primordial Gas} 
\label{sssec:chemistry}
%-------------------------------------------------%

%---------------------------------------------------------------------------------%
\begin{table}[t]
\label{tb:chem}
\begin{center}
Table 2. Chemical Reactions Included \\[3mm]
\begin{tabular}{llc}
\hline
\hline
No. & Reactions & References \\
\hline
R1 &  H      +  e  $\rightarrow$ H$^+$ + 2 e        &  1 \\
R2 &  H$^+$  +  e  $\rightarrow$ H     + $\gamma$   &  2 \\
R3$^\dag$ &  H$^-$  +  H  $\rightarrow$ H$_2$ + e   &  3 \\
R4 &  H$_2$  +  H$^+$ $\rightarrow$ H${_2}^+$ + H   &  4 \\
R5 &  H$_2$  +  e     $\rightarrow$ 2 H + e         &  4 \\
R6$^\dag$ &  H$_2$  +  H  $\rightarrow$  3 H        &  5 \\
R7$^\dag$ &  3 H          $\rightarrow$  H$_2$ + H  &  6 \\
R8 &  2 H  + H$_2$ $\rightarrow$  2 H$_2$           &  7 \\
R9 &  2 H$_2$      $\rightarrow$  2 H   + H$_2$     &  7 \\
R10 & H    + e     $\rightarrow$  H$^-$ + $\gamma$  &  4 \\  
R11 & H + $\gamma$ $\rightarrow$  H$^+$ + e         &  2 \\
R12 & H$_2$ + $\gamma$ $\rightarrow$ 2 H            &  8, 9 \\ 
R13 & 2 H          $\rightarrow$  H$^+$ + e + H     &  7 \\ 
R14 & H$^-$ + e    $\rightarrow$  H + 2 e           &  1 \\
R15 & H$^-$ + H$^+$ $\rightarrow$ 2 H               &  4 \\
R16$^*$ & H$^-$ + H$^+$ $\rightarrow$ H$_2^+$ + e       & 4 \\
R17$^*$ & H$^-$ + $\gamma$ $\rightarrow$ H + e          & 10 \\
R18$^*$ & H + H$^+$ $\rightarrow$ H${_2}^+$ + $\gamma$  & 4 \\
R19$^*$ & H${_2}^+$ + H $\rightarrow$ H$_2$ + H$^+$     & 4 \\
R20$^*$ & H${_2}^+$ + e $\rightarrow$ 2 H               & 4 \\
R21$^*$ & H${_2}^+$ + H$^-$ $\rightarrow$ H$_2$ + H     & 11 \\
\hline
\end{tabular}
\noindent
\end{center}
$^\dag$ reaction rates updated with respect to HR14, \\
$^*$ additional reactions added with respect to HR14 \\
\\
REFERENCES: (1) \citealt{Abel97}, (2) \citealt{OF06}, 
(3) \citealt{Kreckel10}, (4) \citealt{GP98}, (5) \citealt{Martin96},
(6) \citealt{Forrey13}, (7) \citealt{PSS83}, (8) \citealt{TH85},
(9) \citealt{DB96}, (10) \citealt{John88}, (11) \citealt{Millar91}
%\tablecomments{}
\end{table}
%------------------------------------------------------------------------------% 

%--------------------- current network -------------------%

We have implemented an updated version of the primordial chemistry
network used in HS11 and HR14 to the {\tt PLUTO} code.
Here, we solve for the non-equilibrium chemical abundances of 
H, H$_2$, H$^+$, e, H$^-$, and H${_2}^+$
without the approximation of using 
the equilibrium value for H$^-$ \citep[as employed by e.g.,][HS11]{Abel97}. 
The reactions included are summarized in Table~2.

%------------------- without D-chemistry --------------------%

As described in Section~\ref{ssec:cases_cosmo}, HD molecules 
will form during the early collapse in some cases. 
The resulting HD molecular cooling reduces the temperature
and density in the accretion envelope, which in turn
leads to a somewhat slower initial protostellar growth
\citep[e.g.,][HS12]{Y07}.
However, this effect ends during the early collapse
by the time $n \sim$$10^8~\cmc$ and is unimportant thereafter.
While we include deuterium chemistry for cosmological simulations
following the early collapse stage (see Section~\ref{ssec:cases_cosmo}
below), we do not in the current network for the accretion stage.  
We have checked with our previous 2D SE-RHD code with the same setting, 
starting after the density exceeds $\sim$$10^{10}~\cmc$
as in HS12, that omitting the deuterium chemistry hardly changes the results.

%++++++++++++++++++++++++++++++++++++++++++%
\subsection{Stellar Evolution Calculations}
\label{ssec:stellar}
%++++++++++++++++++++++++++++++++++++++++++%

%--------------------------- basic eqs ---------------------------%

We follow the evolution of the central protostar by solving
the four stellar structure equations taking account of 
mass accretion \citep[e.g.,][]{SST80, SPS86}:
\begin{equation}
\left( \frac{\partial r}{\partial M} \right)_t = \frac{1}{4 \pi \rho r^2},
\label{eq:con} 
\end{equation}
\begin{equation}
\left( \frac{\partial P}{\partial M}  \right)_t = - \frac{GM}{4 \pi r^4}, 
\label{eq:mom}
\end{equation}
\begin{equation}
\left( \frac{\partial L}{\partial M} \right)_t 
= \epsilon - T \left( \frac{\partial s}{\partial t}  \right)_M ,
\label{eq:ene}
\end{equation}
\begin{equation}
\left( \frac{\partial s}{\partial M} \right)_t
= \frac{G M}{4 \pi r^4} \left( \frac{\partial s}{\partial p} \right)_T
  \left( \frac{L}{L_s} - 1  \right) C ,
\label{eq:heat}
\end{equation}
where $M$ is the Lagrangian mass coordinate, $\epsilon$ the specific energy
production rate by nuclear fusion, $s$ the specific entropy,
and $L_s$ the radiative luminosity for an adiabatic temperature
gradient. The coefficient $C$ in equation (\ref{eq:heat}) is given by the
mixing-length theory for convective zones and is unity otherwise. 
Hydrogen burning via pp1-, pp2- and pp3-chains and via the CNO-cycle 
as well as helium burning via triple-$\alpha$ reactions are included.

%-------------------------- Henyey method ----------------------------%

In order to construct stellar models by solving the above equations,
we adopt the numerical code {\tt STELLAR},
originally developed by \citet{YB08}, which uses the Henyey method.
This differs from codes we used in HS11 and HR14, which
employ shooting methods to solve the same equations
\citep[e.g.,][]{OP03, HO09}. 
We have confirmed that, despite a number of technical differences, 
both numerical codes provide essentially the same results 
for various different accretion histories
\citep[e.g.,][]{HYO10,Hosokawa13}.
Here, we adopt {\tt STELLAR}, because the code offers
relatively stable convergence properties
for highly variable accretion rates. 
\citet{KY13} implemented {\tt STELLAR} as their stellar evolution
module in {\tt PLUTO} to study present-day high-mass star 
formation assuming 2D axial symmetry. 
We make use of their implementation by suitably modifying
the material functions (e.g., EOS, chemistry module, and opacity tables) for primordial 
star formation.

%------------------ evolutionary calculations ---------------------%

We start the stellar evolution calculations
after the sink mass exceeds $1~\msun$ via mass accretion,
assuming an initial $1~\msun$ polytropic star model
obtained by solving the Lane-Emden equation with the index $n=1.5$.
A time sequence of stellar models are constructed as the stellar mass 
subsequently increases at the rate provided by
the 3D RHD simulations.
Varying the characteristics of the initial model (mass, radius, entropy, etc.) 
only affects the early evolution long
before stellar UV feedback begins to operate 
\citep[e.g.,][]{HYO10}. 

%-------------- averaged accretion rates for SE --------------------%

For the SE calculations, 
we use accretion rates averaged over a short duration: 
$300$~years for the default cases and $30$~years for cases 
with enhanced spatial resolution.
This ``smearing'' of the instantaneous 
accretion rate measured at $r = r_{\rm min}$ simulates the
transport of material through the innermost part of the disk
enclosed within the sink. Such a smearing and delaying
timescale is given by the local viscous timescale, presumably
several times longer than the Kepler orbital
period,
\begin{equation}
 P_{\rm Kepler} \simeq 16.4~{\rm yr}
\left( \frac{r}{30~{\rm AU}} \right)^{3/2} 
\left( \frac{M_*}{100~\msun} \right)^{-1/2} .
\label{eq:pkep}
\end{equation}
Note that accretion variability considered below occurs
on much longer timescales than our assumed averaging timescale.

%------------------------ boundary condition -------------------------%

Mass accretion onto a protostar generally occurs through a circumstellar
disk with finite angular momentum. 
Effects of disk accretion for the stellar evolution 
calculations can be modeled by considering 
proper outer boundary conditions 
\citep[e.g.,][]{PS92}.
A simple way is adding the accreted gas to 
the stellar outermost zone with the same thermal state 
as in the atmosphere. 
This supposes an extreme ``cold'' disk accretion, for which 
the gas lands on the stellar surface gently, having
enough time to adjust its thermal state to that in the stellar atmosphere.
The flow from the disk only covers a small part of the surface, 
and the other part freely radiates into a vacuum. 
In reality, however, the gas newly joining the star
will carry some additional heat into the opaque interior. 
We approximate this ``warm'' accretion by allowing a fraction
of the accretion luminosity to be advected into the star 
\citep[e.g.,][]{Siess97}, 
\begin{equation}
L_{\rm *, acc} \equiv \eta L_{\rm acc} 
= \eta \frac{G M_* \dot{M_*}}{R_*} , 
\end{equation}
where $\eta$ is a parameter, which we normally set to $\eta = 0.01$. 
As shown in \citet{Hosokawa13}, however, the resulting stellar evolution
is almost independent of varying $\eta$, except at early stages.
Increasing $\eta$ from 0 to 1 simply leads to earlier bloating of the star.
Even for extreme ``cold'' disk accretion ($\eta = 0$), bloating 
occurs long before radiative feedback becomes significant.

%---------------------- calculating UV luminosity --------------------%

The stellar EUV and FUV emissivities are 
calculated assuming a black-body spectrum 
\begin{equation}
S_{\rm EUV} = 4 \pi R_*^2 \int_{13.6~{\rm eV}}^\infty
\frac{\pi B(T_{\rm eff})}{h \nu}~d \nu, 
\end{equation}
\begin{equation}
S_{\rm FUV} = 4 \pi R_*^2 \int_{11.2~{\rm eV}}^{13.6~{\rm eV}}
\frac{\pi B(T_{\rm eff})}{h \nu}~d \nu, 
\end{equation}
\citep[][HS11]{MT08}, where the effective temperature is given by
\begin{equation}
T_{\rm eff} =  \left( \frac{L_*}{4 \pi R_*^2 \sigma}  \right)^{1/4},
\label{eq:teff}
\end{equation}
where $\sigma$ is Stefan-Boltzmann constant.
We do not include the accretion luminosity
$(1 - \eta) L_{\rm acc}$ (see equation \ref{eq:teff}), which presumably
has spectral characteristics differing from a black-body at $T_{\rm eff}$.
This omission has little effect on our results, because the accretion
luminosity $L_{\rm acc}$ is generally much smaller than the stellar luminosity 
$L_*$ at the stage when UV feedback becomes effective. 
The contribution of the accretion luminosity may become large 
when the accretion rate suddenly rises as in a burst event. 
For these cases, however,
the protostar also rapidly inflates for peak rates exceeding
$\sim$$10^{-2}~\msunyr$. The stellar UV emissivity accordingly drops,
which diminishes the UV feedback
(also see Section~\ref{ssec:VMstar} below). For accretion rate
spikes below $\sim$$10^{-2}~\msunyr$, the accretion luminosity
does not significantly affect the total luminosity.

%%%%%%%%%%%%%%%%%%%%%%%%%%%%%%%%%%%%%%%%%%%%%%%%
\section{Cases Considered}
\label{sec:cases}
%%%%%%%%%%%%%%%%%%%%%%%%%%%%%%%%%%%%%%%%%%%%%%%%

%--------------------------------------------------------------------%
%%% Fig.1 %%%
\begin{figure*}
  \begin{center}
\epsscale{1.0}
\plotone{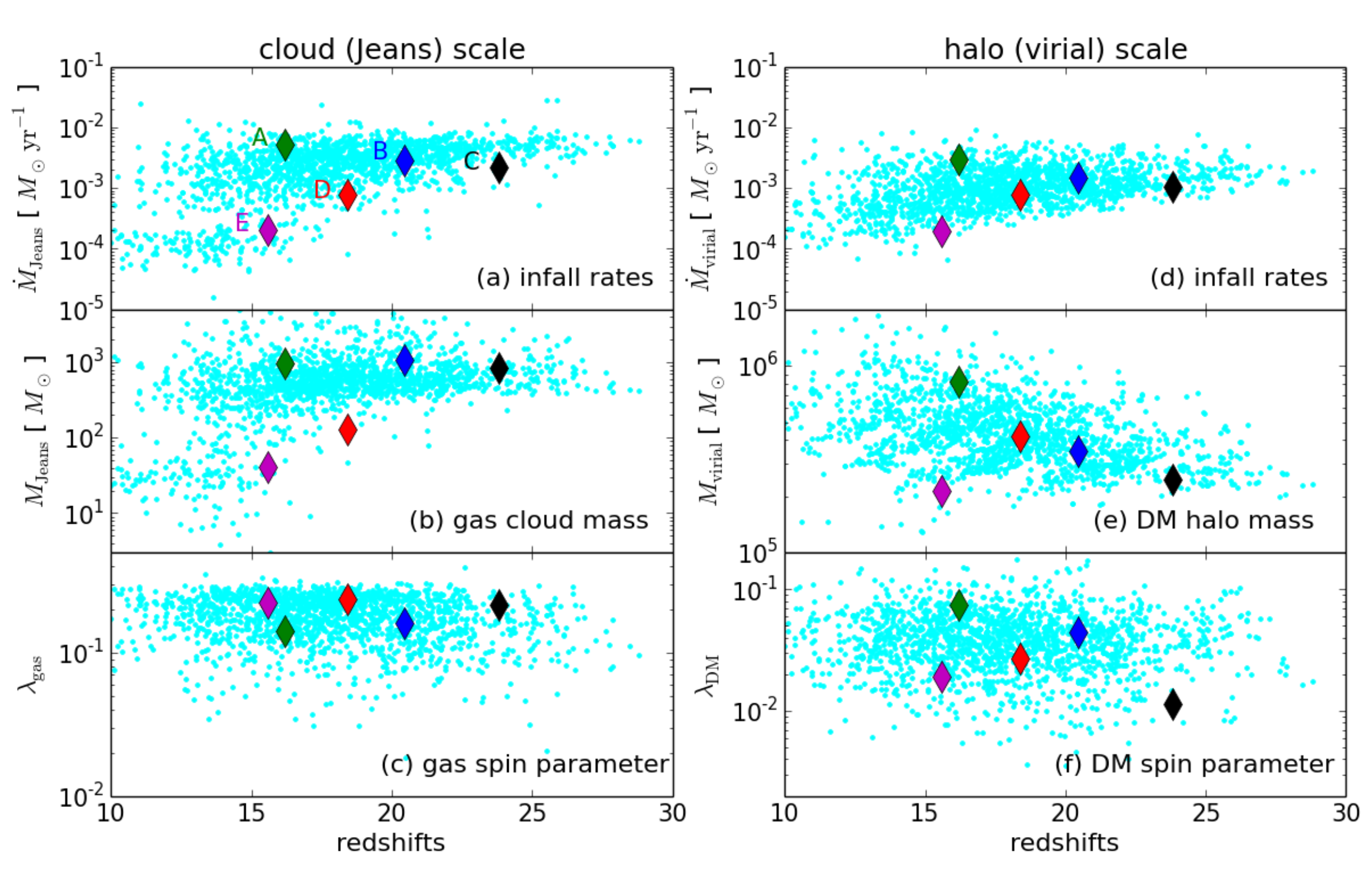}
\caption{
Physical properties of the cosmological cases
considered here (colored rhombuses denoting cases
A--E; see also Table~1) compared to the 
cosmological samples analyzed in \citet{Hirano15} (cyan dots).
{\sl Left column:} physical properties of gas clouds
measured at the self-gravitating cloud (Jeans) scale,
(a) infall rates, (b) masses, and (c) spin parameters 
as a function of the formation redshifts. 
{\sl Right column:} same as the left column but for the dark 
matter halo (virial) scale. 
The masses and spin parameters presented in (e) 
and (f) are only for the DM components.}
\label{fig:ini}
  \end{center}
\end{figure*}
%--------------------------------------------------------------------%

%++++++++++++++++++++++++++++++++++++++++++++++++++++++++++++++++++++++%
\subsection{Primordial Cloud Collapse in Cosmological Simulations}
\label{ssec:cases_cosmo}
%++++++++++++++++++++++++++++++++++++++++++++++++++++++++++++++++++++++%

We study the formation of primordial stars in a full
cosmological context. To this end, we first follow the formation
and gravitational collapse of a number of primordial clouds utilizing 
a modified version of 3D SPH/N-body code {\tt GADGET}-2 
\citep[e.g.,][]{Springel05, Y06, Hirano13}.
The procedures for this step are well described in our 
previous investigations HR14 and HR15, 
who in total have investigated the evolution of more than 
1600 clouds in a standard $\Lambda$CDM universe.
The hierarchical zoom-in and particle-splitting techniques have
been used to follow the cloud evolution with sufficiently high
spatial resolution in cosmological volumes.

%---------------------- remapping cosmological data ----------------------%

We focus on five representative cases A--E taken from 
the cosmological samples of HR14 and HR15. 
For each case, we followed 
the run-away collapse until the central density reaches 
$\simeq 10^{13-14}~\cmc$ (HR14). 
Here, we study the subsequent evolution 
in 3D, using the method described in Section~\ref{sec:method}.
For initial conditions, we remap the particle-based 
cosmological simulation data onto our 3D
spherical grid. The coordinate origin is set at the density maximum 
and we allow a single star to grow within the assumed sink,
normally $r < 30$~AU.  
The direction of the polar axis is taken to be that of the
angular momentum vector for the gas enclosed in the
inner region $r \leq 0.01 - 0.1$~pc.
With this choice, the circumstellar disk that is formed is
nearly perpendicular to the polar axis
(but see discussion on fluctuations of the disk's orientation
in Section~\ref{ssec:VMstar}). 
Table~1 summarizes several important parameters defining
the examined cases. 
In addition to the default cases A--E,
we also consider additional cases identified by suffixes 
for a number of different parameter settings and/or different 
grid resolutions.

%-----------------------------------------------------------------------%

HR14 have previously studied the evolution of the same cases 
with 2D SE-RHD simulations, modifying the initial conditions
to fit their axisymmetric grid configuration.
Table~1 summarizes their final stellar masses attained after 
UV feedback halted the mass accretion. 
HR14 demonstrate that the final stellar masses 
tend to be somewhat higher for higher accretion rates
averaged over $\sim$$10^5$~years after the formation of the protostars.
Since accretion histories are predestinated by
the envelope structure set in the early collapse stage,
higher stellar masses result from
a higher global infall rate $<$$4 \pi r^2 \rho v_r$$>$ 
during the collapse, 
where $v_r$ is the radial component of the velocity.
In particular, a good estimator of the final stellar mass can
be constructed using the infall rate $\dot{M}_{\rm J}$ 
measured at the self-gravitating
cloud (Jeans) scale, defined as the radius $r_J$ where
the ratio of the enclosed gas mass $M_{\rm enc}(r)$
to the the local Bonner-Ebert mass $M_{\rm BE}$ reaches its
maximum as a function of radius $r$, i.e., 
$M_{\rm enc}(r_J) / M_{\rm BE} = {\rm max}(M_{\rm enc}(r) / M_{\rm BE})$.
HR15 have derived a fitting formula relating the final
stellar masses and cloud-scale infall rates,
\begin{equation}
M_* = 250~\msun
\left( 
\frac{\dot{M}_{\rm J}}{2.8 \times 10^{-3}~\msunyr}  
\right)^{0.8},
\label{eq:mjcorr}
\end{equation}
where the infall rates are measured when the cloud central density
reaches $10^7~\cmc$.
Figure~\ref{fig:ini} shows the diversity of the cloud-scale infall
rates and relevant quantities: the masses and spin parameters 
at the cloud (Jeans) and halo (virial) scales.
We see that the five cases A--E considered here well cover the 
distribution of $\sim$$1500$ cases studied in HR15.

%---------------------------------------------------------------------%

The cloud-scale infall rates for cases A--E decrease in alphabetical order, 
as do the final stellar masses obtained in the 2D SE-RHD simulations 
(see Table 1).
HR15 show that, despite such diversity, the distribution of
stellar masses evaluated for the $\sim$$1500$ cases has a pronounced peak 
at $M_* \simeq 250~\msun$. 
This suggests the presence of a fiducial initial condition
for primordial star formation. Among our five cases, 
cases A--C are more representative of such a fiducial condition 
than cases D and E. 

%-------------------------- HD-cooling mode -------------------------------%

Case E provides the lowest final stellar mass 
$M_* \simeq 24.4~\msun$ in 2D. 
For this case, HD line cooling also operates in addition to H$_2$
cooling in the collapse stage. 
The temperature consequently drops to a few 
$\times~10$~K, limited by the
cosmic microwave background floor. 
HR14 show that the HD-cooling mode can 
be triggered when the collapse is somewhat slower than 
for purely spherical free-fall. Such a slow collapse can occur, for example, 
when rotational support decelerates the collapse. 
In the protostellar accretion stage, 
UV feedback typically limits the stellar masses to
a few tens of~$\msun$ (e.g., HS12; HR14).

%--------------------------------------------------------------------%
%%% Fig.2 %%%
\begin{figure}
  \begin{center}
\epsscale{1.15}
\plotone{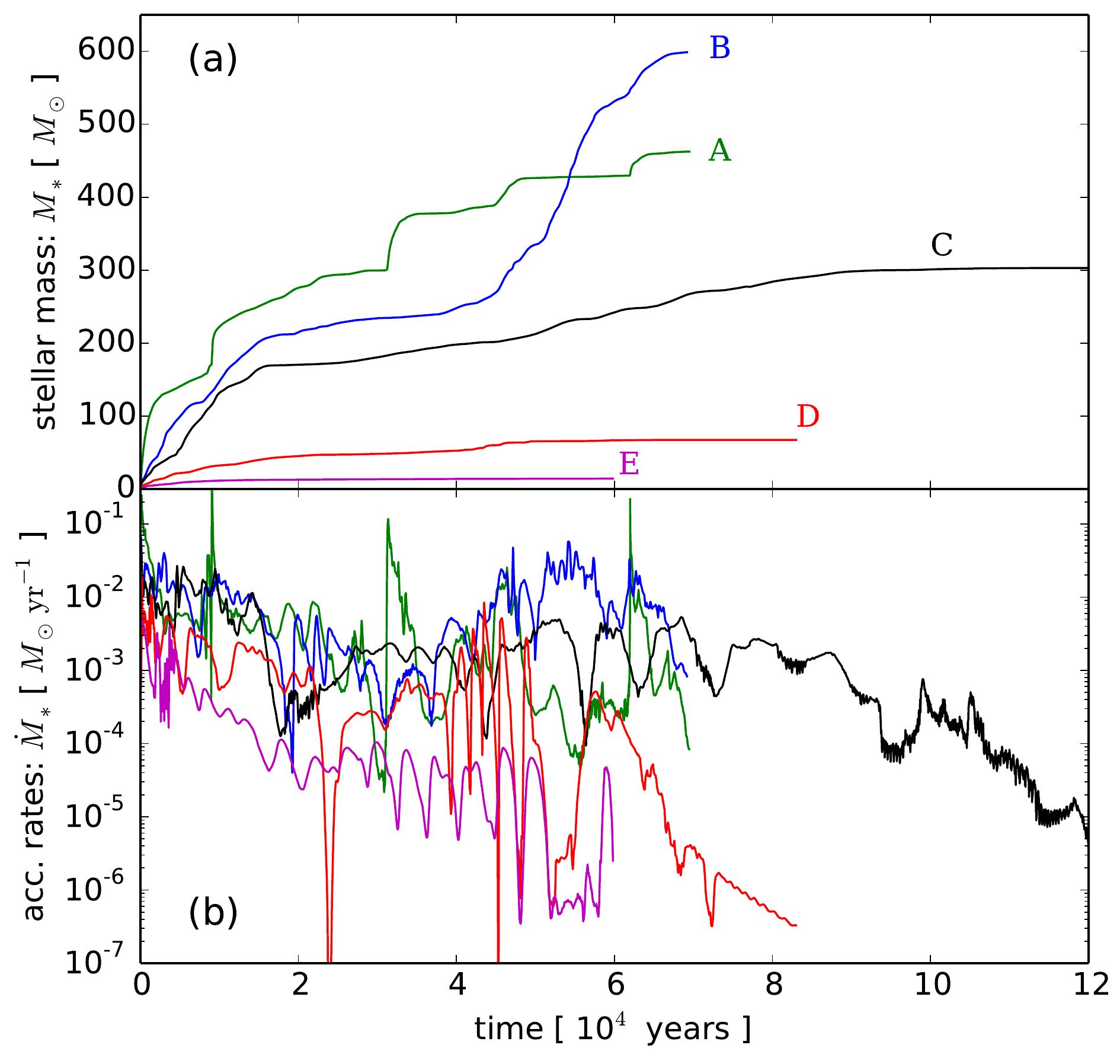}
\caption{Time evolution of the stellar masses (panel a)
and accretion rates (panel b) for cases A (green), B (blue), 
C (black), D (red), and E (magenta). 
The time $t = 0$ corresponds to the formation of embryo protostars.}
\label{fig:xm_xmdot_t}
  \end{center}
\end{figure}
%--------------------------------------------------------------------%

%------------------------ III.2 not examined ----------------------------%

We assume that the primordial clouds for cases
A--E do not experience any external feedback.
In reality, however, FUV radiation from nearby 
stars can alter the thermal evolution during
collapse and consequently the resulting final stellar masses 
\citep[e.g.,][]{OH02,OShea08}. 
HR15 show that, by considering the spatial distribution of
FUV fields created by Pop III stars at
different redshifts, these so-called Pop III.2 stars
 \citep[e.g.,][]{Bromm09,OShea08b} become comparable in number to 
the so-called Pop III.1 primordial stars, for which the external 
feedback is negligible. 
Interestingly, HR15 show that the final stellar masses
and cloud-scale infall rates still have the same correlation
among the different sub-populations of III.1 and III.2 stars.
We examine how this correlation can differ in 3D, 
focusing on the interplay between 
variable mass accretion and UV radiative feedback.

%----------------------------------------------------%

In addition to the cosmological cases described above, 
we perform simulations starting with idealized initial 
conditions: a homogeneous, zero-metallicity gas cloud in solid body rotation 
with constant density, temperature, and chemical composition.
The results of these idealized simulations are discussed in the Appendix.
These more simplified initial conditions allow us to test 
the effects of varying numerical settings in a controlled manner.

%%%%%%%%%%%%%%%%%%%%%
\section{Results}
\label{sec:results}
%%%%%%%%%%%%%%%%%%%%%

%-------------------------------%
\subsection{Overall Results}
\label{ssec:overall}
%-------------------------------%

%--------------------------------------------------------------------%
%%% Fig.3 %%%
\begin{figure}
  \begin{center}
\epsscale{1.0}
\plotone{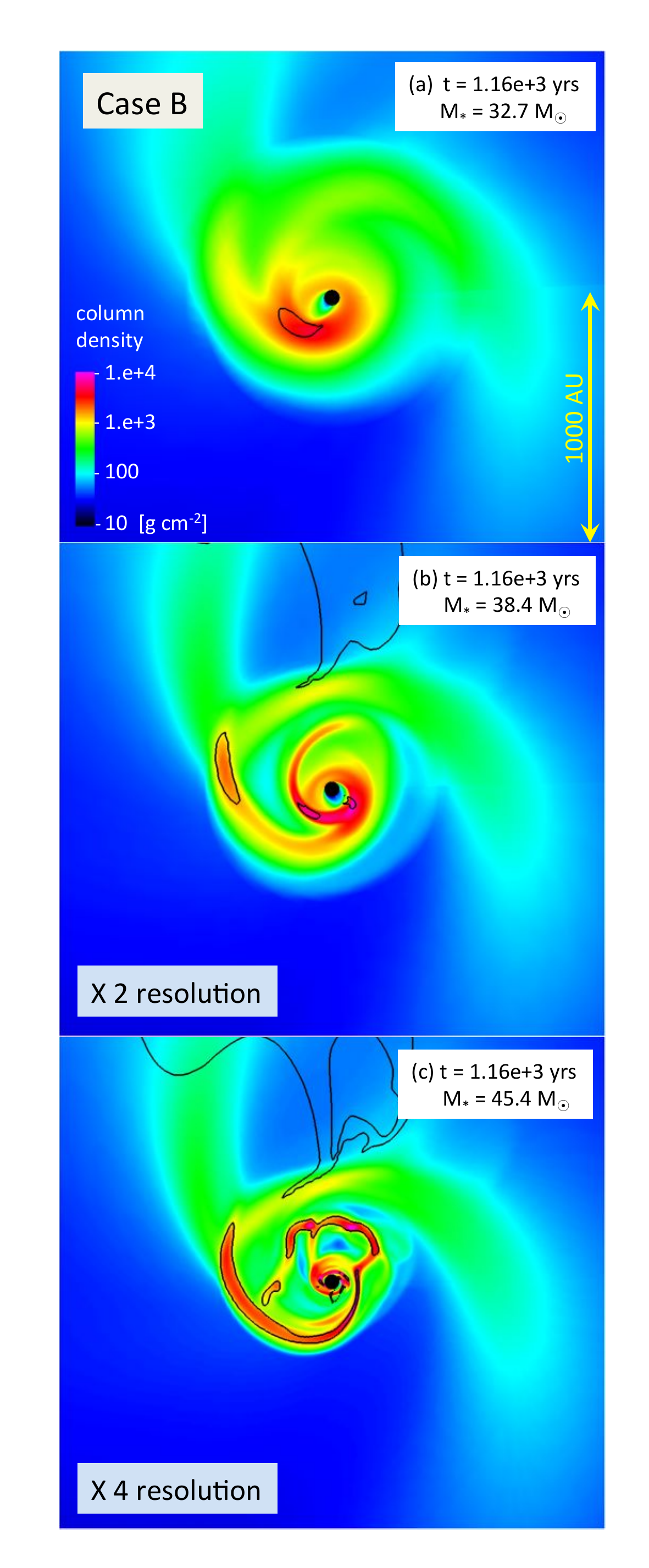}
\caption{Gas column density along the polar direction
of the self-gravitating circumstellar disks obtained
with different spatial resolution 
(cases B-NF, B-NF-HR2-m0, B-NF-HR4-m20 for 
panels a, b, and c, respectively; see text and Table 1).
The snapshots are shown at the same epoch 
$t = 1.16 \times 10^3$ years, but the accreted stellar
masses differ due to different prior accretion histories.
In each panel, the black contours mark the positions where the Toomre-Q
parameter takes the value of $Q = 0.3$.
}
\label{fig:caseB_disk_rescomparison}
  \end{center}
\end{figure}
%--------------------------------------------------------------------%
%--------------------------------------------------------------------%
%%% Fig.4 %%%
\begin{figure*}
  \begin{center}
\epsscale{1.0}
\plotone{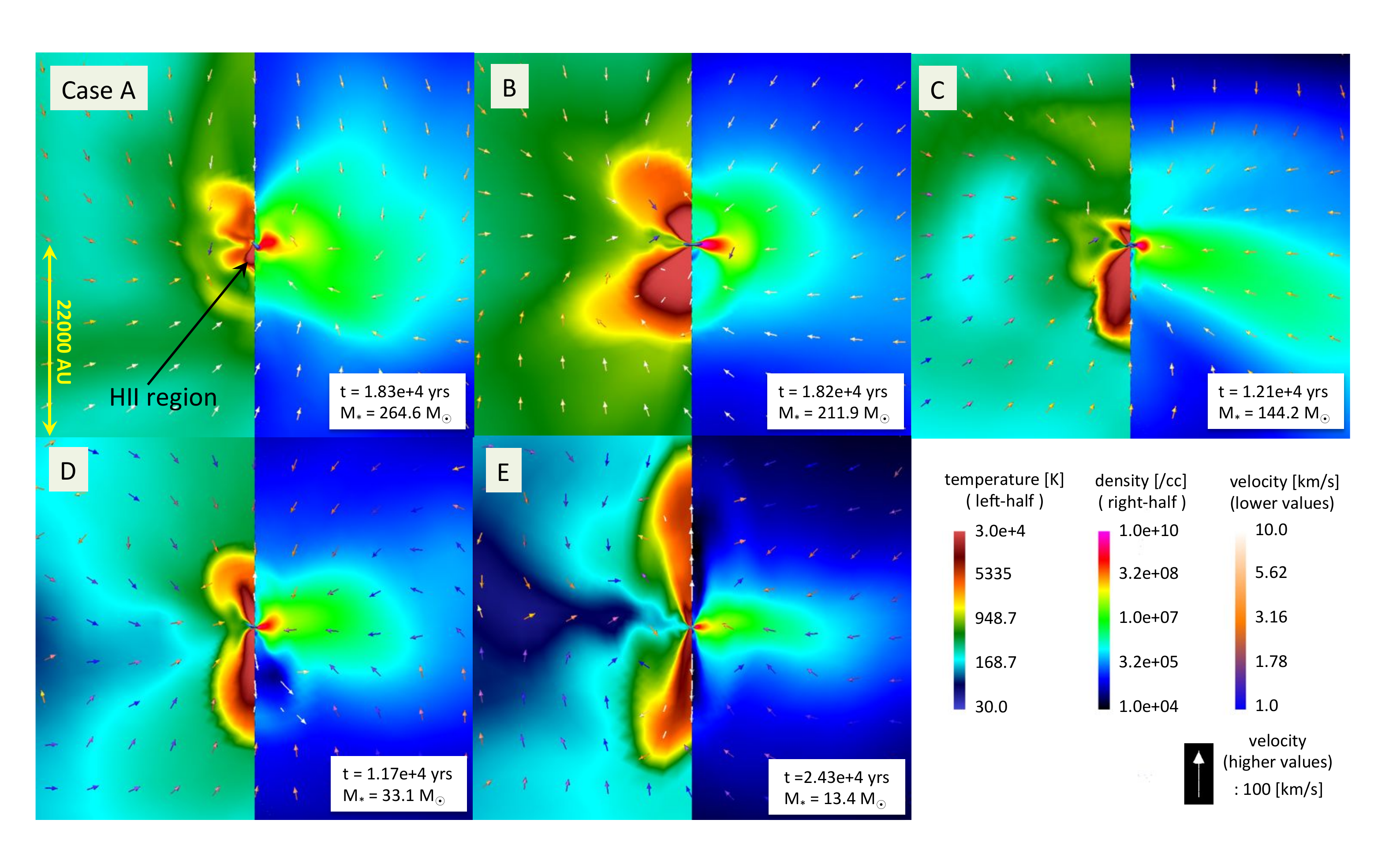}
\caption{Temperature, density, and velocity structure of cases A--E
in a plane containing the polar axis at the time when the polar 
extension of \hii regions first reaches $10^4$~AU. 
In each panel, the gas temperature is shown to the left and 
the density distribution to the right.
Velocity vectors $v < 10~{\rm km~s^{-1}}$ 
are depicted by small blue and orange arrows,
whereby the color gradation represents the magnitude.
Velocities $v > 10~{\rm km~s^{-1}}$ are depicted
by white arrows with lengths varying in proportion 
to the magnitude (see the legend in the lower-right panel). 
The small 3D magenta contour at the panel center delineates the
iso-density surface for $10^{10}~\cmc$, which roughly traces
an accretion disk around the central star. 
Each panel also shows the elapsed time and stellar mass.
}
\label{fig:HIIr1e4AU_A2E}
  \end{center}
\end{figure*}
%--------------------------------------------------------------------%
%--------------------------------------------------------------------%
%%% Fig.5 %%%
\begin{figure*}
  \begin{center}
\epsscale{1.0}
\plotone{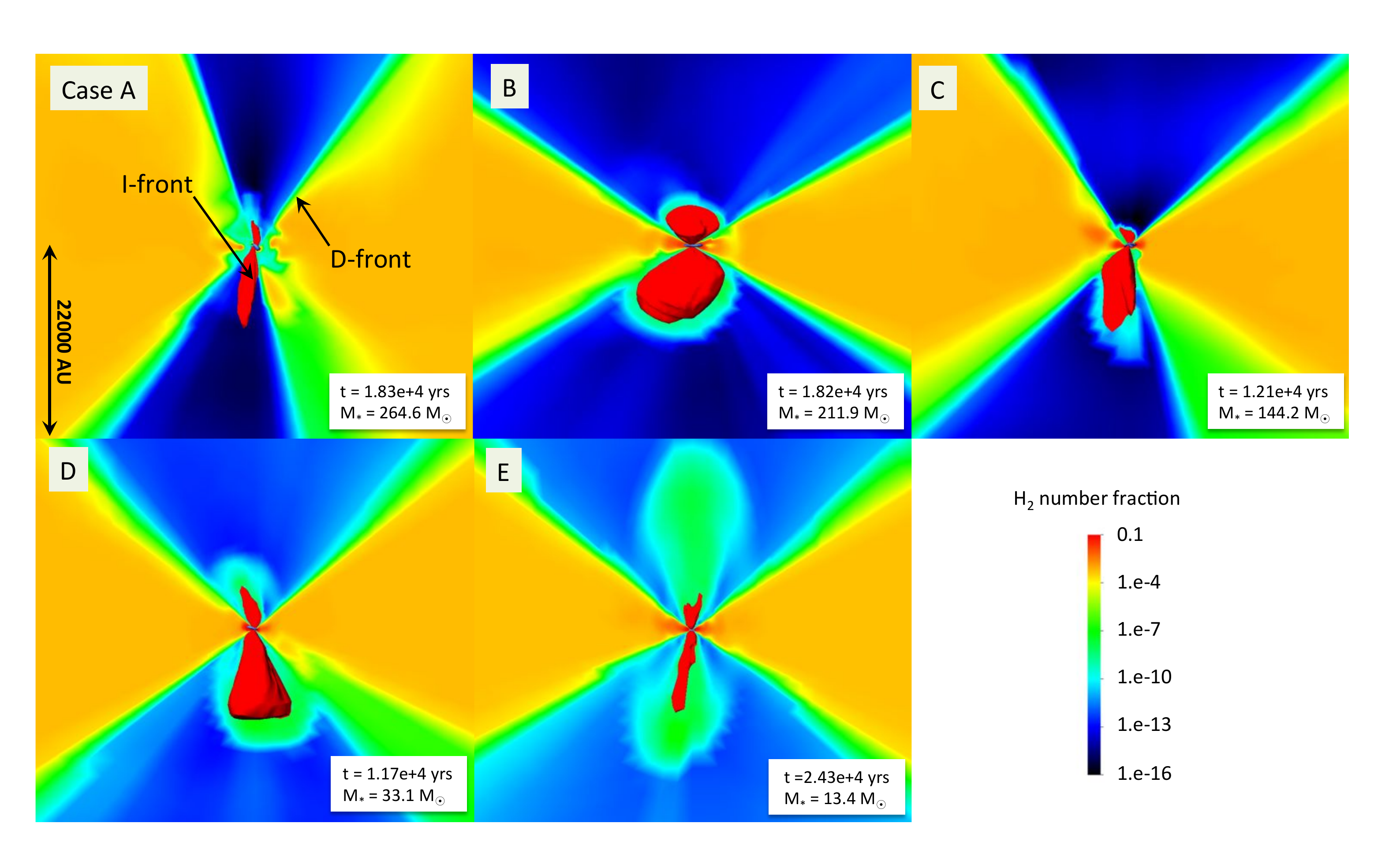}
\caption{
Same as Figure~\ref{fig:HIIr1e4AU_A2E} but for the
chemical structure in the accretion envelope.
In each panel, the color image shows the distribution of the 
hydrogen molecular number fraction
in the same vertical plane as in Figure~\ref{fig:HIIr1e4AU_A2E}. 
The 3D red contour delineates the structure of ionization
fronts, within which the hydrogen atomic fractions
is below 0.5.
}
\label{fig:HIIr1e4AU_A2E_C}
  \end{center}
\end{figure*}
%--------------------------------------------------------------------%
%--------------------------------------------------------------------%
%%% Fig.6 %%%
\begin{figure}
  \begin{center}
\epsscale{1.1}
\plotone{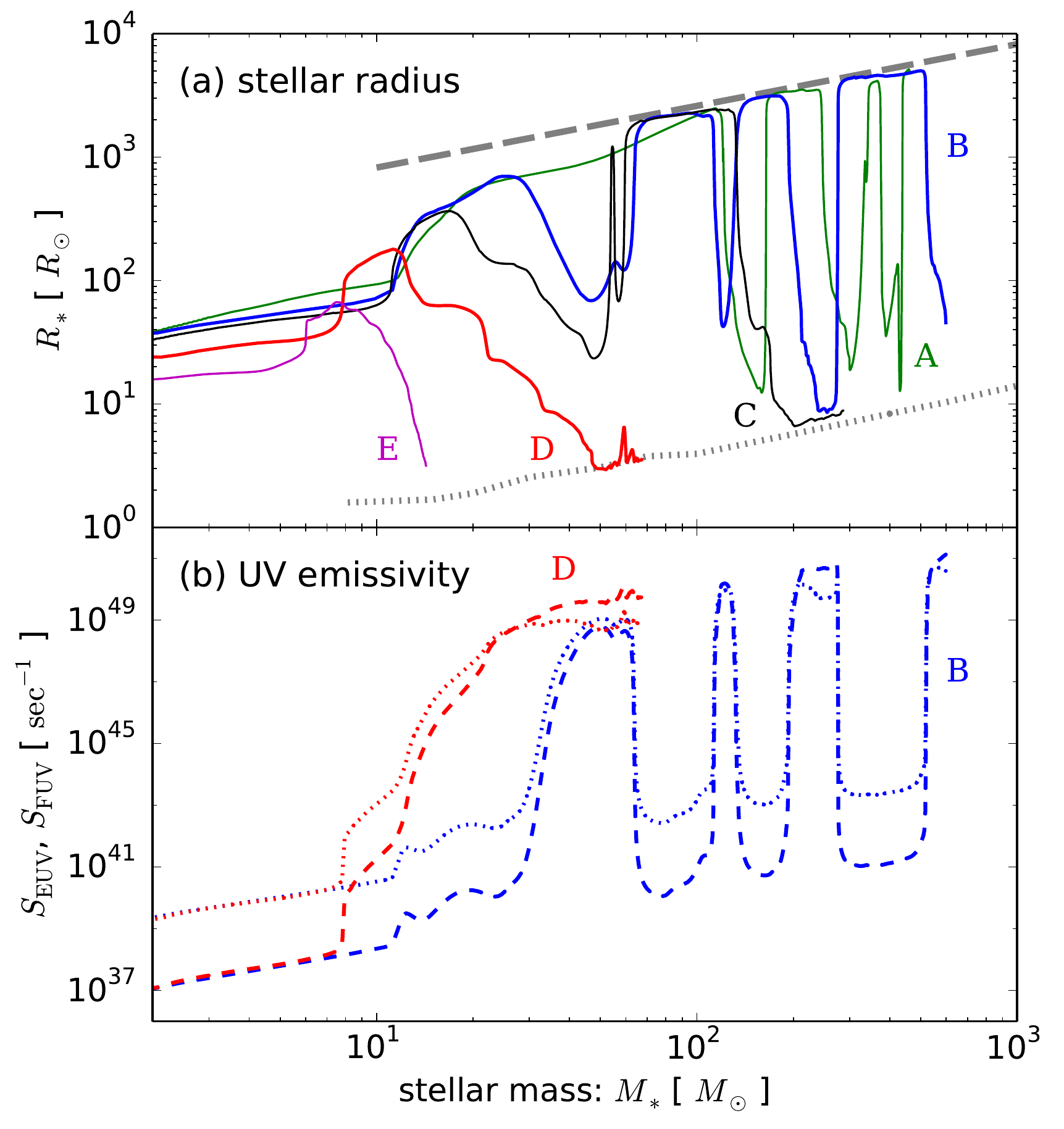}
\caption{Evolution of the stellar radius (panel a) and
UV emissivities (panel b) as a function of accumulated stellar mass.
The different colors represent the same
cases as in Figure~\ref{fig:xm_xmdot_t}.
In panel (a), the thick gray dotted and long-dashed lines represent
the mass-radius relationships for zero-age main-sequence stars
and for supergiant protostars (equation~\ref{eq:mrsgp}).
In panel (b), the dashed line represents the
stellar ionizing (EUV; $h \nu \geq 13.6$~eV) flux and the dotted line the
dissociating (FUV; $11.2~{\rm eV} \leq h \nu \leq 13.6~{\rm eV}$) radiation.
For clarity, we only show the UV emissivities for the two representative 
cases B and D.}
\label{fig:xr_suv_xm}
  \end{center}
\end{figure}
%--------------------------------------------------------------------%

%-------------------------- Fig.2-(a) --------------------------%

We first summarize the simulation results for our fiducial cases A--E. 
As seen in Figure~\ref{fig:xm_xmdot_t}-(a), the stellar mass
growth histories show a great variety in the $\sim$$10^5$ years
after the formation of embryo protostars. 
Ordinary massive stars with $M_* < 100~\msun$ result
in cases D and E. In particular, the final stellar mass is only
$\sim$$15~\msun$ for case E. For cases A--C, on the other hand, 
the stellar masses reach several hundred solar masses
at the end of the simulations.

%--------------------------------------------------------------------%
%%% Fig.7 %%%
\begin{figure*}
  \begin{center}
\epsscale{1.0}
\plotone{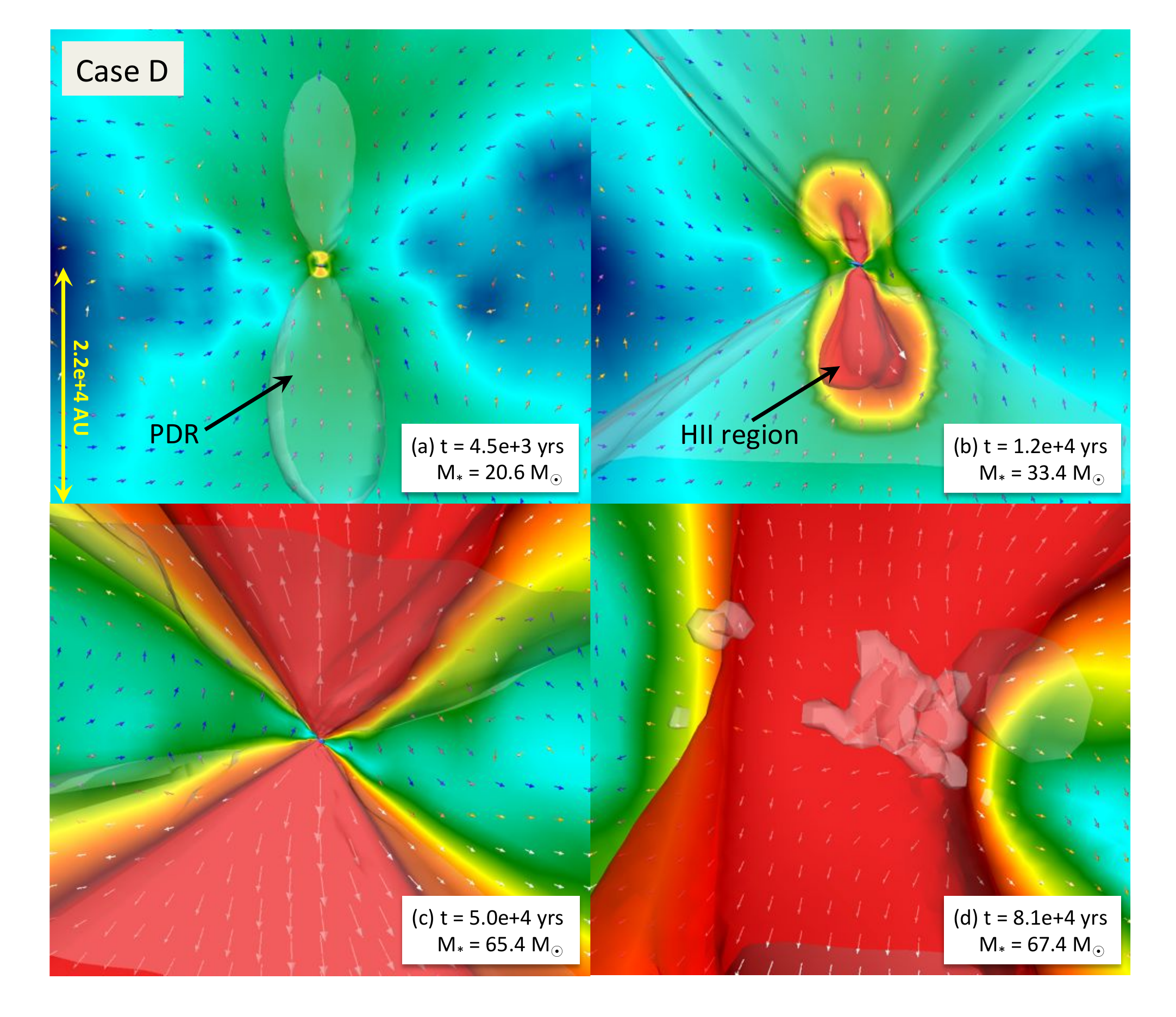}
\caption{Time evolution of the temperature distribution 
in a plane containing the polar axis for case D. The stellar
mass and evolutionary time are given in the insets of panels (a)--(d). 
The color scale for temperature is the same
as in Figure~\ref{fig:HIIr1e4AU_A2E}.
The 3D white and red contours delineate the 
structure of dissociation and ionization
fronts, within which the hydrogen atomic and molecular fractions
are below 0.5 and $10^{-4}$, respectively.
}
\label{fig:caseD_snapshots}
  \end{center}
\end{figure*}
%--------------------------------------------------------------------%

%-------------------------- Fig.2-(b) --------------------------%

Figure~\ref{fig:xm_xmdot_t}-(b) displays the highly
variable accretion histories for all cases. 
Such a high degree of variability is
a well-known feature of mass and angular momentum 
transport through self-gravitating circumstellar disks. 
%------------------------------- Fig. 3 -----------------------------%
Figure~\ref{fig:caseB_disk_rescomparison} presents 
examples of the density structure of self-gravitating disks seen in our simulations. 
The disk has non-axisymmetric spiral arms, which produce 
the gravitational torques leading to variable mass accretion. 
In Figure~\ref{fig:xm_xmdot_t}-(b), we see that
the mass accretion finally ceases for cases C-E;
the mean accretion rates for the last $10^4$ years 
having fallen down to $\sim$$10^{-5}~\msunyr$ 
due to UV feedback. 
For cases A and B, however, the accretion rates are still 
higher than $10^{-4}~\msunyr$ at $t = 7 \times 10^4$ years. 
Although the simulations end at this point due to high
computational costs, 
the stellar masses will continue to increase
via accretion, because UV feedback has not yet become fully effective.

%---------------- Fig.4 (formation of HII region) ---------------------%

Figures~\ref{fig:HIIr1e4AU_A2E} and \ref{fig:HIIr1e4AU_A2E_C}
shows that, for all cases, 
bipolar \hii regions form and grow within the larger H$_2$ photodissociation 
regions (PDRs) within a few $\times~10^4$ years 
after stellar birth.
Although the snapshots look quite similar, the stellar masses at the
evolutionary stage of \hii region formation differ significantly among the cases; the stellar 
mass is lower in alphabetical order for cases A through E.   
For cases A--C, it is not until the stellar masses exceed 
$100~\msun$ that the \hii regions appear. For cases D and E, 
however, only a few $\times~10~\msun$ stars can create \hii regions.

%--------------------------- Fig.5 -----------------------------%

The diversity of stellar masses which first form an \hii region
comes from the different protostellar evolution with
different accretion histories (Fig.~\ref{fig:xr_suv_xm}).
For cases D and E, the protostar 
contracts from its bloated state when $M_* \gtrsim 10~\msun$; 
this is the so-called 
Kelvin-Helmholtz (KH) contraction phase \citep[e.g.,][]{OP03}.
The stellar UV emissivities dramatically increase during this
stage as the effective temperature rises as
$T_{\rm eff} \propto R_*^{-1/2}$.
Stellar contraction continues until the star arrives
at the zero-age main-sequence (ZAMS). 
As a result, the \hii region emerges for $M_* \simeq 10~\msun$.
For cases B and C, however, KH contraction is interrupted by
the abrupt stellar expansion occurring around $M_* \simeq 50~\msun$. 
For case A, there is no contraction stage until the
stellar mass exceeds $100~\msun$.
It is only after the stellar mass reaches a few 
hundred solar masses that the protostar substantially contracts. 
Since the stellar UV emissivity sensitively depends on 
the effective temperature, the epoch of stellar 
contraction actually controls when the \hii region first appears.

%------------------------------------------------------------------%

For cases A--C, the protostar enters 
the supergiant stage with rapid mass accretion $\mdot \gtrsim 0.01~\msunyr$ 
before KH contraction begins (Section~\ref{sec:method}).
Several studies presuppose that supergiant protostars
only appear in extreme cases of primordial star formation,
for which very high accretion rates are available
\citep[e.g., direct-collapse model, see][]{BL03}. 
However, HR14 and HR15 show with 2D SE-RHD simulations that
the supergiant stage can appear even for more typical Pop III.1 cases 
when accretion rates are relatively high but not necessarily extremely high.
By contrast, our 3D SE-RHD simulations now demonstrate that the supergiant stage 
is a common feature of primordial star formation. 
This is due to the highly variable mass accretion onto primordial protostars
caused by self-gravitating
disks, for which the accretion rates temporarily and recurrently exceed
the critical rate for expansion to the supergiant stage.
As a result of such variable mass accretion, in fact,
stellar contraction and re-expansion are repeated several times for cases A and B.
The volatile evolution of the stellar radius results in strong
fluctuations of the stellar UV emissivity. 
This variability causes recurrent extinction and re-formation of 
\hii regions (see Section~\ref{ssec:VMstar} below). 
UV feedback only intermittently operates and cannot
efficiently stop mass accretion. 
Very massive stars can potentially form by this mechanism.

%------------------- resolution dependence ---------------------------%

Although our results are resolution dependent
because of the ansatz explained in Section~\ref{ssec:flimit}, 
the basic features described become more prominent 
at higher spatial resolution (Section~\ref{ssec:sres}). 
This is demonstrated in Figure~\ref{fig:caseB_disk_rescomparison}, 
showing that higher resolution results in sharper and finer 
density substructure within the disk. 
The resulting mass accretion is more variable with increased spatial
resolution, bringing the central protostar 
to the supergiant stage more often. 
Disk fragmentation (e.g., Fig.~\ref{fig:caseB_disk_rescomparison}-c) 
does not necessarily reduce the final stellar masses.
The emerging fragments cause repeated accretion bursts when 
falling onto the star, which can extinguish the stellar UV emission 
and thus ultimately enhances stellar mass growth.

%+++++++++++++++++++++++++++++++++++++++++++++++++++++++%
\subsection{Mass Accretion under the Stellar UV Feedback}
\label{ssec:basic_results}
%+++++++++++++++++++++++++++++++++++++++++++++++++++++++%

%---------------------------------------------------------%
\subsubsection{Massive Stars with Tens of Solar Masses: \\
UV Feedback Halts the Mass Accretion}
\label{ssec:OMstar}
%---------------------------------------------------------%

%--------------------------------------------------------------------%
%%% Fig.8 %%%
\begin{figure}
  \begin{center}
\epsscale{1.0}
\plotone{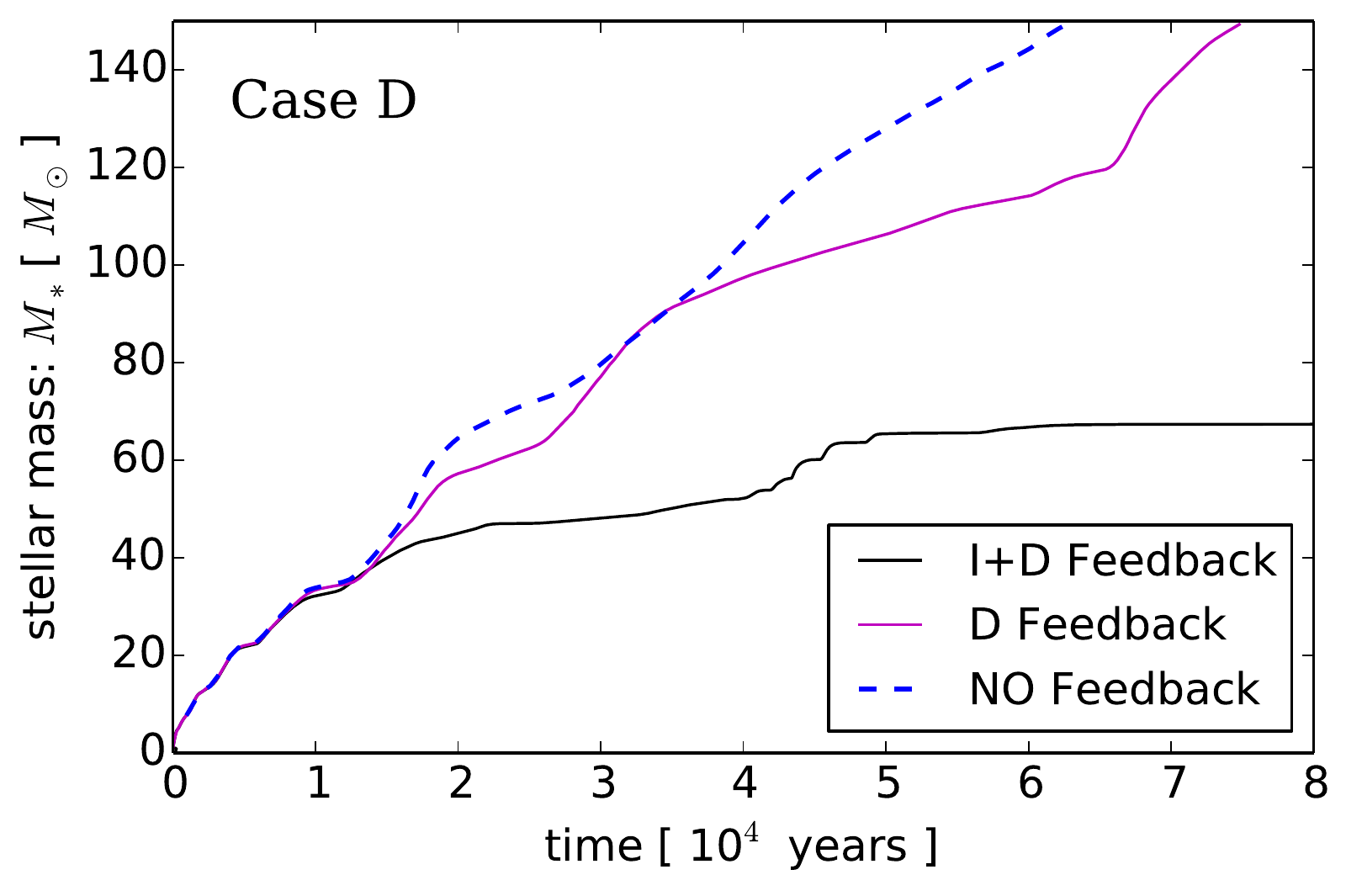}
\caption{Example of the impact of differing stellar UV feedback on
stellar mass growth. 
The black solid (case D), magenta solid (case D-DF), 
and blue bashed lines (case D-NF) represent
the cases including both the ionizing and dissociating feedback,
with the dissociating feedback only,
and without UV feedback, respectively.} 
\label{fig:xm_t_D}
  \end{center}
\end{figure}
%--------------------------------------------------------------------%
%--------------------------------------------------------------------%
%%% Fig.9 %%%
\begin{figure}
  \begin{center}
\epsscale{1.0}
\plotone{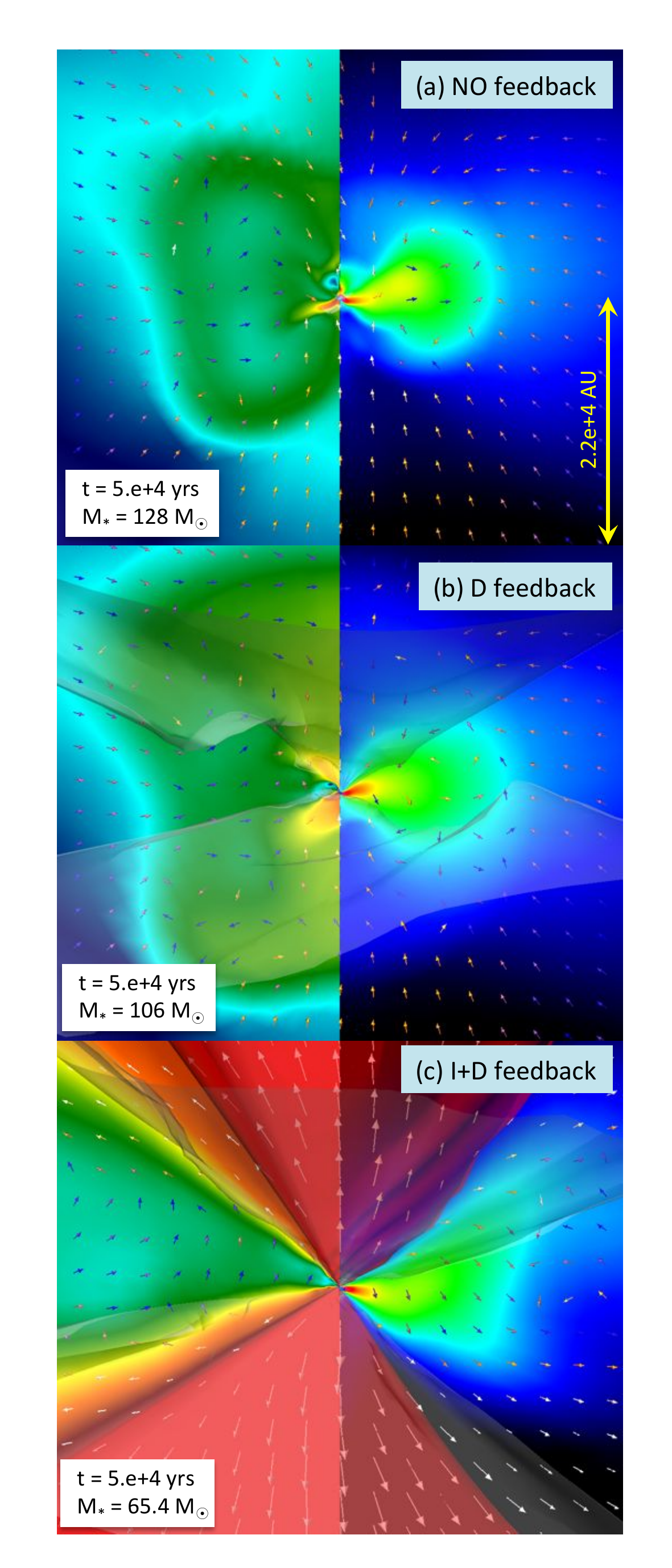}
\caption{Impact of differing stellar UV
feedback on the accretion envelope:
(a) no feedback (case D-NF),
(b) only photo-dissociating feedback (case D-DF), 
and (c) both photo-ionizing and dissociating feedback (case D). 
The temperature, density and velocity distributions are shown at the same epochs,  
$5 \times 10^4$ years after the birth of the central protostar. 
Note that the stellar masses differ among the cases
because of the different mass accretion histories 
(also see Fig.~\ref{fig:xm_t_D}).
The illustration is in the same style 
as in Figure~\ref{fig:HIIr1e4AU_A2E}.
}
\label{fig:caseD_comparison}
  \end{center}
\end{figure}
%--------------------------------------------------------------------%

We first describe in detail the evolution of cases D and E, which
produced ordinary massive stars with tens of solar masses.
As shown below, the evolution for these cases is 
similar to that found in our previous 2D SE-RHD simulations
(HS11; HS12).

%---------------------------- Fig.6 --------------------------------%

\paragraph{Case D}

Figure~\ref{fig:caseD_snapshots} presents the evolution of the temperature
structure for case D as an H$_2$ PDR and an \hii region expand
into the accretion envelope. 
After the protostar enters the KH contraction stage (see Fig.~\ref{fig:xr_suv_xm}), 
the stellar FUV emissivity rises and forms a bipolar H$_2$ PDR 
(Fig.~\ref{fig:caseD_snapshots}-a). At the same time,
the stellar EUV emissivity increases and slightly later creates
a bipolar \hii region within the larger PDR (Fig.~\ref{fig:caseD_snapshots}-b).
The PDR and \hii region continue to grow and expand as shown for the epoch
$t = 5 \times 10^4$ years (Fig.~\ref{fig:caseD_snapshots}-c),
but these regions are still centrally ``pinched'' because a
circumstellar disk blocks the stellar UV radiation close to the disk plane. 
However, even the gas protected from the 
UV photons is disturbed by shocks that propagate into
the envelope ahead of the ionization fronts. 
Comparing Figure~\ref{fig:caseD_snapshots}-(b) and (c), we
see that even the gas outside the PDR is heated up
by shock compression. 
HS11 have shown that the mass influx from the envelope
onto the disk is hindered by this effect.  
Moreover, the disk loses mass via photoevaporation, being
exposed to the stellar EUV radiation.
The \hii region is filled by the photoevaporating outflow 
launched from the disk surface. 
Ultimately, the circumstellar disk is destroyed due to 
photoevaporation, allowing the \hii region to expand 
in equatorial directions (Fig.~\ref{fig:caseD_snapshots}-d),
and accretion onto the star is effectively shut off 
for $M_* \simeq 67.4~\msun$ at $t = 8 \times 10^4$ years.

%------------------------------- Fig.7  -----------------------------%

In Figure~\ref{fig:xm_t_D}, we plot the evolution of the stellar mass
for case D (solid black line). For comparison, we show the stellar
mass growth for case D-NF (blue dashed line), for which all UV feedback effects had been
turned off. We see that mass accretion onto the protostar is
significantly reduced by UV feedback effects.
%----------------- strength of dissociation feedback -----------------%
This figure also displays the stellar mass growth
for a test case (case D-DF) that only included photodissociation (FUV) feedback
(solid magenta line). Comparing these mass growth histories we conclude that
FUV feedback does hinder mass accretion somewhat, 
but its impact is much weaker than EUV feedback.  

%------------------------------- Fig.8  -----------------------------%

Figure~\ref{fig:caseD_comparison} shows the variation of the envelope 
temperature - density - velocity
structure with differing UV feedback effects. 
For case D-NF with neither EUV nor FUV feedback 
(Fig.~\ref{fig:caseD_comparison}-a), the disk is surrounded by a warm 
($T \simeq 300 - 500$~K) envelope, where compressional heating 
due to gas infall is balanced by H$_2$ molecular line cooling.
The gas temperature is higher ($T \simeq 5000$~K)
in a central region $r \lesssim 10^3$~AU, where H$^-$ 
free-bound emission is the main cooling process.
For case D-DF with only FUV feedback
(Fig.~\ref{fig:caseD_comparison}-b), the warm envelope is larger,
because FUV radiation reduces the abundances of efficient coolants 
via H$_2$ photodissociation and H$^-$ photodetachment.
However, this only occurs in polar regions. 
Because the bulk of stellar FUV radiation is blocked by the disk,
the envelope structure is hardly affected in equatorial directions, 
where the disk mostly accretes gas from the envelope.
As seen in Fig.~\ref{fig:caseD_comparison}-c
EUV feedback dramatically changes the envelope structure.
In addition to the bipolar photoevaporation flow within the \hii region, 
portions of the envelope outside the \hii region are expelled from the
computational grid as shocks propagate through the envelope and 
spread the pressure excess of ionized gas.

%--------------------- difference from Susa+ -----------------------%

Our results demonstrate that FUV feedback does reduce mass accretion rates, 
which qualitatively agrees with 3D RHD simulations performed by \citet{Susa13}.
However, the FUV feedback seen in our simulations seems to be weaker 
than in \citet{Susa13}, who claims that
FUV feedback is largely responsible for terminating 
mass accretion onto primordial protostars. 
We discuss possible causes of this difference 
in Section~\ref{ssec:2Dcomp} below.

%------------------------------ case E --------------------------------%

\paragraph{Case E}

The evolution of case E is qualitatively similar to that for case D,
except that HD molecular line emission also
contributes to the cooling in the early collapse stage.
During collapse, in general, the thermal evolution of the central 
homogeneous core strongly affects the radial structure of the accretion envelope. 
Since the radius of the core is well approximated by the
Jeans length, the envelope temperature and density at a given 
radius $r$ obey $r = \lambda_{\rm J} \propto T^{1/2} \rho^{-1/2}$,
i.e., a lower temperature implies a lower density. 
In case E the accretion envelope thus has relatively 
low temperatures and densities,
which can been seen in Figure~\ref{fig:HIIr1e4AU_A2E}. 
KH contraction begins at lower stellar masses with correspondingly
lower accretion rates (Fig.~\ref{fig:xr_suv_xm}).  
The final stellar mass is also lower than for case D, $M_* \simeq 14.3~\msun$.

%-------------------------------------------------------------------------%
\subsubsection{Massive Stars with Hundreds of Solar Masses: \\
Intermittent UV Feedback with Variable Mass Accretion}
\label{ssec:VMstar}
%-------------------------------------------------------------------------%

%--------------------------------------------------------------------%
%%% Fig.10 %%%
\begin{figure*}
  \begin{center}
\epsscale{1.1}
\plotone{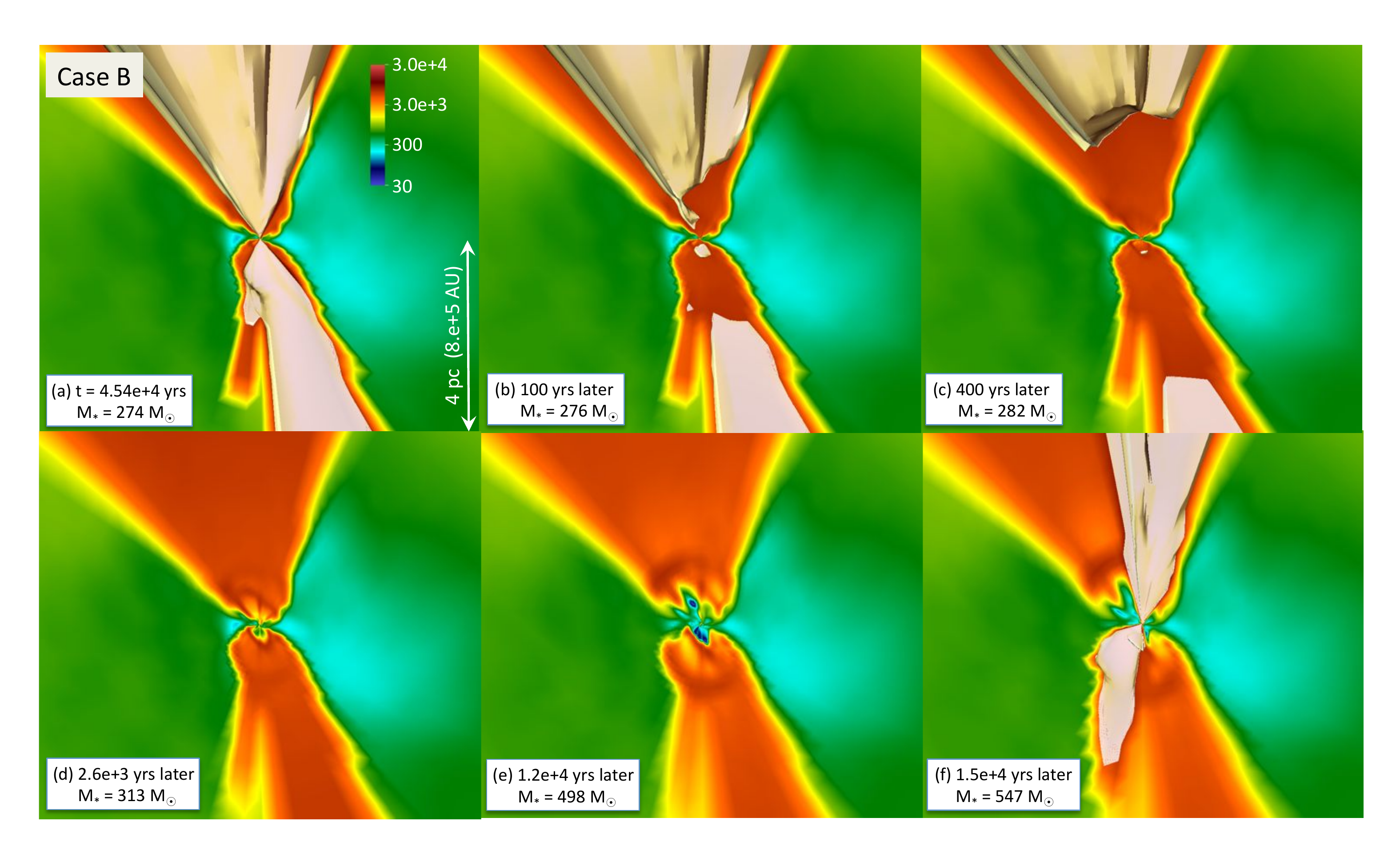}
\caption{The extinction and re-formation 
of \hii regions for case B. 
In each panel, the 3D yellow contour delineates the structure of the
ionization front (as in Figures~\ref{fig:HIIr1e4AU_A2E_C} 
and \ref{fig:caseD_snapshots} with the red contours),
and the color image shows a 2D slice of the temperature
distribution (color scale as in Figure~\ref{fig:HIIr1e4AU_A2E}).
The panels (a)--(f) show a time sequence for
$t > 4.54 \times 10^4$ years after the birth of the protostar;
the elapsed time after the epoch of panel (a) is given for
(b)--(f). Note that the box size is more than
50 times larger than in Figure~\ref{fig:caseD_snapshots}.}
\label{fig:caseB_HIIdetachment_y}
  \end{center}
\end{figure*}
%--------------------------------------------------------------------%
%--------------------------------------------------------------------%
%%% Fig.11 %%%
\begin{figure}
  \begin{center}
\epsscale{1.1}
\plotone{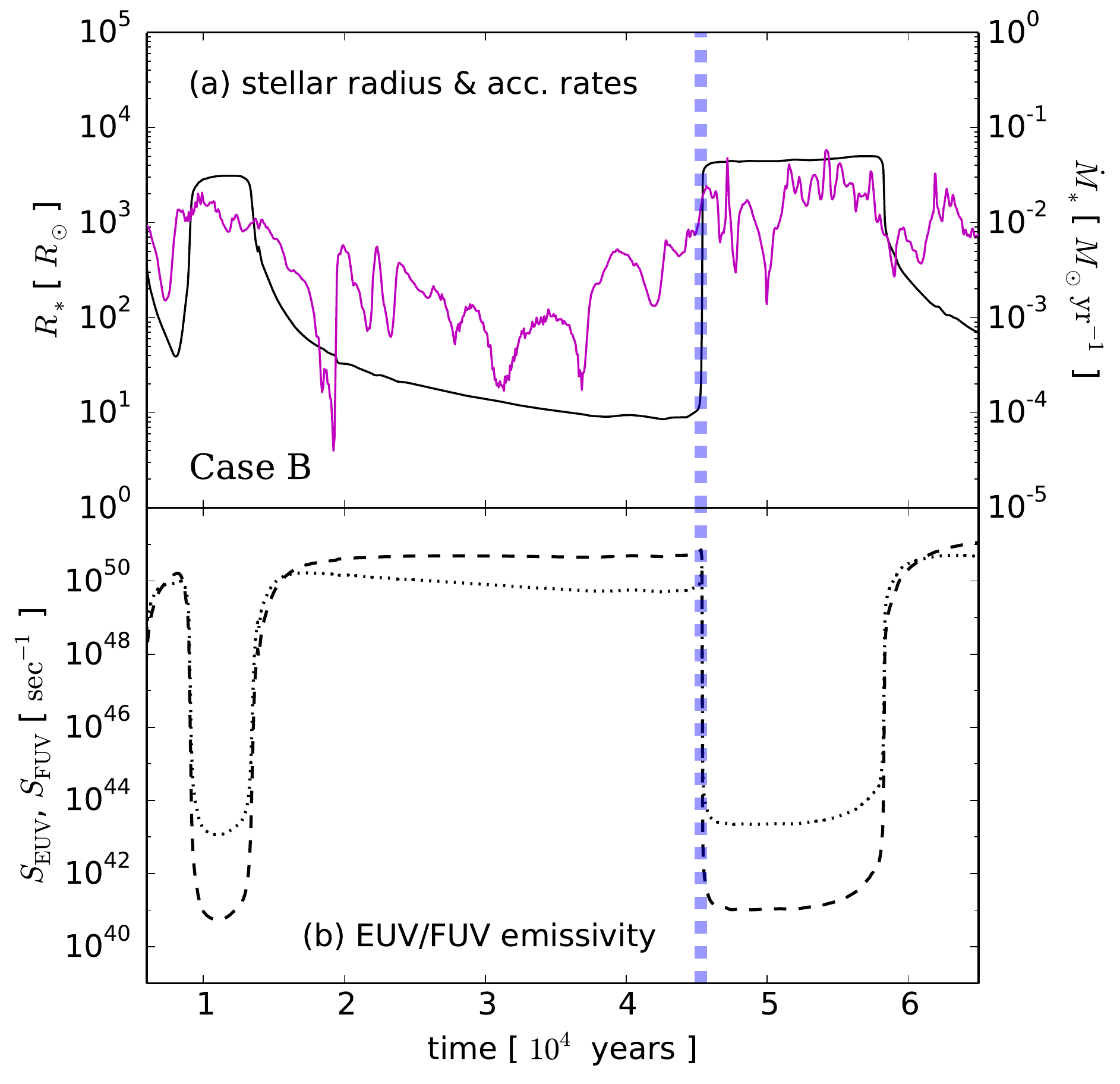}
\caption{Effects of the variable mass accretion on the 
evolution of the stellar radius (panel a) and UV emissivities 
(panel b) for case B. In panel (a), the black and magenta lines 
show the time evolution of the stellar radius and accretion rates 
respectively. In panel (b), the dashed and dotted lines represent 
the stellar EUV and FUV emissivities as 
in Figure~\ref{fig:xr_suv_xm}-(b). 
The vertical blue dotted line marks the epoch 
of the stellar inflation at $t = 4.53 \times 10^4$~years 
(also see Fig.~\ref{fig:structureB}). 
}
\label{fig:burstsB}
  \end{center}
\end{figure}
%--------------------------------------------------------------------%
%--------------------------------------------------------------------%
%%% Fig.12 %%%
\begin{figure}
  \begin{center}
\epsscale{1.23}
\plotone{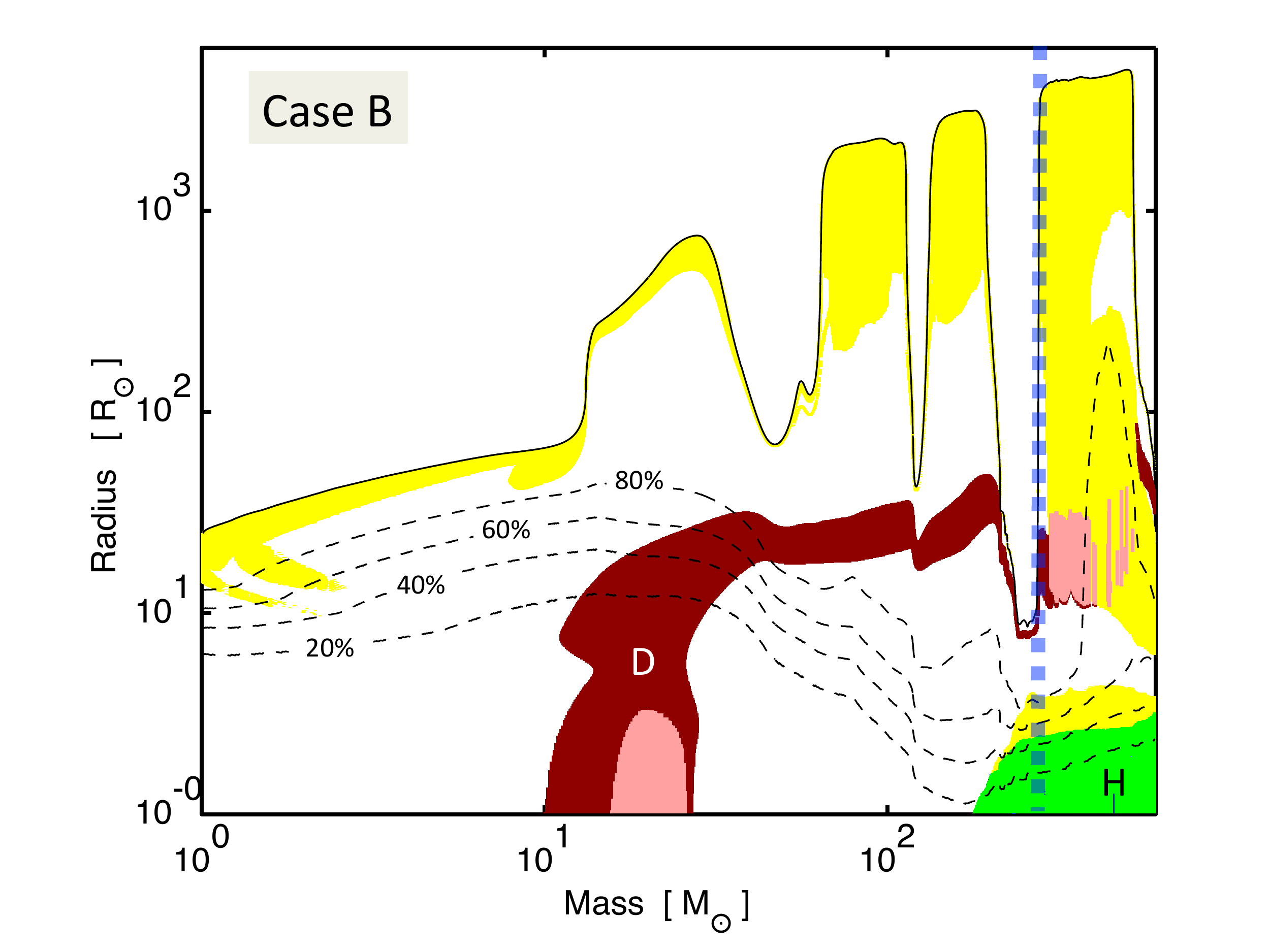}
\caption{Evolution of the stellar interior structure 
as a function of increasing mass for case B. The black
solid and dashed lines represent the radial positions
of the photosphere and mass coordinates 80\%, 60\%, 40\%, and 20\%
of the total stellar mass.
The yellow regions denote convective zones and white regions
radiative zones, each without active nuclear burning.
The pink and brown stripes denote convective and radiative
deuterium-burning zones, respectively and green depicts the central
convective hydrogen-burning zone.
Layers with active nuclear burning are indicated if the 
depletion time of the ``burning'' species is shorter than the lifetime of a
main-sequence star with the equivalent mass.
The vertical blue dotted line marks the same epoch
as indicated in Figure~\ref{fig:burstsB}.
}
\label{fig:structureB}
  \end{center}
\end{figure}
%--------------------------------------------------------------------%
%--------------------------------------------------------------------%
%%% Fig.13 %%%
\begin{figure*}
  \begin{center}
\epsscale{1.0}
\plotone{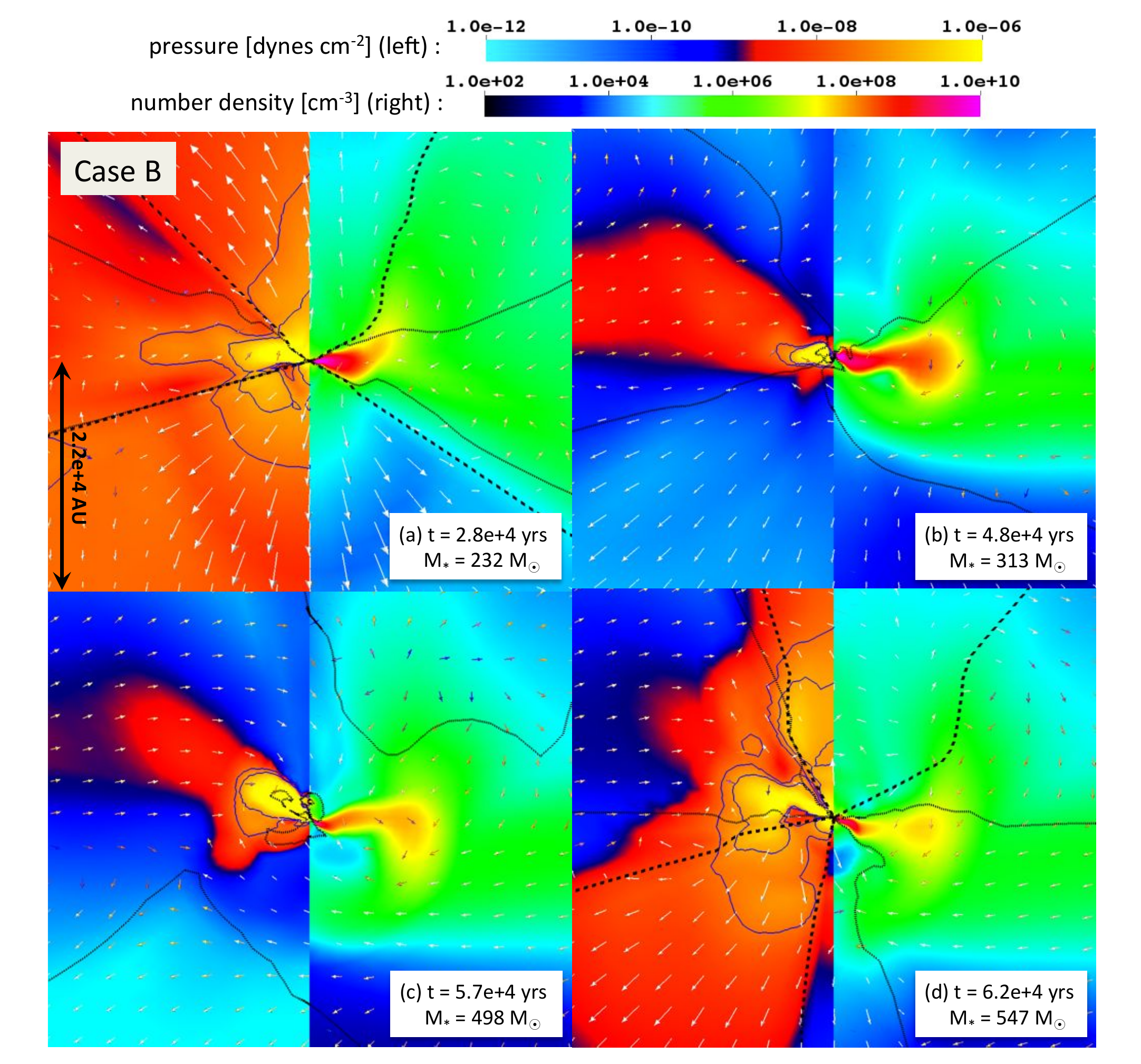}
\caption{The evolution of pressure 
(left-half of each panel) and density (right-half) under the influence of 
intermittent UV feedback for case B.
Panels (b)--(d) show the same
epochs as in Figure~\ref{fig:caseB_HIIdetachment_y} (d)--(f)
but on a smaller physical scale.
The 2D velocity vectors are presented in the same style 
as in Figure~\ref{fig:HIIr1e4AU_A2E}.
In each panel, the short dashed line delineates the location of the
hydrogen ionization front and the solid gray line depicts the dissociation 
front, whereby the fractional atomic and molecular abundances 
are 0.5 and $10^{-4}$, respectively.
The thin blue lines in the left panels show the pressure 
contours for $P = 1$ and $0.3 \times 10^{-7}~\press$.
}
\label{fig:caseB_snapshots}
  \end{center}
\end{figure*}
%--------------------------------------------------------------------%
%--------------------------------------------------------------------%
%%% Fig.14 %%%
\begin{figure}
  \begin{center}
\epsscale{1.0}
\plotone{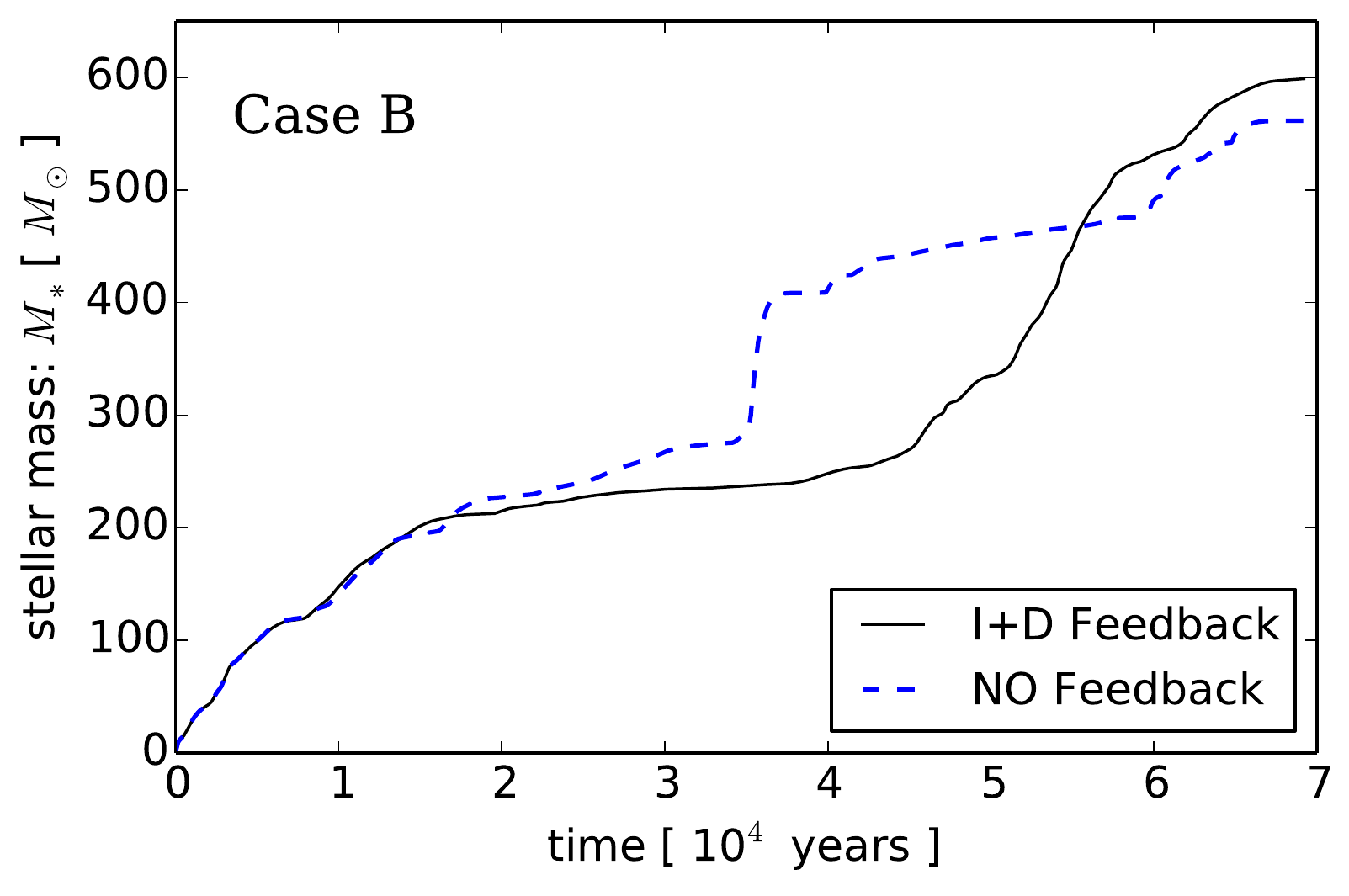}
\caption{Same as Figure~\ref{fig:xm_t_D} but for case B,
i.e., the black solid and blue dashed lines represent the
evolution for cases B and B-NF.} 
\label{fig:xm_t_B}
  \end{center}
\end{figure}
%--------------------------------------------------------------------%

Next we describe the results for cases A--C, for which very massive 
($M_* \gtrsim 250~\msun$) stars form that pass through 
a supergiant evolutionary stage. 

%As briefly reviewed in Section~\ref{ssec:overall}, 
%the variable mass accretion causes the violent evolution of the
%stellar radius, which results in the recurrent extinction
%and re-formation of \hii regions.

%--------------------------- Fig.10-a, 11 --------------------------------%

\paragraph{Case B}

Figures~\ref{fig:caseB_HIIdetachment_y} and \ref{fig:burstsB}
show that for case B,  
variable mass accretion leads to abrupt and radical changes of the
stellar radius and the associated recurrent extinction
and re-formation of \hii regions.
A bipolar \hii region, which have formed prior to $t \simeq 4.5 \times 10^4$ years
(Fig.~\ref{fig:caseB_HIIdetachment_y}-a),
begins to retreat in a direction away from the star a few 
hundred years later (Fig.~\ref{fig:caseB_HIIdetachment_y}-b,c),
shortly after the accretion rates rise above
$\sim$$10^{-2}~\msunyr$ and the protostar 
rapidly inflated to a supergiant (Fig.~\ref{fig:burstsB}-a). 
Due to the protostar's expansion
the stellar EUV emissivity accordingly drops by almost ten orders 
of magnitude (Fig.~\ref{fig:burstsB}-b), initiating the extinction 
of the \hii region.

%----------------------------- Fig. 12 ------------------------------%

Figure~\ref{fig:structureB} depicts the evolution 
of the stellar interior structure for this case B. 
Although the star initially begins KH contraction when
$M_* \simeq 30~\msun$, the contraction is interrupted by
a phase of rapid accretion at $M_* \simeq 45~\msun$.
The star quickly and substantially bloats up again, 
because the accretion rate exceeds $\sim$$10^{-2}~\msunyr$, 
and only begins to contract after the accretion rate falls 
during a subsequent quiescent phase (also see Fig.~\ref{fig:burstsB}-a). 
The abrupt expansion depicted in detail in Figures~\ref{fig:caseB_HIIdetachment_y}
and \ref{fig:burstsB} occurs for $M_* \simeq 274~\msun$; it is
the third of such events. 
Figure~\ref{fig:structureB} shows that, even during the supergiant stage, 
an inner core with a large fraction of the total mass 
continues to contract while releasing gravitational energy. 
Only a surface layer with a tiny fraction of the stellar mass
participates in the bloating up, trapping a part of 
the energy transferred from the contracting core. 
We note that hydrogen burning occurs in the
stellar center for $M_* \gtrsim 180~\msun$ and
conclude that rapid mass accretion will even trigger the bloating of an
H-burning star and extinguish its stellar UV emissivity.

%-------------------------------- Fig.10-(b+c) ---------------------------------%

The \hii region begins to recombine each time it loses its exciting UV source.
Because the recombination timescale varies as $t_{\rm rec} \propto 1/n_{\rm HII}$,
hydrogen recombination occurs in the densest region first.
Within the bipolar \hii region, the radial density gradient is established
by the photoevporation flow with geometrical dilution, 
which roughly follows 
\begin{eqnarray}
n_{\rm HII} \sim 10^4~\cmc \times
\left( \frac{r}{10^4~{\rm AU}}  \right)^{-2} \nonumber \\
\left( \frac{v_{\rm evp}}{100~{\rm km~sec^{-1}}}  \right)^{-1}
\left( \frac{\dot{M}_{\rm evp}}{10^{-3}~\msunyr}   \right) ,
\end{eqnarray}
where $\dot{M}_{\rm evp}$ and $v_{\rm evp}$ are the photoevaporation
rate and flow velocity, normalized with typical values 
\citep[e.g.,][]{HJLS94,TKei13}. 
The structure of recombining ionized gas thus resembles a ``recombination front'', 
starting at the dense inner regions and propagating outward 
through the bipolar \hii region (Fig.~\ref{fig:caseB_HIIdetachment_y}-b,c). 
The recombination timescale for the current case is 
\begin{equation}
t_{\rm rec} \sim 100~{\rm years} 
 \left( \frac{n_{\rm HII}}{10^4~\cmc} \right)^{-1},
\end{equation}
which matches the fast propagation of the recombination front shown
in Figure~\ref{fig:caseB_HIIdetachment_y}-(b-c).

%----------------------------- Fig.10-(d-e) -------------------------------%

Even after the ionized gas has recombined, 
the hot atomic gas still remains as
a ``remnant'' of the \hii region (or fossil \hii region)
for a long time (Fig.~\ref{fig:caseB_HIIdetachment_y}-c,d). 
This recombined atomic gas still flows outward
through inertia, but it is no longer
replenished via photoevaporation of the disk.
The temperature of the expanding recombined atomic gas 
gradually decreases due to adiabatic cooling 
and Ly-$\alpha$ cooling. 
About $1.2 \times 10^4$ years after the 
extinction of the UV source (Fig.~\ref{fig:caseB_HIIdetachment_y}-d),
for instance, the remnant neutral gas has $\simeq 1000 - 5000$~K 
over $\sim$${\rm pc}$ scales. 
The central regions $r \lesssim 10^5$~AU have the coldest
($T \lesssim 100$~K) gas because of strong adiabatic cooling. 

%------------------------------ Fig.13 ------------------------------%

Whereas the signature of a fossil \hii region remains over
$\sim$pc scales for many thousands of years, 
influences of the stellar UV radiation
in the stellar vicinity $r \lesssim 10^4$~AU
disappear much quicker. 
Figure~\ref{fig:caseB_snapshots}-(a) shows that 
the pressure structure of the accretion envelope is greatly 
modified by a bipolar \hii region if present. 
The radial pressure gradient created by the photoevaporation
outflow spreads over the envelope even beyond the \hii region
(also see Section~\ref{ssec:OMstar}). 
However, after the stellar UV emissivity shuts off, these
pressure structures quickly disappear. 
At the epoch depicted in Figure~\ref{fig:caseB_snapshots}-(c),
almost no signatures of UV feedback are present.
The $M_* \simeq 500~\msun$ protostar quiescently accretes the gas 
through the circumstellar disk with negligible stellar radiative feedback. 

%-------------------------------- Fig. 14 ------------------------------------%

Figure~\ref{fig:xm_t_B} shows the stellar mass growth history
for this case. We see that UV feedback does hinder the
mass accretion for $2 \lesssim  (t /10^4~{\rm years}) \lesssim 4.5$, 
when a bipolar \hii region emerges 
(also see Figs.~\ref{fig:burstsB} and \ref{fig:caseB_snapshots}-a). 
However, the accretion flow is no longer halted
once the \hii region disappears at $t \simeq 4 \times 10^4$ years. 
The gas accumulated in the envelope 
thus far begins to fall back toward the star. 
As a result, the stellar mass rapidly increases during the period
$4.5 \lesssim (t/10^4~{\rm years}) \lesssim 6$, during which
the protostar remains in the supergiant stage emitting few
ionizing photons. 

%---------------------- reformation of HII region -------------------------%

After the star has accreted most of the gas accumulated 
in the envelope, the accretion rate gradually decreases
for $t \gtrsim 6 \times 10^4$ years.
The protostar then begins to contract, and the stellar UV
emissivity dramatically increases again (Fig.~\ref{fig:burstsB}).  
Figure~\ref{fig:caseB_HIIdetachment_y}-(f) shows that a bipolar \hii region 
begins to grow again within the hot ($T \simeq 5000$~K) atomic envelope, 
the remnant of the previous \hii region that disappeared 
after the star had inflated.
Figure~\ref{fig:caseB_snapshots}-(d) shows that
the photoevaporation outflow refills the \hii region. 
The pressure structure of the accretion envelope is accordingly modified
by the expanding \hii region. At this point
UV feedback begins to regulate the mass accretion as before.

%---------------------------------------------------------------------------%

For this case B, however, UV feedback only intermittently operates
during the $7 \times 10^4$ years considered,
so that the stellar mass growth is not efficiently reduced.
Figure~\ref{fig:xm_t_B} shows that, in fact, UV feedback
only temporarily postpones stellar mass growth. 
At the end of the simulation, the stellar mass is slightly
higher than for the comparison case B-NF,\footnote{
For case B-NF, there is a big jump of the stellar mass
at $t \simeq 3.5 \times 10^4$ years. Although we do not 
resolve those additional individual infalling clumps 
that would appear at higher resolution (see Figure~\ref{fig:caseB_disk_rescomparison}), 
variable accretion is a general feature even at the default resolution.
Interestingly, the inner part of the B-NF disk is greatly tilted at 
$t \simeq 3.5 \times 10^4$ years.
It is known that external tidal perturbations 
from the surrounding non-axisymmetric envelope could excite the ``bending
wave'', which enhances the disk inclination and angular momentum
transport \citep[e.g.,][]{PT95}.
}
for which UV feedback has been turned off.
This slightly accelerated stellar mass growth can
be understood as follows. As described above, the gas accumulated 
in the envelope outside the disk rapidly 
falls back toward the star-disk system
once the \hii region disappears for $t \gtrsim 4 \times 10^4$ years. 
With the resulting rapid mass supply from the envelope,
the disk becomes highly gravitationally unstable.
This promotes the gravitational torque operating in the disk,
which consequently helps the stellar mass growth.

%--------------------------------------------------------------------%
%%% Fig.15 %%%
\begin{figure}
  \begin{center}
\epsscale{1.1}
\plotone{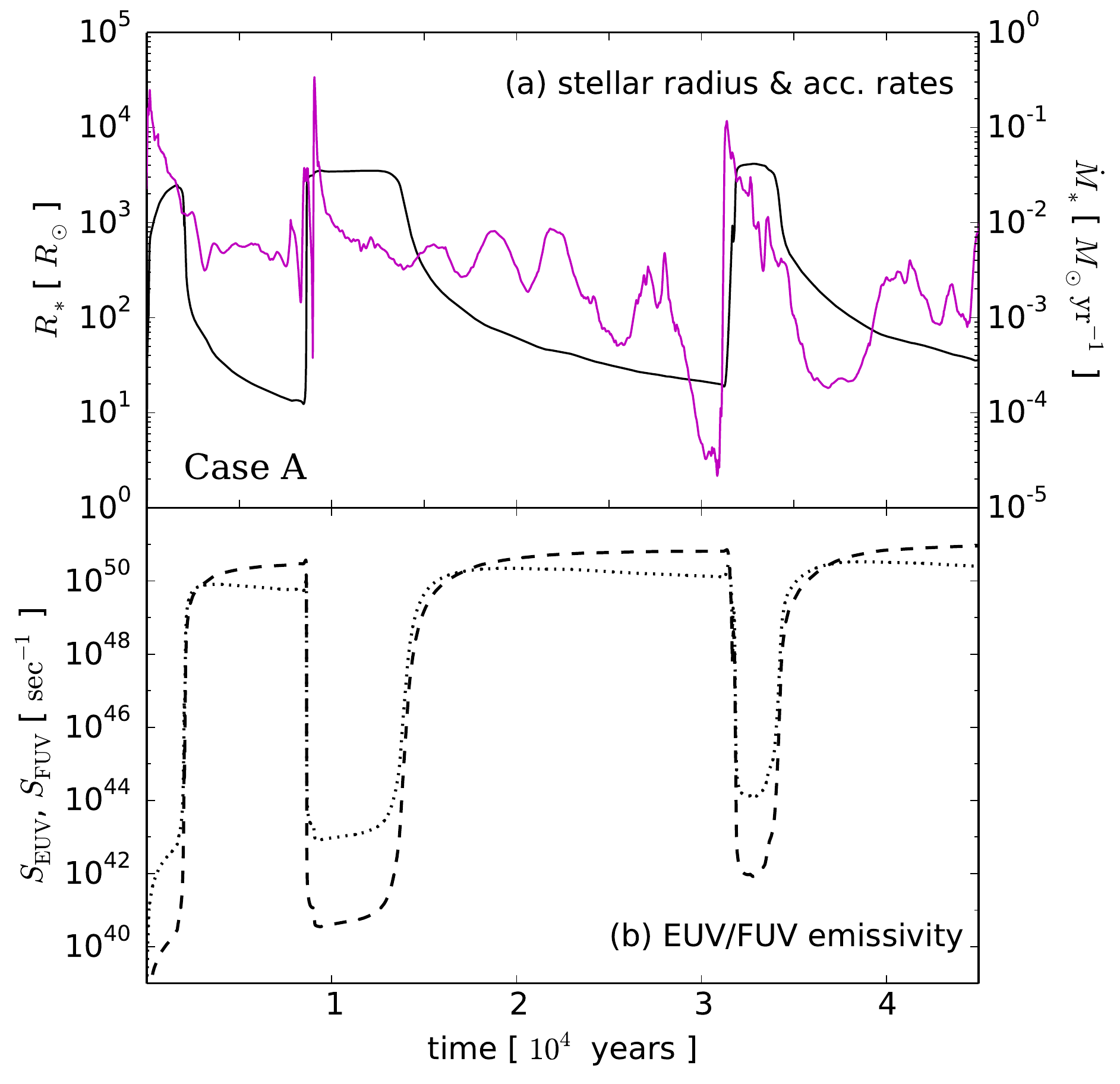}
\caption{Same as Figure~\ref{fig:burstsB} but for case A.} 
\label{fig:burstsA}
  \end{center}
\end{figure}
%--------------------------------------------------------------------%
%--------------------------------------------------------------------%
%%% Fig.16 %%%
\begin{figure}
  \begin{center}
\epsscale{1.0}
\plotone{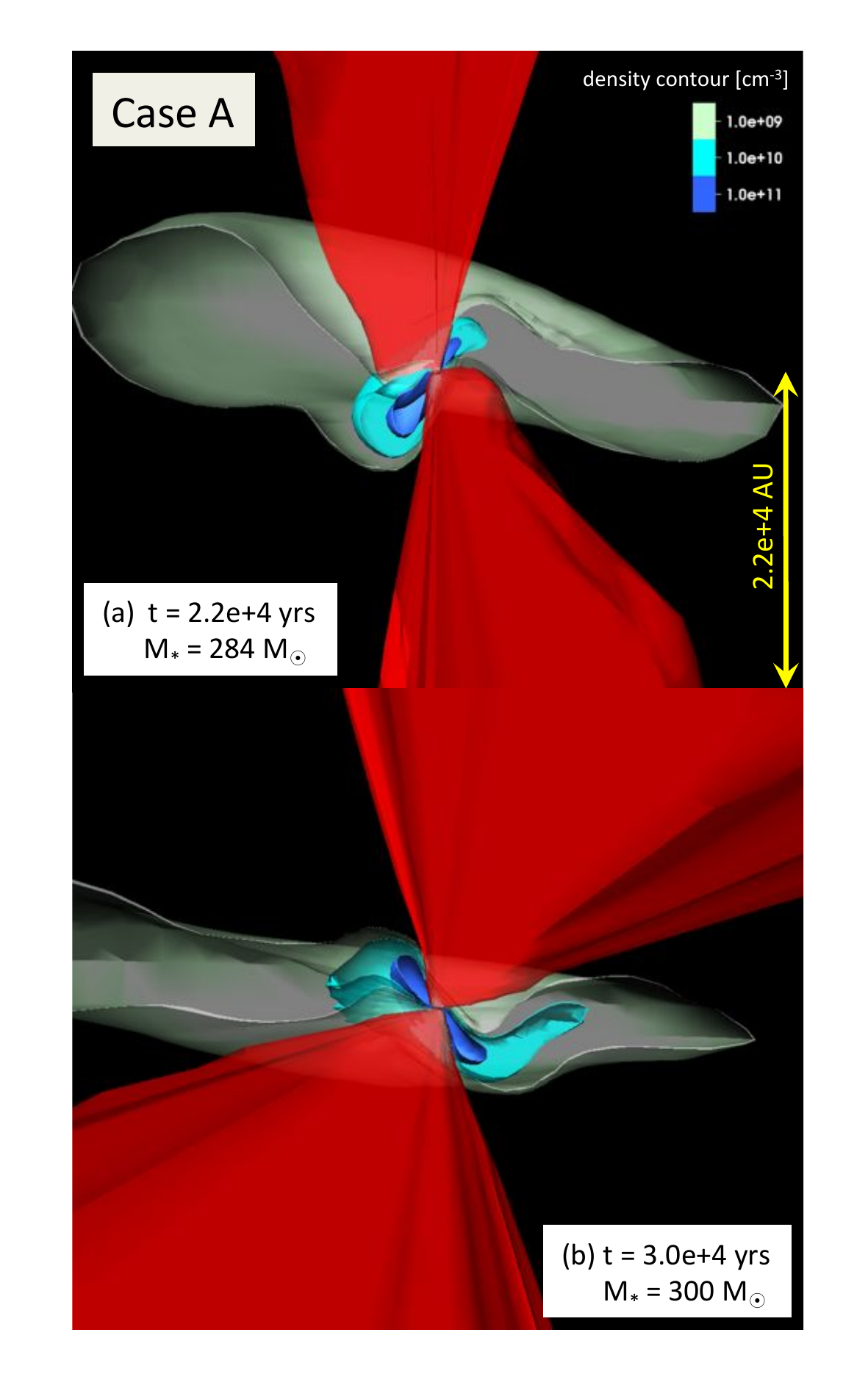}
\caption{Example of a precessing accretion disk with a bipolar \hii region
as seen for case A. The panels (a) and (b) depict the density and ionization
structure as viewed from the same angle for the evolutionary times
$2.2 \times 10^4$ years and $3 \times 10^4$ years after the birth
of the protostar. The dark blue, cyan, and green-gray
contours represent cut-outs of the iso-density 
surfaces for $n_{\rm H} = 10^{11}$, $10^{10}$, and $10^9~\cmc$, respectively.
The red contours delineate the \hii region.
}
\label{fig:caseA_disk_precession}
  \end{center}
\end{figure}
%--------------------------------------------------------------------%

%------------------------------ case A ----------------------------------%

\paragraph{Case A:}

%--------------------------- Fig. 15 -------------------------------%

Intermittent UV feedback is also seen in case A. 
Figure~\ref{fig:burstsA} shows that variable accretion 
causes the abrupt expansion of the stellar radius and 
rapid drop of UV emissivities as in case B.
This case exemplifies a characteristic feature of
stellar evolution with accretion bursts. 
For instance, let us focus on the short accretion burst
of duration $\sim$$10^2$ years occurring at $t \simeq 10^4$ years.
The high rate of accretion for this burst causes the protostar to inflate, 
terminating KH contraction.
The protostar remains swollen in the supergiant stage for $\sim$$10^4$ years,
long after the accretion rate had dipped below $\mdot \lesssim 10^{-2}~\msunyr$. 
This is because, whereas the protostar can react almost immediately to
a high accretion rate and quickly expands, it can contract only on a KH
timescale even if the accretion drops to zero \citep[e.g.,][]{Sakurai15}. 
Thus, even short episodes of accretion bursts can have a
long-lasting impact on the star's evolution. 
A similar feature is also seen for the second burst event 
at $t \simeq 3 \times 10^4$ years.
This effect can only be correctly simulated by solving for the stellar
interior structure simultaneously with time-dependent accretion histories,
as provided by SE-RHD simulations. 

%-------------- disk precession (moved from Appendix A) --------------%

%--------------------------- Fig. 15 -------------------------------%

In our simulations, circumstellar disks normally form
perpendicular to the polar axis of the spherical
coordinate system 
because of our choice of the polar axis orientation
(see Section~\ref{ssec:cases_cosmo}).
The orientation of the disk occasionally fluctuates 
with time, however, especially for cases A and B,
for which the spin parameters measured at the cloud scale 
are relatively low (Fig.~\ref{fig:ini}). 
For instance, Figure~\ref{fig:caseA_disk_precession} shows the
density contours of the disk and ionization contours 
of the bipolar \hii region at different epochs for case A. 
Although viewing from the same angle, the orientations of the
\hii region as well as the disk changes with time. 
Since the primordial cloud forms through
cosmological structure formation and 
is not necessarily perfectly axisymmetric, material falling onto the
disk has different mean angular momentum vectors at different times. 
Similar phenomena have been also reported for
present-day star formation, whereby turbulent motions in 
molecular clouds easily twist the distribution of the angular momentum
\citep[e.g.,][]{Bate10,Fielding15}.
Due to this disk ``precession'', the stellar EUV radiation
can ionize the envelope gas over rather wide 
opening angles. 
The orientation of the disk at any particular time also
depends on whether or not UV feedback
is included in the simulations. 
This is not surprising because UV feedback blows away the 
gas in the polar regions and modifies the pressure structure
even near the equator.
As a result, UV feedback can modify the accretion history
of both angular momentum and mass from the envelope onto
the disk.

%------------ case C: end of the accretion after quitting supergiant stage ---------%

\paragraph{Case C:}

For our case C, which also shows the intermittent UV feedback
in an early stage, we follow the much longer evolution than
the above.
For cases A and B, mass accretion is still on-going under
intermittent UV feedback for $7 \times 10^4$ years, 
because episodes of
rapid mass accretion are able to interrupt the KH contraction sufficiently often. 
Although it is uncertain how long this evolution can continue for these
cases, UV feedback will likely fix the final stellar masses,
if KH contraction remains uninterrupted for a sufficiently long time. 
Our case C exemplifies this evolution. 
After the protostar enters the supergiant stage for $M_* \gtrsim 50~\msun$,
the star again begins KH contraction, after the stellar mass 
reaches $\sim$$130~\msun$ at $t \simeq 10^4$ years
(Fig.~\ref{fig:xr_suv_xm}).
The bipolar \hii region quickly expands as the star approaches the ZAMS. 
Mass accretion is not immediately shut off 
by UV feedback, however, but continues at a much reduced rate. 
As a result, the star's mass almost doubles by 
$t \simeq 1.2 \times 10^5$ years, when the accretion
rate falls below $10^{-5}~\msunyr$.

%-----------------------------------------------%
\subsubsection{Comparison to 2D SE-RHD Results}
\label{ssec:2Dcomp}
%-----------------------------------------------%

%--------------------------------------------------------------------%
%%% Fig.17 %%%
\begin{figure}
  \begin{center}
\epsscale{1.0}
\plotone{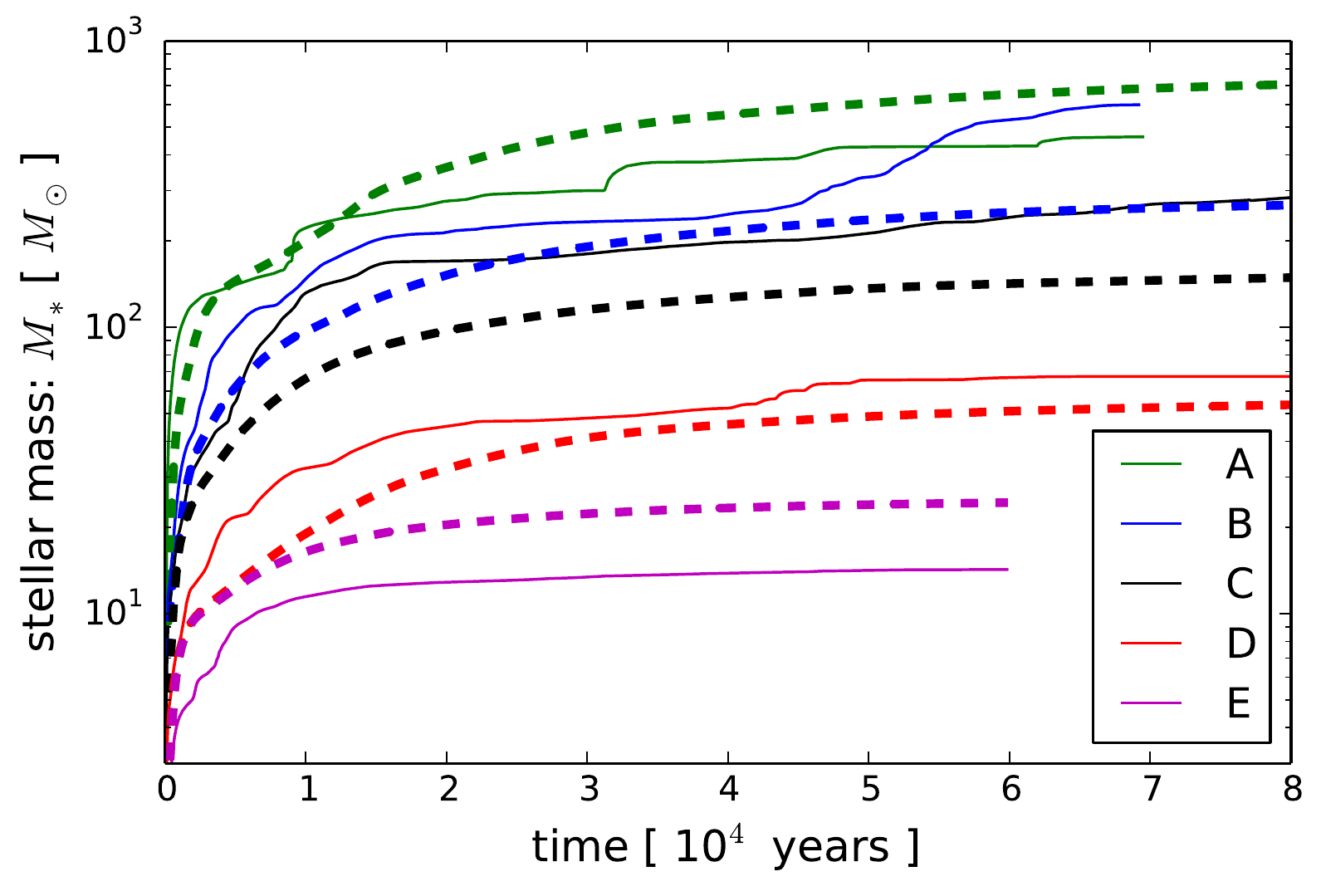}
\caption{Stellar mass growth histories obtained in 
the current 3D SE-RHD simulations compared to our previous 
2D results \citep{Hirano14}.
The solid lines represent the 3D cases with the same colors
as in Figure~\ref{fig:xm_xmdot_t}-(a); 
dashed lines depict the 2D results for the same clouds.
}
\label{fig:3Dto2D}
  \end{center}
\end{figure}
%--------------------------------------------------------------------%
%--------------------------------------------------------------------%
%%% Fig.18 %%%
\begin{figure}
  \begin{center}
\epsscale{1.1}
\plotone{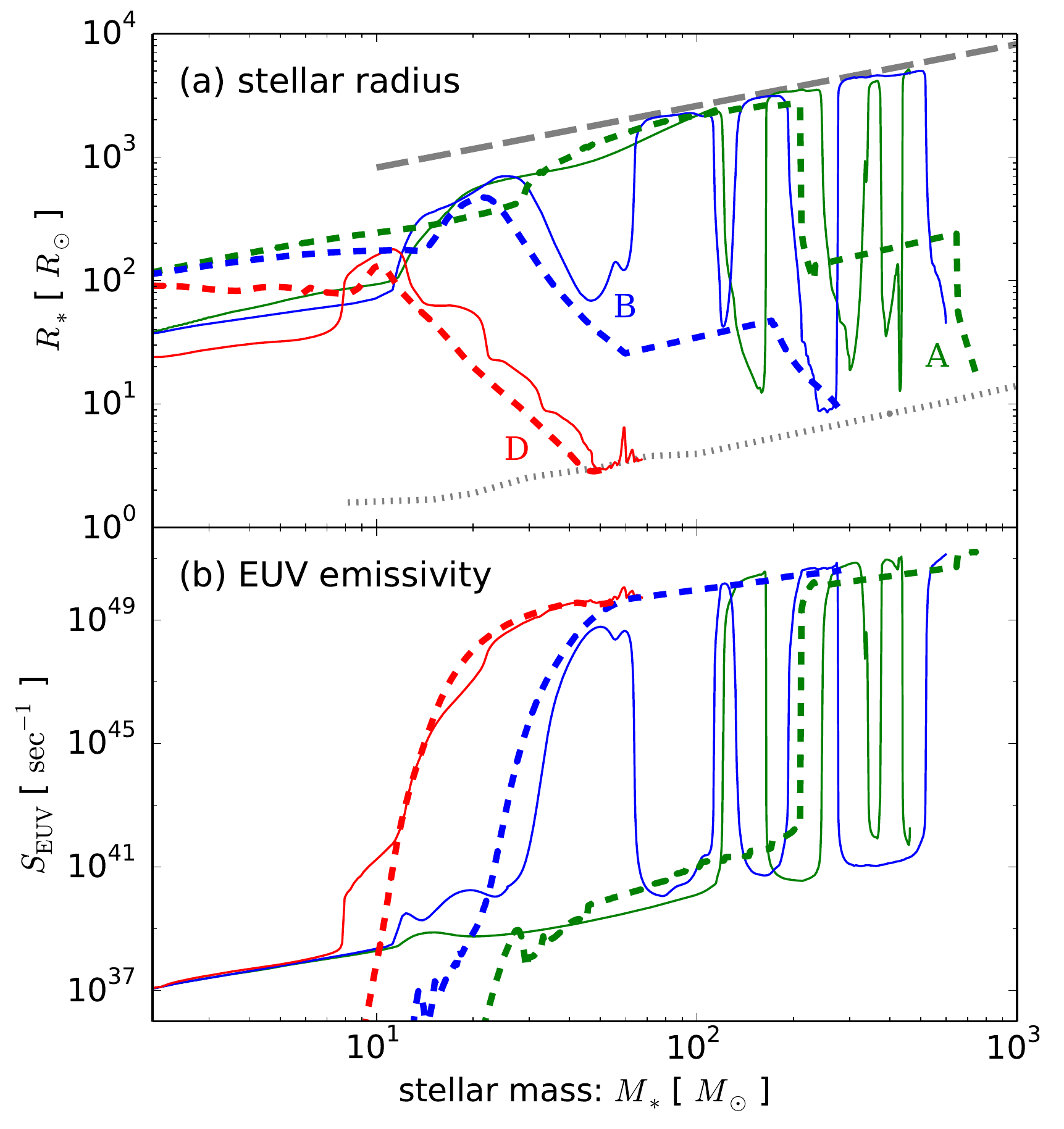}
\caption{
The protostellar evolution of radius and UV emissivities
for the 3D cases considered here in comparison 
to those of our previous 2D calculations \citep{Hirano14}.
This illustration is almost the same as Figure~\ref{fig:xr_suv_xm};
the panels (a) and (b) show the evolution of the stellar radius
and EUV emissivity with increasing mass.
For clarity only cases A, B, and D are presented.
The thick dashed lines represent the 2D results 
for the same clouds as in Figure~\ref{fig:3Dto2D}.
}
\label{fig:3Dto2D_stellar}
  \end{center}
\end{figure}
%--------------------------------------------------------------------%
%--------------------------------------------------------------------%
%%% Fig.19 %%%
\begin{figure}
  \begin{center}
\epsscale{1.0}
\plotone{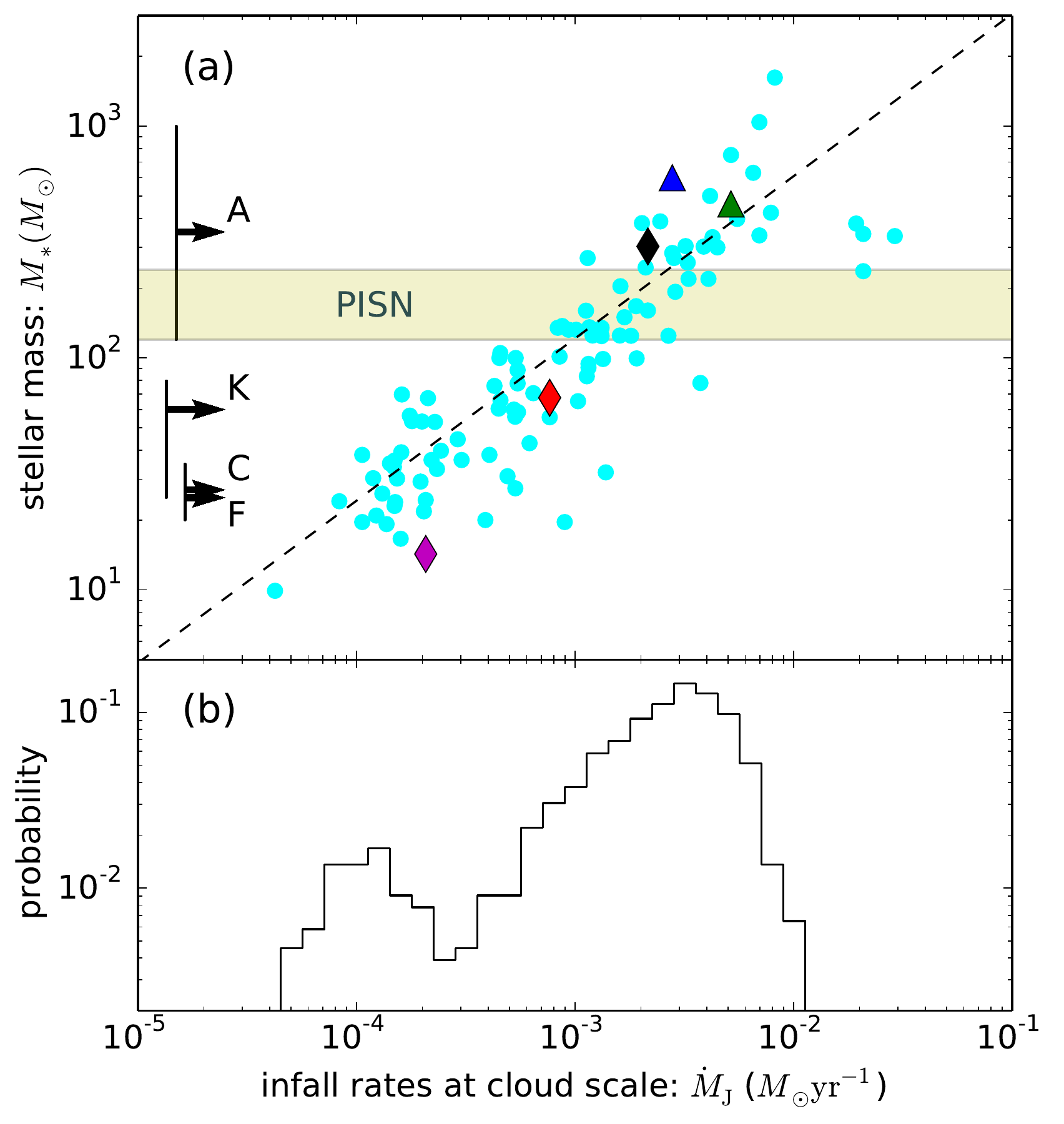}
\caption{{\sl Panel (a)}: correlation between the final stellar masses 
and infall rates measured at the cloud (Jeans) scale
when the cloud central density is $\simeq 10^7~\cmc$. 
The rhombuses and triangles denote the current 3D results 
for cases A (green), B (blue), C (black), D (red), 
and E (magenta). 
The triangles are lower limits to the mass, indicating that mass accretion 
continues beyond
$t = 7 \times 10^4$ years, the end of the simulations.
The cyan dots represent the 2D SE-RHD results for 
about 110 cosmological cases studied by \citet{Hirano14}.  
The dashed line represents the fitting formula for these
110 cases given by equation~\ref{eq:mjcorr}.
The horizontal yellow bar denotes the expected mass
range of progenitors of pair-instability supernovae,
$120~\msun \lesssim M_* \lesssim 240~\msun$ 
\citep[][]{YDL12}.
The arrows mark the estimated stellar masses of
Pop III supernova progenitors, which have been 
estimated from the abundance patterns of
recently-discovered Galactic metal deficit stars:
(A) SDSS J001820.5-093939.2 \citep{Aoki14},
(K) SMSS J031300.36-670839.3 \citep{Keller14},
(C) SDSS J102915+172927 \citep{Caffau11}, and
(F) SD 1313-0019 \citep{Frebel15}.
The vertical bar associated with each arrow represents
the possible mass range discussed in the literature
\citep[e.g.,][]{Schneider12,Ishigaki14}. 
{\sl Panel (b)}: probability distribution of the cloud-scale
infall rates in the 1540 cosmological cases studied by
\citet{Hirano15} (also see their Fig.~17). 
}
\label{fig:xmdotj_xm}
  \end{center}
\end{figure}
%--------------------------------------------------------------------%

%--------------------------- Fig. 17 -------------------------------%

As mentioned in Section~\ref{ssec:cases_cosmo}, HR14 have 
previously studied the evolution of protostellar accretion 
for the same primordial clouds considered above, but
under the assumption of 2D axial symmetry. 
Figure~\ref{fig:3Dto2D} presents the stellar mass growth histories for both
the current 3D and the previous HR14 2D results in comparison.  
We see that the 3D results roughly match the overall trends of 
the 2D results, in spite of a number of technical differences
between the adopted numerical codes.
Comparing the 2D and 3D results for each case, the stellar masses 
at a given epoch differ by a factor of a few at most. 
Some differences come from the fact that, 
for a 2D simulation, the original 3D structure of 
a natal cloud is modified to fit the 2D axisymmetric grid by 
averaging physical quantities over the azimuthal direction.
Moreover, the 2D results also depend on the somewhat ad-hoc use of an
$\alpha$-viscosity to model the gravitational torque operating
in self-gravitating disks.

%--------------------------- Fig. 17 -------------------------------%

Figure~\ref{fig:3Dto2D_stellar} shows the protostellar
mass and UV emissivities as a function of stellar mass 
for several representative cases. 
For case D, the evolution is quite similar between 3D and 2D cases. 
The qualitative differences appear for cases A and B, where
both the stellar radius and EUV emissivity abruptly change
for $\msun \gtrsim 100~\msun$ because of the variable accretion 
encountered in 3D
(also see Section~\ref{ssec:VMstar}).
In our previous 2D cases, by contrast, there was an
intermediate stage between the supergiant and ZAMS stage, 
where the stellar radius gradually increases with the mass
(named the ``P2'' path in HR14). 
Stellar evolution calculations demonstrate that this behavior
is generally present, when the accretion rates are in the
range of $4 \times 10^{-3}~\msunyr \lesssim \mdot \lesssim 10^{-2}~\msunyr$
\citep[e.g.,][]{OP03}.
For 2D cases, this stage appears because the accretion rates 
smoothly evolved with time. In 3D, however, the highly variable
accretion rates do not remain in this narrow range very long.
The intermediate P2 stage thus does not occur in our 3D cases.
Instead, the protostar evolves back and forth between the supergiant and ZAMS 
stages depending on details of the variable accretion history.
This effect reduces the stellar EUV emissivity for 
$M_* \gtrsim 100~\msun$ in comparison with the 2D cases, 
as episodes of rapid accretion bring the star
back to the supergiant stage several times.

%---------------------------- Fig. 18a ------------------------------%

Figure~\ref{fig:xmdotj_xm}-(a) displays the stellar masses at the end of
the 3D simulations as a function of the infall rates measured at
the cloud (Jeans) scale (see Section~\ref{ssec:cases_cosmo}),
demonstrating that the 3D results still show a similar correlation
as found in 2D.
Considering the fact that the masses for cases A and B 
are only lower limits, 
the correlation found in 3D appears to have a steeper slope
than for the 2D results (equation~\ref{eq:mjcorr}),  
albeit with much poorer statistical significance (five cases in 3D).
This does not necessarily agree with \citet{Susa14}, who do not find 
this correlation in their 3D RHD simulations.
We discuss this issue in Section~\ref{ssec:multi} separately. 

%---------------------------- Fig. 18b ------------------------------%

As Figure~\ref{fig:xmdotj_xm}-(b) shows, primordial clouds taken
for cases A--C offer rather typical initial conditions for primordial
star formation. 
Our results suggest that the very massive stars, even above the 
mass range of the pair-instability supernovae (PISN) progenitors 
$120~\msun \lesssim M_* \lesssim 240~\msun$ \citep[][]{YDL12},
can form for such typical cases.
A significant number of $\sim$$10^3~\msun$ BHs might be produced 
as remnants of these very massive primordial stars. 
We shall discuss possible observational signatures of
such massive primordial stars in Section~\ref{ssec:obsig}.

%++++++++++++++++++++++++++++++++++++++++++++++%
\subsection{Varying Spatial Resolution}
\label{ssec:sres}
%++++++++++++++++++++++++++++++++++++++++++++++%

%------------------------------------------------------%
\subsubsection{Disk Fragmentation and Accretion Bursts}
\label{sssec:diskfrag_sres}
%------------------------------------------------------%

Although our 3D simulations show a new 3D effect of the
intermittent UV feedback, the ansatz described in 
Section~\ref{ssec:flimit} can bring some 
resolution-dependence which can affect the evolution 
quantitatively. 
It is thus necessary to explore how resolution
affects the results of our simulations. 
First, we shall focus on case B described in Section~\ref{ssec:VMstar},
for which UV feedback becomes important when
the stellar mass exceeds $200~\msun$ at 
$t \gtrsim 2 \times 10^4$ years (Fig.~\ref{fig:xm_t_B}).
For this case, we study effects of increasing the grid resolution 
during early times before UV feedback begins to operate.  

%----------------------------------------------------------------%

We first recalculate the evolution of case B with our standard resolution, 
turning off UV feedback (case B-NF; see Table~1).
We use four model configurations of case B-NF at different epochs for 
$M_* \simeq 0$, $40$, $70$, and $120~\msun$ as
the starting configurations for calculations at higher spatial resolution. 
Next, we interpolate each of the four starting model configurations
onto a grid with double the resolution for each spatial coordinate
and follow the evolution for 3000 years. 
These test cases are named as B-NF-HR2-mX, where the suffix
``mX'' represents the restarting point at $M_* \simeq X~\msun$
(see Table~1).
The duration of 3000 years is comparable
to the Kepler orbital period
$P_{\rm Kepler} \simeq 1600$~yr for
typical values of stellar mass $M_* \simeq 50~\msun$ 
and disk radius $\sim$500~AU (c.f. equation \ref{eq:pkep}).
The accretion envelope surrounding the disk typically
has a density $\lesssim 10^9~\cmc$, for which the
free-fall timescale is $\gtrsim 3 \times 10^3$ years.
Mass infall from the envelope onto the disk at the
higher resolution hardly affects the evolution, whereas
cooling, fragmentation, and the associated mass and 
angular momentum transport within the disk occur on
much shorter timescales and therefore should be 
affected by varying spatial resolution.

%----------------------------- Fig.19 --------------------------------%

\begin{figure}
  \begin{center}
\epsscale{1.0}
\plotone{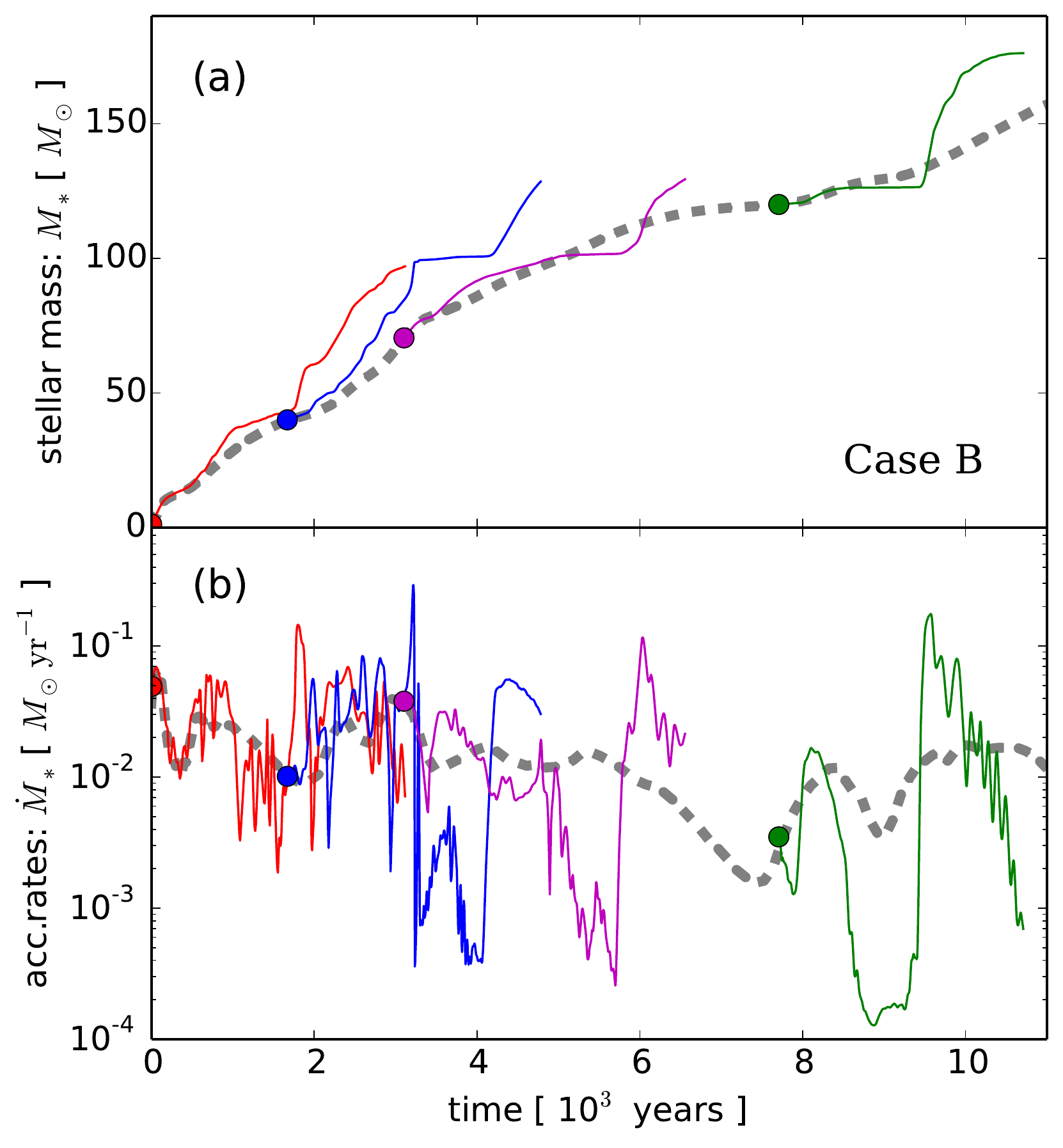}
\caption{Comparison of the mass growth histories with different
spatial resolutions in halo B cases. The panels (a) and (b) show
the time evolution of the stellar masses and accretion rates onto
the protostar. In both panels, the thick dashed lines 
represent case B-NF. The solid lines represent the cases
with $2$ times higher spatial resolution:
B-NF-HR2-m0 (red), B-NF-HR2-m40 (blue), B-NF-HR2-m70 (magenta), 
and B-NF-HR2-m120 (green).
The filled circles mark the starting points of these 
doubled-resolution cases, i.e., when the stellar mass is 
$\simeq 1~\msun$, $40~\msun$, $70~\msun$, and $120~\msun$ for
case B-NF (also see text).
}
\label{fig:hn1_B}
  \end{center}
\end{figure}

Figure~\ref{fig:hn1_B} compares the stellar mass growth 
and accretion histories for the examined cases.  
We see that mass accretion becomes more strongly variable with higher
resolution; i.e., the accretion rates vary over shorter timescales and
with much higher variations: 
$10^{-4}~\msunyr \lesssim \mdot \lesssim 0.1~\msunyr$ (Fig.~\ref{fig:hn1_B}-b).
Furthermore, the mass that the star accretes during the $3 \times 10^3$ years
is somewhat larger with higher resolution, independent of the different 
starting points of the doubly-resolved cases (Fig.~\ref{fig:hn1_B}-a). 
This can be understood with the help of Figure~\ref{fig:caseB_disk_rescomparison}, 
which displays the face-on disk structure at $t \simeq 1.2 \times 10^3$ years
for the different resolutions.
The disk clearly has sharper and finer density structure
at higher resolution. 
This is because radiative cooling is less restricted by our ansatz
and the disk is thus more gravitationally unstable.
Since gravitational torques are caused by the non-axisymmetric
disk structure such as spiral arms, 
increasing the resolution can lead to more efficient mass and
angular momentum transport through the disk,
which results in a higher stellar mass.

%----------------------------- Fig.20, 21 --------------------------------%

\begin{figure}
  \begin{center}
\epsscale{1.0}
\plotone{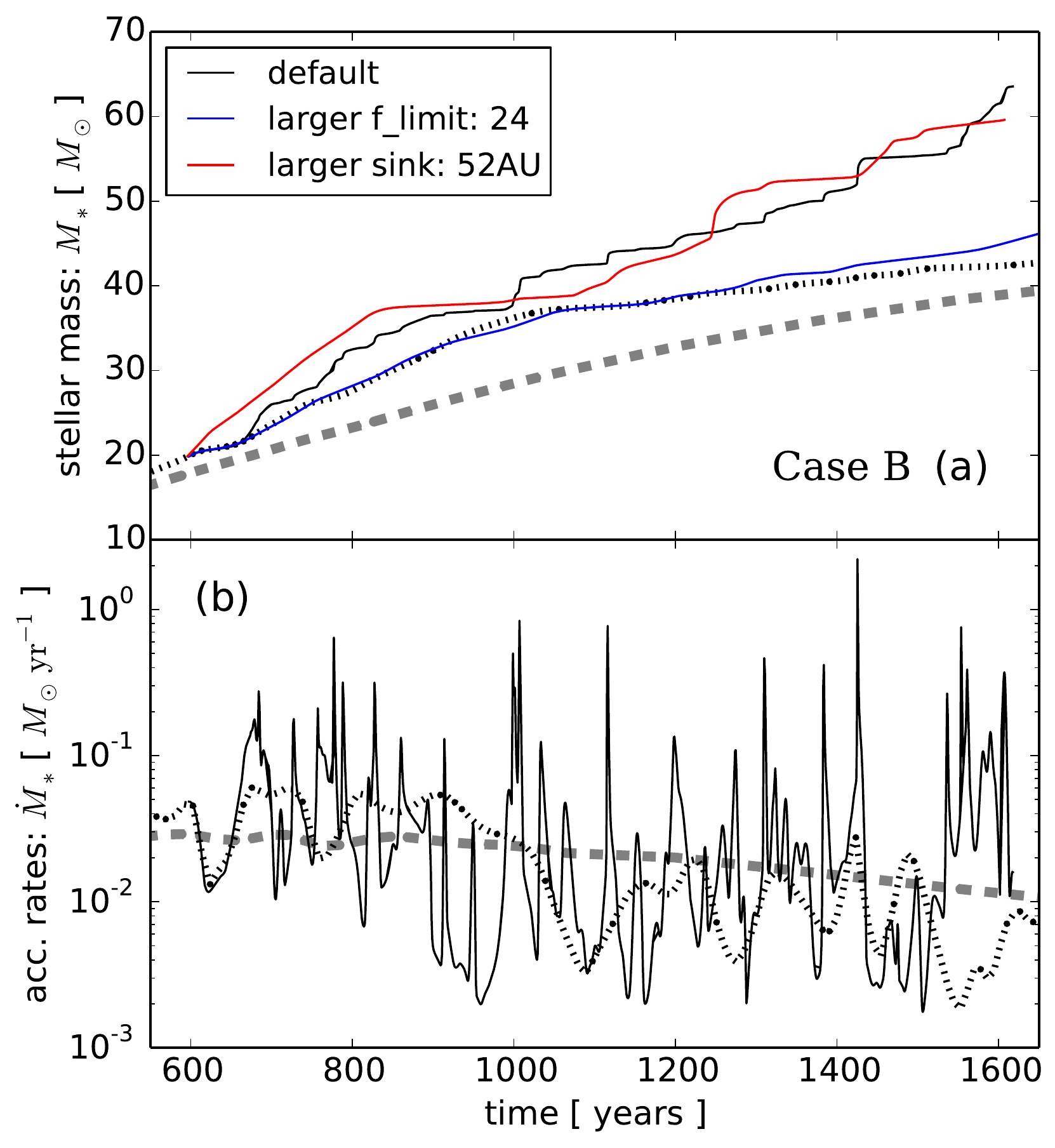}
\caption{Same as Figure~\ref{fig:hn1_B} but including
cases at $4$ times higher spatial resolution 
than case B-NF (thick dashed lines) and $2$ times higher
resolution than case B-NF-HR2-m0 (dotted lines). 
The solid lines represent cases (see text and Table 1)
B-NF-HR4-m20 (black), B-NF-HR4-m20-fc (blue), 
and B-NF-HR4-m20-lsk (red).
In panel (b), accretion histories for cases 
B-NF, B-NF-HR2-m0, and B-NF-HR4-m20 only are shown for clarity.
}
\label{fig:hn1p2_B}
  \end{center}
\end{figure}

We once again double the resolution using case 
B-NF-HR2-m0 at $t \simeq 600$ years as a starting model, a time
at which $M_* \simeq 20~\msun$, and recalculate the evolution 
(case B-NF-HR4-m20). We repeat these higher
resolution simulations with a larger cooling limit  $f_{\rm limit}=24$
(case B-NF-HR4-m20-fc; see Section~\ref{ssec:flimit}) 
and a larger sink cell radius $r_{\rm min} = 52$~AU (case B-NF-HR4-m20-lsk).
The resolution is now $4$ times higher than the original case B-NF.
Because of the high computational cost, the highest resolution
simulations are evolved for only $10^3$~years.
Figure~\ref{fig:hn1p2_B} compares the stellar mass growth and
accretion histories of these quadruple resolution cases
with the results of the lower resolution simulations.
The resolution dependences discussed above appear again
more prominently.

%------------------------------------------------------------------%

\begin{figure}
  \begin{center}
\epsscale{1.0}
\plotone{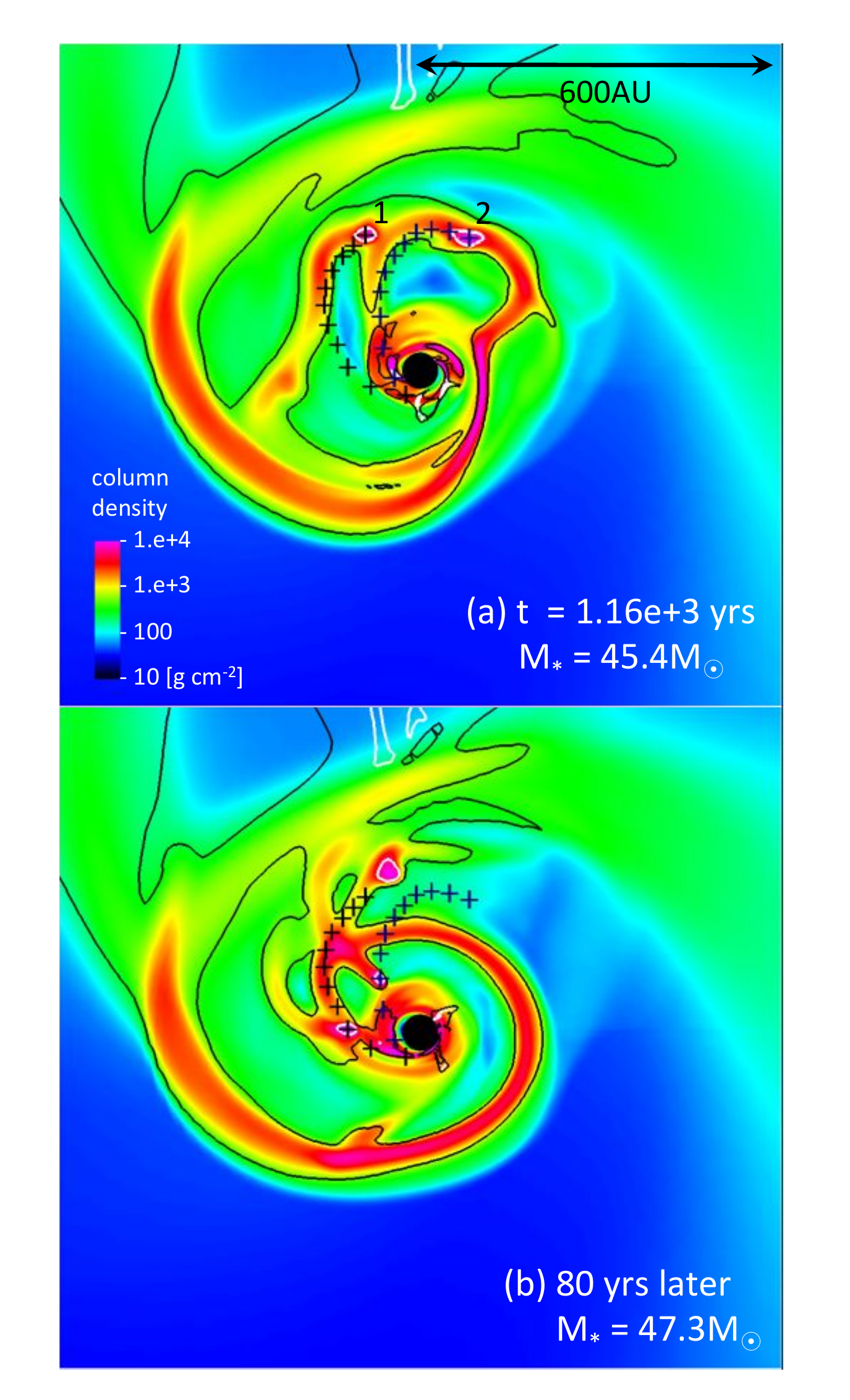}
\caption{Inward migration of two fragments (crosses)
for case B-NF-HR4-m20. 
As in Figure~\ref{fig:caseB_disk_rescomparison}, 
colors represent the gas column density
but for a smaller region around the central star.  
The black (white) contours delineate the boundaries 
where the Toomre-Q parameter takes the value
$Q=1.0$ ($Q=0.1$). Spiral arms and fragments consistently
have $Q<1$.
{\sl Panel (a):} Evolutionary age
$t = 1.16 \times 10^3$ years (the same moment as in 
Fig.~\ref{fig:caseB_disk_rescomparison})
{\sl Panel (b):} 80 years later.
The black and dark-blue crosses mark the positions 
of the two fragments every 10 years.
}
\label{fig:caseB_TMRQ_v}
  \end{center}
\end{figure}

\begin{figure}
  \begin{center}
\epsscale{1.0}
\plotone{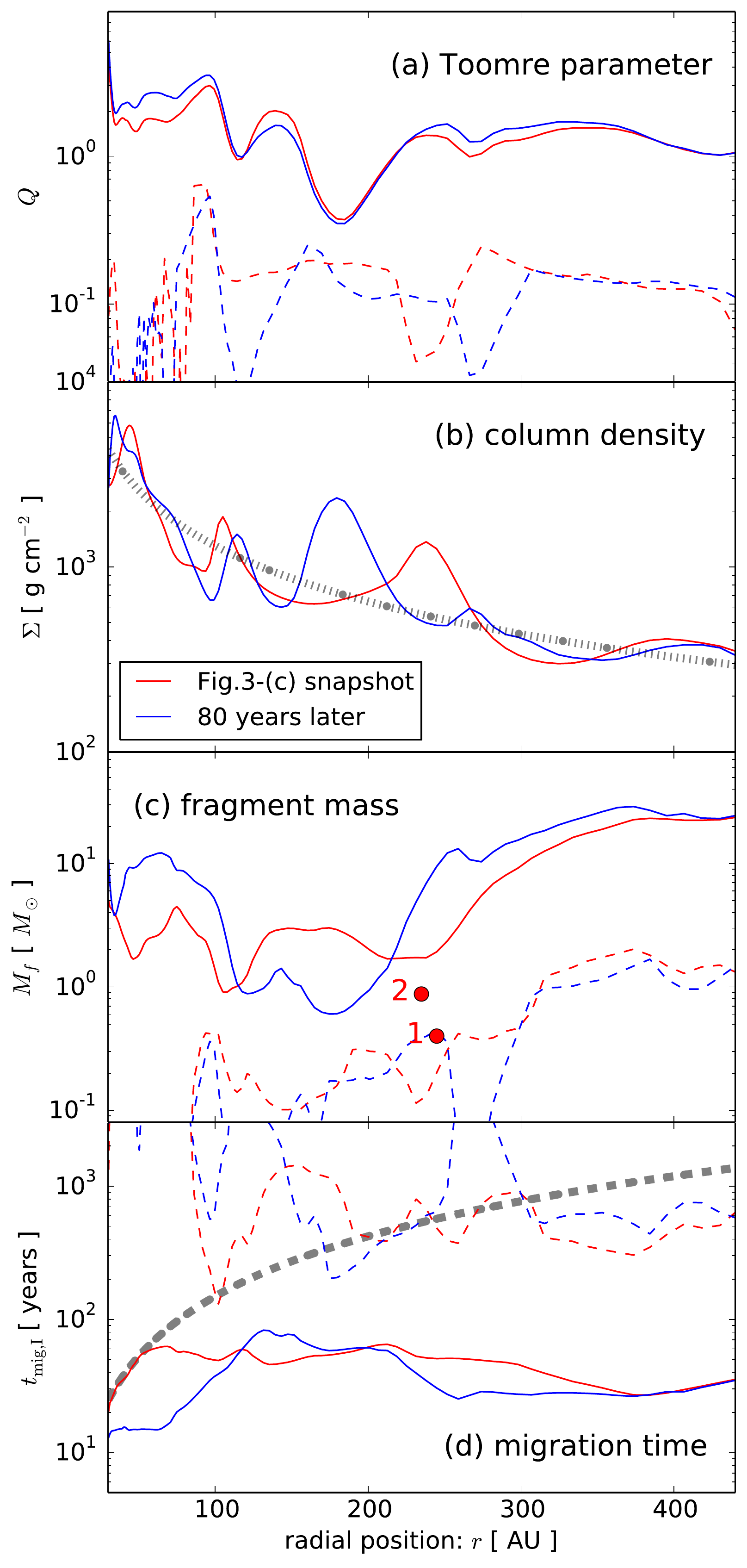}
\caption{
Radial distributions of physical quantities
for the circumstellar disks shown in Figure~\ref{fig:caseB_TMRQ_v}: 
(a) Toomre $Q$-parameter, (b) vertical column density $\Sigma$, 
(c) fragment mass $M_f$ (equation~\ref{eq:mfrag}), 
and (d) type I migration timescale (equation~\ref{eq:tmigI}).
In each panel, the red curves pertain to the evolutionary state depicted
in Figure~\ref{fig:caseB_disk_rescomparison}-(c) and 
Figure~\ref{fig:caseB_TMRQ_v}-(a), whereas the blue curves correspond to
80 years later (see Figure~\ref{fig:caseB_TMRQ_v}-(b)).
These two epochs are also displayed in Figure~\ref{fig:caseB_xMfrag_v}.
In each panel, solid curves depict quantities averaged over the 
azimuthal $\phi$ direction.
The dashed lines in panel (a) represent the minimum values $Q_{\rm min}(r)$
in an annulus of radius $r$, which roughly trace the $Q$ values
through either spiral arms or fragments.
In panel (b), the thick dotted line shows
the profile $\Sigma \propto r^{-1}$ for reference.
The colored dashed lines in panels (c) and (d) represent the mininum
values of the estimated fragment mass and corresponding migration
timescale in the annulus of radius $r$.
In panel (c), the red filled circles indicate the masses of
the two fragments marked in Figure~\ref{fig:caseB_TMRQ_v}-(a).
In panel (d), we also plot the Kepler orbital period
$P_{\rm Kepler} = 2 \pi \sqrt{r^3/G M_*}$ for $M_* = 45.4~\msun$
with a black dashed line.
}
\label{fig:diskprof}
  \end{center}
\end{figure}

Case B-NF-HR4-m20's accretion history displays
lots of spiky features, representing multiple accretion burst
events (Fig.~\ref{fig:hn1p2_B}-b). 
The stellar mass growth over $10^3$ years is correspondingly
greater at higher resolution (Fig.~\ref{fig:hn1p2_B}-a).
Figures~\ref{fig:caseB_disk_rescomparison}-(c) and
\ref{fig:caseB_TMRQ_v} show that,
at this high resolution, one of the spiral arm 
breaks up and creates two fragments. 
These and Figure~\ref{fig:diskprof}-(a) show the 
distributions of the Toomre $Q$-parameter on the equator,
defined by
\begin{equation}
Q = \frac{c_s \Omega}{\pi G \Sigma},
\end{equation}
where $c_s$ is the local sound speed and 
$\Sigma$ is the vertical column density. 
Because the spiral arms have $Q \lesssim 1$, 
gravitational instability makes them prone to fragmentation.
This feature is not seen
at lower resolution for the same epoch 
(Fig.~\ref{fig:caseB_disk_rescomparison}-a, b).
The evolution for this case is quite similar to
that shown in \citet{Vorobyov13}, who study
disk fragmentation using 2D face-on (thin disk) simulations 
at even higher resolution than ours.

%--------------------------------------------------------------------%
%%% Fig.20 %%%
%--------------------------------------------------------------------%
%--------------------------------------------------------------------%
%%% Fig.21 %%%
%--------------------------------------------------------------------%
%--------------------------------------------------------------------%
%%% Fig.22 %%%
%--------------------------------------------------------------------%
%--------------------------------------------------------------------%
 %%% Fig.23 %%%
%--------------------------------------------------------------------%
%--------------------------------------------------------------------%
%%% Fig.24 %%%
\begin{figure}
  \begin{center}
\epsscale{1.0}
\plotone{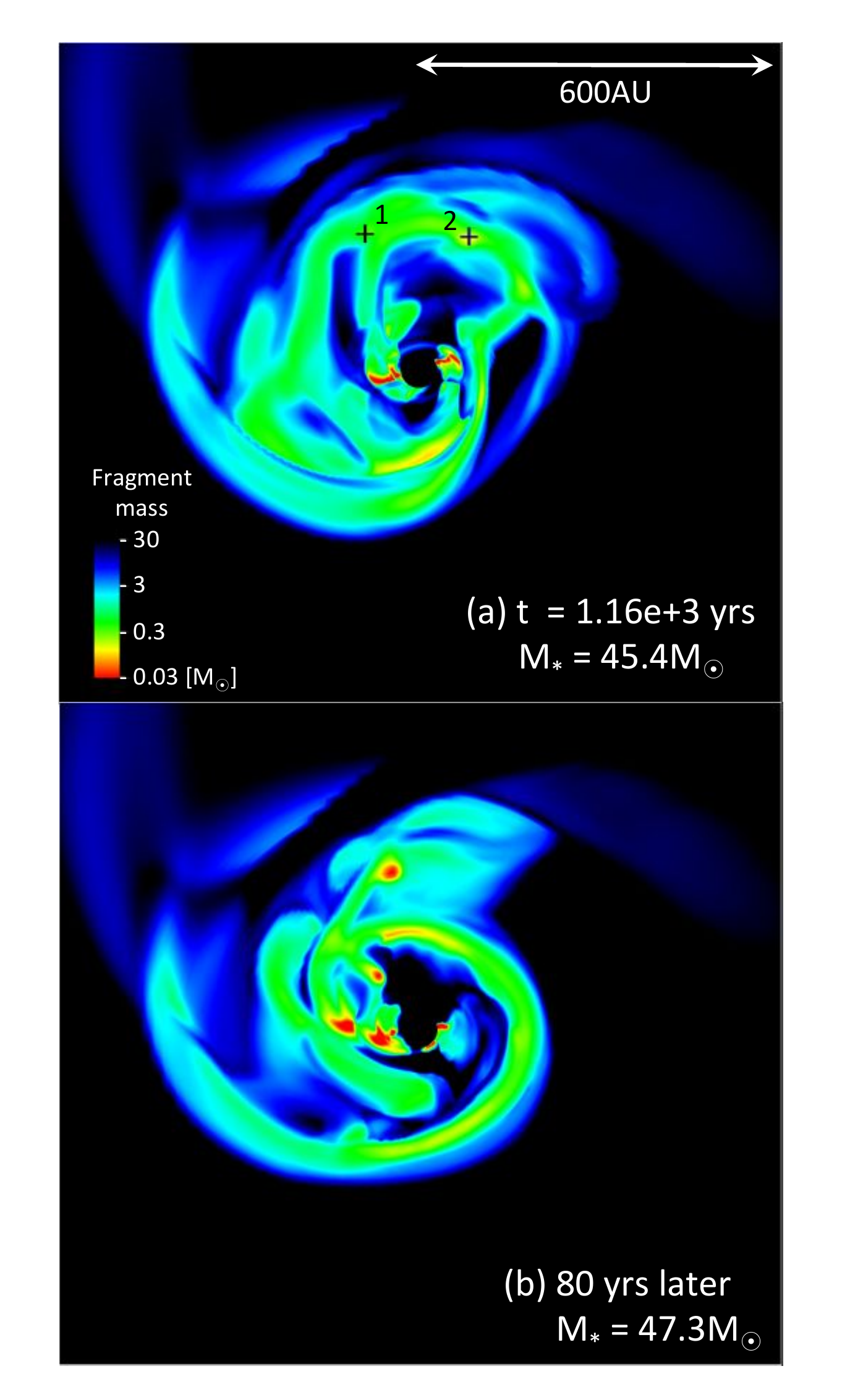}
\caption{Same as Figure~\ref{fig:caseB_TMRQ_v} but for the fragment
masses evaluated locally with equation (\ref{eq:mfrag}).
In panel (a), we only mark the positions of the two fragments
referred to in Figures~\ref{fig:caseB_TMRQ_v} and 
\ref{fig:diskprof} for this epoch.}
\label{fig:caseB_xMfrag_v}
  \end{center}
\end{figure}
%--------------------------------------------------------------------%

%--------------------- Fig.22 (migrating inward) ---------------------%

Our results suggest that the stellar mass growth is not hindered,
but rather accelerated due to disk fragmentation. 
This is because newly created fragments rapidly migrate inward 
through the disk and ultimately accrete onto the star.
This behavior is shown in Figure~\ref{fig:caseB_TMRQ_v},
tracking the orbital motion of the fragments seen in 
Figure~\ref{fig:caseB_disk_rescomparison}-(c). 
Both fragments rapidly migrate inward and reach the
star in less than 100 years.
Recent 3D simulations by \citet{Greif12} also report a
similar rapid migration in highly gravitationally unstable disks. 
\citet{Greif12} conclude that the typical timescale for such migrations 
is roughly the local free-fall timescale, the shortest
timescale for any physical process driven by gravity.
For our case B-NF-HR4-m20, the density near the disk midplane
is $n \gtrsim 10^{11}~\cmc$, for which the free-fall 
timescale is $t_{\rm ff} \lesssim 2.5 \times 10^2$ years. 
The migration timescale is therefore also of the order of the local
free-fall timescale in our calculations. 
The rapid inward migration of the fragments in gravitationally
unstable disks is also commonly seen in simulations of present-day 
star and planet formation
\citep[e.g.,][]{VB06,Baruteau11,Machida11,Zhu12}.
These studies show that rapid migration occurs
roughly over the so-called type I migration timescale 
\begin{equation}
t_{\rm mig, I} = \frac{h^2}{4 C \mu q} \frac{2 \pi}{\Omega}
\label{eq:tmigI}
\end{equation}
\citep{Tanaka02}, where $C = 2.7 + 1.1 \xi$, 
$\xi$ is the radial gradient of the column
density $\Sigma \propto r^{- \xi}$, $h$ is the disk aspect ratio, 
$\mu = \pi \Sigma r^2 / M_*$, and $q = M_f/M_*$, where
$M_f$ is the fragment mass. 
Figure~\ref{fig:diskprof}-(b) shows that the column density profile
in our disk is well approximated by $\Sigma \propto r^{-1}$, i.e.,
$\xi = 1$. The fragment mass $M_f$ is approximated by the mass enclosed
within a spatial scale of the most unstable wavelength 
$\lambda = 2 c_s^2 / G \Sigma$,
\begin{equation}
M_f = \pi (\lambda/2)^2 \Sigma = \frac{\pi c_s^4}{G^2 \Sigma} .
\label{eq:mfrag}
\end{equation} 
Figure~\ref{fig:diskprof}-(c) shows that the $\phi$-averaged 
fragment masses estimated by equation (\ref{eq:mfrag}) span
the range from $\sim$$1~\msun$ to several $10~\msun$.
Because there are large density fluctuations in the disk,
the locally-estimated fragment mass varies considerably.
Figure~\ref{fig:caseB_xMfrag_v} shows 
$M_f \sim 0.1 - 1~\msun$ along the spiral arms, 
where the fragmentation often occurs.
We also directly calculate the masses of the two fragments seen in
Figure~\ref{fig:caseB_TMRQ_v}-(a) (marked as 1 and 2),
by summing up the masses in the cells with the column density
higher than $5000~{\rm g~cm^{-2}}$ around each fragment.
The resulting values,
$M_{f1} \simeq 0.4~\msun$ and $M_{f2} \simeq 0.9~\msun$, 
are in rough agreement with the estimated values in 
Figure~\ref{fig:caseB_xMfrag_v} (see also Figure~\ref{fig:diskprof}-(c)).

%------------------------------------------------------------------%

In Figure~\ref{fig:diskprof}-(d), we plot the 
type I migration timescale given by equation (\ref{eq:tmigI}). 
Considering the fragment masses discussed above, 
we estimate that the corresponding
migration timescales should be around $\sim$100 years, which are
shorter than or comparable to the Kepler orbital periods.
The rapid inward migration seen in our simulations can be well
explained by the above evaluation.
In addition to the two infalling fragments
shown in Figures~\ref{fig:caseB_disk_rescomparison}-(c) and
\ref{fig:caseB_TMRQ_v}, several similar events were
observed during the simulation duration of $10^3$ years. 
Some of the accretion bursts seen in Figure~\ref{fig:hn1p2_B}, 
although not all, are caused by such events.

%-------------------------- experimental cases -------------------------%

As mentioned earlier, we have also performed quadruple resolution
simulations with different numerical parameters.
For case B-NF-HR4-m20-fc, for instance,
%ne case marked with the suffix ``fc'', for instance, 
we adopt $f_{\rm limit} = 24$ instead of the fiducial 
value $f_{\rm limit} = 12$, where $f_{\rm limit}$ 
sets the cooling limit threshold (equation~\ref{eq:flimit}).
As described in Section~\ref{ssec:flimit}, a larger value for 
$f_{\rm limit}$ sets a more stringent limit on cooling.
Figure~\ref{fig:hn1p2_B} shows that the evolution for this case 
is quite similar to that for the double resolution case B-NF-HR2-m0, 
verifying that the resolution dependences examined above really come from 
the more efficient cooling allowed with higher resolution.
%%%
For case B-NF-HR4-m20-lsk, on the other hand,  
we adopt a larger sink $r_{\rm min} = 52$~AU 
instead of the fiducial value $r_{\rm min} = 30$~AU.
The stellar mass growth history for this 
case roughly traces that for case B-NF-HR4-m20,
although with less variable accretion rates.
The shortest timescale of accretion variability is roughly
given by the Kepler orbital timescale near the radial
inner boundary at $r = r_{\rm min}$.

%----------------------------------------------------------------------%
\subsubsection{Stellar UV Feedback with Variable Accretion}
\label{sssec:fdbk_sres}
%----------------------------------------------------------------------%

%--------------------------------------------------------------------%
%%% Fig.25 %%%
\begin{figure}
  \begin{center}
\epsscale{1.0}
\plotone{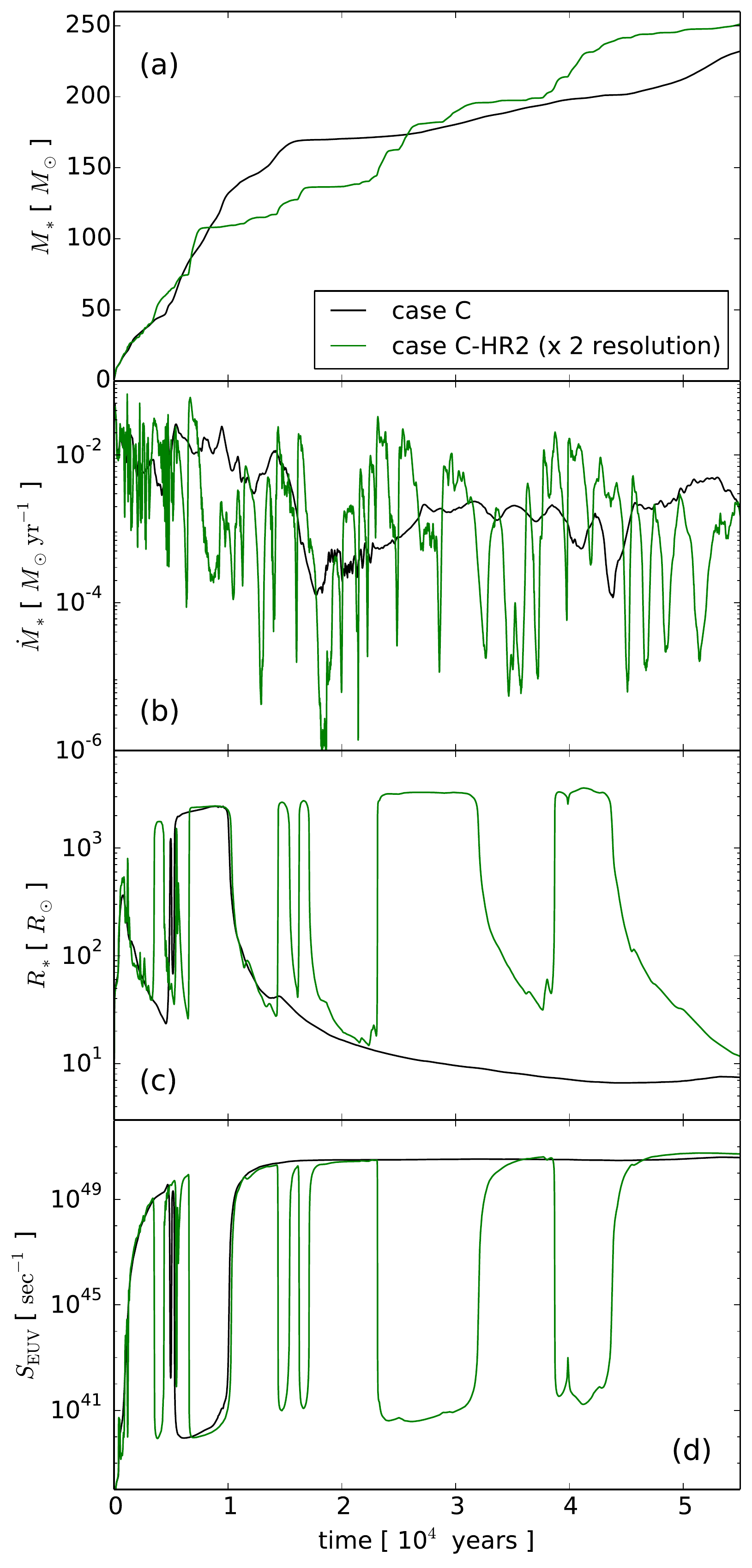}
\caption{Effects of doubling the spatial resolution for a long-term 
evolution under the influence of UV feedback for case C. 
The evolution of (a) stellar mass, (b) accretion rate, 
(c) stellar radius, and (d) emissivity of ionizing photons
is presented.
In each panel, the black and green lines represent the fiducial
case C and a test case with $2$ times higher 
spatial resolution, C-HR2 (also see Table~1). 
}
\label{fig:rescompC}
  \end{center}
\end{figure}
%--------------------------------------------------------------------%

Next we investigate the effects of varying the grid resolution
for evolutionary stages when UV feedback becomes important.
For this purpose, we take case C (see Section~\ref{ssec:VMstar})
as the basis for comparison.
Unlike case B considered above, UV feedback 
operates earlier and at lower stellar masses for case C
(e.g., Figs.~\ref{fig:HIIr1e4AU_A2E} 
and \ref{fig:HIIr1e4AU_A2E_C}). 
This enables us to follow the evolution with UV feedback 
for a longer period of time at a reasonable computational cost. 
We double the resolution for the initial configuration, 
$M_* = 0~\msun$, and follow the evolution for
$5.5 \times 10^4$ years (case C-HR2).

%----------------------------- Fig. 23 --------------------------------%

Figure~\ref{fig:rescompC} displays the stellar mass (a) accretion
rate (b), stellar radius (c), and UV emissivity (d) as a function of time
for both case C and C-HR2. In agreement with what we have shown 
in Section~\ref{sssec:diskfrag_sres}, the accretion rate is more highly variable 
at the higher resolution (Fig.~\ref{fig:rescompC}-b).
After the different accretion histories over 
$5.5 \times 10^4$ years, however, the final masses differ only by $\sim$$10$~\%.
This is because the current evolutionary timescale is much longer 
than the local Kepler timescale within the disk
(equation~\ref{eq:pkep}), so that we are no longer considering
the short term evolution of a nearly isolated disk; 
the mass supply from the envelope ultimately controls stellar 
growth via mass accretion through the disk.
Since the stellar radius abruptly changes with the variable
mass accretion, the UV emissivity also displays very large
fluctuations (Fig.~\ref{fig:rescompC}-c, d).
With the doubled resolution, in particular, the protostellar 
bloating and subsequent KH contraction are repeated several
times because of multiple accretion bursts.  
The extinction and re-formation of bipolar \hii
regions also occurs accordingly.
In comparison with the original case C, UV feedback
begins to limit stellar mass growth earlier at 
$M_* \simeq 100~\msun$, but the star nevertheless accretes a
larger amount of gas ($\simeq 150~\msun$) after the onset of 
significant UV feedback.
The overall result is a slightly higher stellar mass at the end of the 
simulation than for the original case C.

%%%%%%%%%%%%%%%%%%%%%%%%%
\section{Discussion}
\label{sec:discussion}
%%%%%%%%%%%%%%%%%%%%%%%%%

%+++++++++++++++++++++++++++++++++++++++++++++%
\subsection{Impact of Dissociation Feedback}
\label{ssec:fuv_fdbk}
%+++++++++++++++++++++++++++++++++++++++++++++%

As described in Section~\ref{ssec:OMstar}, the accretion limiting effect of 
stellar dissociating (FUV) radiation is modest in our simulations;
FUV feedback slightly reduces mass accretion onto the 
star, but does not completely stop it, as opposed to ionizing 
(EUV) feedback (Figs.~\ref{fig:xm_t_D} and \ref{fig:caseD_comparison}). 
This is in contrast to the conclusions of \citet{Susa13} and
\citet{Susa14}, who show that mass accretion can be
halted by FUV feedback alone. 
EUV feedback is not included in their SPH/N-body
simulations because of the limited spatial resolution, especially
in polar regions where an \hii region should first appear. 
To examine the roles of FUV feedback for our simulations
in comparison with theirs, we return to our case D-DF, for which 
EUV feedback had been artificially turned off.

%------------------- H2 self-shielding near the disk inner-edge? ------------------%

As described in \citet{Susa13}, a circumstellar disk of 
density $n \gtrsim 10^{11}~\cmc$ is ultimately dispersed
by FUV feedback in their simulations. 
This is because H$_2$ formation heating mainly due to the three-body
reaction efficiently heats the gas 
when the dense gas ($n \gtrsim 10^{10}~\cmc$) is exposed to FUV radiation. 
As the disk disperses, the gas density around stars rapidly
falls to $n \lesssim 10^7~\cmc$, resulting in low accretion rates
onto the stars. 
For our case D-DF, however, FUV radiation heats up 
the gas primarily in the polar regions throughout the course of 
$8 \times 10^4$ years evolution. 
The dense disk with $n~\gtrsim~10^{11}~\cmc$ survives in our
case D-DF simulation, being protected against the FUV radiation
by efficient H$_2$ self-shielding in the inner regions of the disk. 
The pressure excess in the polar PDR is much less than
what an \hii region can provide when
stellar EUV radiation is present. 
Unlike the effects of EUV feedback, the pressure structure in the 
envelope is hardly modified outside the PDR.
Therefore mass accretion onto the disk continues from the envelope
as well as from the disk onto the star
with little modification by
FUV feedback.

%----------------------------------------------------------------------%

It seems that the disagreement outlined above stems from
different efficiencies of H$_2$ self-shielding 
in the vicinity of the star. 
Both simulations suffer from the inability
to exactly model this effect in the innermost regions $r \lesssim 30$~AU,
currently masked by the sink. 
Numerical simulations devoted to resolving small-scale
structure near the stellar surface will be needed in the future.

%+++++++++++++++++++++++++++++++++++++++++++++++++++++++++++++%
\subsection{Formation of Binary and Multiple Stellar Systems}
\label{ssec:multi}
%+++++++++++++++++++++++++++++++++++++++++++++++++++++++++++++%

Our current simulations do not show clear 
indications for the formation of binary or multiple stellar systems.
As explained in Section~\ref{ssec:sres}, disks will readily
fragment, especially in simulations with high spatial 
resolution, but fragmentation actually enhances 
angular momentum transport in the disk and mass accretion 
onto the central stars via tidal torques and the rapid inward migration 
of the fragments.
This is in disagreement with recent
simulations claiming the formation of multiple stellar
systems (see Section~\ref{sec:intro}). 

%--------------------------------------------------------------%

First of all, our inability to form binary and multiple stellar
systems may be partly due to a selection bias;
we did not choose primordial clouds that have strongly 
asymmetric or complex structure as starting configurations. 
Among HR14's cosmological sample, for instance, 
some clouds have multiple density
peaks emerging before the formation of embryo protostars
\citep[see also][]{Turk09}. 
With typical separations of $0.1 - 1$~pc,
they may evolve to form wide-binary systems.
We note that the number fraction of such clouds is
only $\simeq 5~\%$ among 110 clouds. 

%--------------------------------------------------------------%

On the other hand, more compact (e.g., $\lesssim 10^3$~AU)
binary or multiple stellar systems could form via 
disk fragmentation.  
Whereas a number of the fragments migrate inward to fall onto the star, 
some could potentially survive orbiting around the star or even 
as ejecta via gravitational interactions with other fragments.
Our numerical method using the spherical coordinates and
a stationary central sink {\it does} allow the formation
of such systems.
In fact, \citet{VB15} report finding such events in their simulations
using the similar numerical settings.
We cannot discount the potential claim that, had we followed the 
long-term evolution at sufficiently high spatial
resolution, we might also observe such events in our cases \citep[e.g.,][]{Stacy13}. 
However, \citet{VB15} also point out that the ejection events are
much rarer than the burst accretion events initiated by the migration
of fragments.
If the fragments grow to become massive stars emitting
a copious amount of UV photons, then our current framework will
have the limitation of not dealing with the multiple light sources.

%------------------------------------------------------------------%

It is also possible that binary or multiple stellar
systems could form within our 30~AU sink. For present-day
massive star formation, close binaries are the rule rather than
the exception \citep[e.g.,][]{ZY07}. For example, the brightest
Trapezium star, Ori $\Theta^1$C has components ``1'' and ``2''
separated by $\sim$10~AU. One of the binary components
of the second brightest star, Ori $\Theta^1$A1, is itself a
spectroscopic binary with a 1~AU separation. Ori $\Theta^1$E
and Ori $\Theta^1$B1 are also spectroscopic binaries with
separations $\sim$$0.2$~AU and $\sim$$0.13$~AU. 
Ori $\Theta^1$D is a binary with a $\sim$10~AU separation. 
All in all, the Trapezium has
at least 14 close components in 5 multiple stellar systems. 
If such systems are also common among primordial stars,
close binary BH systems could form after a common envelope phase 
and eventually merge emitting gravitational waves. 
Interestingly, the recent detection GW150914 reported by 
\citet{Abbott16} may really represent the signature 
of primordial close binaries \citep{Kinugawa14}.

%---------------------- formation rate of the fragments ----------------------%

In general, the formation efficiency of binary or multiple
systems depends on the production and survival rates of the 
fragments and perhaps on fission of rapidly rotating protostars, 
of which we know very little.
As demonstrated in Section~\ref{ssec:sres}, the fragmentation
rates in our simulations clearly depend on the spatial 
resolution used to follow the long-term evolution. 
\citet{MD14} investigate the spatial resolution
necessary to capture disk fragmentation with numerical
convergence, meaning that the results no longer depend on
the resolution. They conclude that numerical codes need to
resolve structure at least down to the 
scale $\sim 0.01 - 0.1$~AU (see their Appendix).
\citet{Greif12} show that about 1/3 of the emerging 
fragments survive in their simulations when they
resolve such small scales without using the sink technique, 
although only the very early evolution for $\sim$$10$ years
could be followed.
Note that even these studies may suffer from limited spatial 
resolution;
the numbers of cells assigned per Jeans length are 8 and 32
in \citet{MD14} and \citet{Greif12}, respectively, 
whereas it is known that significantly higher resolution is required
to also capture the turbulent motions expected in the
early collapse stage \citep[e.g.,][]{Turk12}.

%------------------------ later stage -----------------------------%

We conclude that with current computational capabilities
it is necessary to use sink cells or sink particles to follow
the long term evolution up to the point that UV feedback 
becomes important.
For such cases, disk fragmentation
and the resulting stellar multiplicity will necessarily depend on the 
adopted spatial resolution.
Stellar multiplicity should affect the evolution.
\citet{Susa13} allow the creation of multiple 
sink (star) particles without further mergers, 
and show that UV feedback from 
multiple protostars jointly halt the mass accretion.
This fundamentally different implementation of sinks may
explain why \citet{Susa14} find no correlation between
the final stellar masses and cloud-scale infall rates,
in contrast to our results (see Section~\ref{ssec:2Dcomp}). 

%---------------------- survival rate of the fragments --------------------%

The evolution of the stellar radius
is important for determining merger rates with other stars or 
fragments as well as the associated UV emissivities.
Once the protostar inflates entering the supergiant stage after an
accretion burst event, the large stellar size will further
enhance the merger rates of nearby fragments and
close companions. 
If a supergiant protostar is in a close binary system, for example, 
mass exchange with the companion star
could cause the companion to inflate and initiate common 
envelope evolution of the primary and companion stars. 
As a protostar inflates, it can consume the gas in the inner 
parts of the circumstellar accretion disk, enhancing
the instantaneous accretion rate even more.
These processes can help the formation of very massive stars, in addition
to the intermittent UV feedback shown in Section~\ref{ssec:VMstar}.

%++++++++++++++++++++++++++++++++++++++++%
\subsection{Effects of Magnetic Fields}
\label{ssec:magnet}
%++++++++++++++++++++++++++++++++++++++++%

We have not included effects of magnetic fields in our current
simulations. However, theoretical studies predict that
the field strength can be significantly amplified by the
turbulent dynamo process during early collapse \citep[e.g.,][]{Dominik10,Turk12}.
Magnetic fields could affect the evolution in the
late accretion stage in a number of ways.
The turbulent motion in accretion disks can further increase
the field strength and add to the magnetic pressure support
of the disk \citep[e.g.,][]{TB05}.
The angular momentum of the accreting material
can be reduced by magnetic breaking \citep[e.g.,][]{MD14}. 
Both effects could potentially suppress disk fragmentation.
Nevertheless, some accretion variability should remain as long as 
the gravitational torque controls the mass and angular 
momentum transfer in disks.

%--------------------------------------------------------------%

Magnetically-driven outflows might appear
intermittently with the variable mass accretion 
\citep[e.g.,][]{Machida11,Machida13}. 
It is interesting to speculate on the potential interaction 
between the stellar UV and outflow feedback mechanisms. 
Whereas the UV feedback is turned off by burst accretion
(Section~\ref{ssec:VMstar}), the outflow power will be enhanced
because a fraction of the accreting gas is generally diverted into 
the outflow \citep[e.g.,][]{Matzner99}. 
The UV and outflow feedback mechanisms may actually complement each other;
the outflow is relatively weak as an \hii region 
expands, and the outflow is activated as the
\hii region disappears.
Future numerical simulations such as those performed for the present-day 
high-mass star formation \citep[e.g.,][]{KYT15} will explore these effects.

%++++++++++++++++++++++++++++++++++++++++++++++++++++++++++++++++++++%
\subsection{Signatures of Primordial Stars in Galactic Metal-poor Stars}
\label{ssec:obsig}
%++++++++++++++++++++++++++++++++++++++++++++++++++++++++++++++++++++%

Although only a handful cases were studied above, it would be useful to 
confront our results with current observational constraints on the
masses of primordial stars. 
Figure~\ref{fig:xmdotj_xm}-(a) includes the 
recently estimated masses of Pop III 
supernova progenitors derived from abundance patterns 
in Galactic metal-poor stars. 
\citet{Caffau11}, \citet{Keller14}, and \citet{Frebel15} have
shown that the abundance patterns 
of extremely metal-poor ([Fe/H] $< -4$) stars can be explained by
chemical yields of core collapse supernovae (CCSN),
whose progenitors are several $\times~10~\msun$ stars.
Among our simulations, such ordinary massive stars ultimately form for 
cases D and E, after the UV feedback efficiently shuts off mass
accretion (Section~\ref{ssec:OMstar}).

%----------------------- VMstars (Aoki+14) -----------------------------%

\citet{Aoki14} recently found a peculiar abundance
pattern in SDSS J001820.5-093939.2, which does not match the yields 
of CCSN but seems to indicate the existence of more massive populations:
progenitors of pair-instability supernovae 
\citep[PISN; $120~\msun \lesssim M_* \lesssim 240~\msun$, e.g.,][]{YDL12}
or the core-collapse of very massive stars
\citep[$M_* \gtrsim 240~\msun$, e.g.,][]{Ohkubo09}.
For our cases A--C, the very massive stars above the PISN range
form due to the process of 
intermittent (burst) accretion and its associated variability of 
UV feedback (Section~\ref{ssec:VMstar}).
Although currently absent among the five cases considered here,
we strongly suspect that we would also have formed stars in the PISN range,
if a greater number of cases had been considered.
Figure~\ref{fig:xmdotj_xm} suggests that such stars will appear 
for the cases that have the cloud-scale infall rates between 
the values for cases A and C. 
However, \citet{Aoki14} actually show that such very massive stars may be
preferred to the PISN progenitors to explain the peculiar abundance
pattern. 
Note that this object has a higher metallicity [Fe/H] $\simeq -2.5$
than the above mentioned very metal-poor stars.
A higher metallicity is expected theoretically, 
because even a single massive explosion
will elevate the metallicity of the polluted gas 
to this level \citep{Karlsson08}.
Our results indicate that $M_* \gtrsim 100~\msun$ 
primordial stars constitute a majority (Fig.~\ref{fig:xmdotj_xm}), so
it might be possible to find similar signatures in other nearby stars.
However, this still may be challenging, because such
energetic supernovae, could potentially expel most of the gas out of a
halo and suppress subsequent star formation \citep{Cooke12}. 
It will be necessary to determine the star formation efficiency after energetic
supernovae in order to assess the mass distribution of massive
primordial stars using stellar archaeology.

%\clearpage
%++++++++++++++++++++++++++++++++++++++++++++++++++++++++++++%
\subsection{Formation of Supermassive Stars: an Extreme Case}
\label{ssec:SMBH}
%+++++++++++++++++++++++++++++++++++++++++++++++++++++++++++++%

Several authors have considered the formation of primordial 
supermassive ($M_* \sim 10^5~\msun$) stars (SMSs) 
as a promising pathway to seed supermassive BHs in the early universe
\citep[e.g.,][]{BL03}.  
A popular model supposes that such an extreme mode of
primordial star formation should occur in the so-called
atomic-cooling haloes exposed to strong FUV radiation.
With only atomic hydrogen as an available coolant, 
the cloud collapse advances almost isothermally at
a higher temperature ($T \simeq 8000$~K) than for
typical primordial cases \citep[e.g.,][]{Omukai01}. 
As a result, very rapid accretion and sufficient material (high Jeans mass)
is expected to be available within the stellar lifetime of 
$\sim$Myr, during which the stellar mass could attain values
$\gtrsim 10^5~\msun$ \citep[e.g.,][]{Regan09,Latif13,IOT14}.
Recent numerical simulations show that, also for such cases,  
circumstellar disks easily fragment due to the gravitational
instability \citep[e.g.,][]{Regan14,Becerra15}. 
It is also expected that the emerging fragments will rapidly
migrate inward through the disk \citep[e.g.,][]{IH14},
as for normal Pop III cases (see Section~\ref{sssec:diskfrag_sres}).
The resulting accretion rates will thus show significant
variability.

%------------------------------------------------------------------%

As discussed above, stellar UV feedback will be
very weak whenever the protostar enters the supergiant stage 
for accretion rates exceeding $0.01~\msunyr$.
However, the protostar will contract toward the ZAMS
under realistic conditions of variable accretion, each time the accretion 
rate falls below $0.01~\msunyr$. 
Stellar evolution calculations show that such
a slowly accreting star can reach the ZAMS, 
if this lower rate of accretion continues for
$\gtrsim 10^3$ years \citep{Sakurai15}. 
As it approaches the ZAMS, the stellar UV emissivity rises sharply,
initiating very powerful UV feedback
(Section~\ref{ssec:VMstar}).
For the normal primordial cases studied above, 
a short burst of rapid mass accretion will cause the star 
to bloat up again and the star can again
accrete more available material. Although case B has reached
nearly 600~$\msun$ through this recurrent accretion process and
is still accreting material after $7 \times 10^4$~years, our
simulations are still far from the mass $10^5\;\msun$ and
average accretion rates $0.1~\msunyr$ necessary
to produce such an SMS within a Myr.
Our newly developed SE-RHD numerical code will be useful to
explore the interplay between variable accretion 
and UV feedback for the formation of SMSs. 

%---------------------------------------------------------------%

Needless to say, other physical processes which have not 
been fully considered, such as stellar multiplicity
(Section~\ref{ssec:multi}) and magnetic fields
(Section~\ref{ssec:magnet}), can also jointly modify the protostellar 
accretion histories. 
Since the formation of SMSs is an extreme
version of the formation of very massive stars studied in Section~\ref{ssec:VMstar},
both cases share many common theoretical issues to be
solved in future studies.

%%%%%%%%%%%%%%%%%%%%%%%%%%%
\section{Conclusions}
\label{sec:conclusions}
%%%%%%%%%%%%%%%%%%%%%%%%%%%

We have studied the long term evolution during the accretion stage 
of primordial star formation, focusing on the interplay between 
highly variable accretion onto the star and stellar UV feedback.
To follow the evolution by numerical simulations, 
we have developed a new 3D SE-RHD
hybrid code to solve for the radiation hydrodynamics of the accretion flow 
simultaneously with the stellar evolution of the accreting protostar.
The evolution of the protostellar accretion flow under the influence of both 
ionizing (EUV) and photodissociating (FUV) radiative feedback
has been followed for over $\sim$$10^5$ years.
We have examined the evolution of five different primordial
clouds occurring in representative dark matter halos found 
in cosmological structure formation simulations. 
Many numerical parameters (e.g., spatial resolution, sink cell size) 
used in our simulations 
have been varied to assess their effects on the results. 
Our findings are summarized as follows:
\begin{enumerate}
 \item The resulting stellar masses $M_*$ show a great diversity:
$10~\msun \lesssim M_* \lesssim 10^3~\msun$, which also has been 
inferred in previous studies \citep[e.g., HR14, HR15;][]{Susa14}. 
The stellar mass tends to be higher for clouds with a higher 
average infall rate, a cloud parameter set during the early collapse stage
{\it before} the formation of an embryo protostar.
This roughly agrees with our previous findings discussed in HR14.
\item 
Ordinary massive ($M_* \lesssim 100~\msun$) stars
form in cases with relatively slow mass accretion.
For these cases, UV feedback efficiently shuts off
accretion once a protostar reaches the late KH contraction
stage, as shown in our previous 2D studies. 
On the other hand, very massive ($M_* \gtrsim 250~\msun$) stars
also form when relatively rapid accretion is available. 
For such cases, the protostar goes through a  ``supergiant protostar''
evolutionary stage, which recurrently appears with the highly 
variable mass accretion associated with self-gravitating circumstellar disks. 
As a result of the fluctuating stellar radius, characterized by
alternating periods of contraction and inflation, 
the formation and extinction of \hii regions also recur repeatedly. 
Over the course of time UV feedback only operates intermittently, and does not efficiently
halt mass accretion onto the star. 
 \item Very massive stars form in three 
out of the five examined cases 
(and two of these three yield only lower limits to
the final stellar masses).
According to a statistical study by
HR15, who have analyzed physical properties of more than 
1500 primordial clouds, the formation of such very massive stars 
can be a major mode of primordial star formation.
\item 
According to our systematic assessment of varying the numerical
parameters, our current simulations 
do not converge with increasing spatial resolution. 
As far as we can tell with our present suite of numerical tools, 
however, variable accretion and the resulting intermittent 
UV feedback becomes even more prominent at higher spatial resolution.
\end{enumerate}

%------------------------------------------------------------------%

Our results show that the interplay between stellar UV feedback 
and variable mass accretion can help the formation of 
very massive primordial stars, making this 
a likely pathway to form a significant number of very massive stars in the
early universe, circumventing potential barriers,
such as stellar UV feedback and disk fragmentation.
The peculiar abundance pattern recently discovered in a
Galactic metal-poor star \citep{Aoki14} could be the 
observational signature indicative of such very massive stars.
With the resolution dependence of our results discussed in 
Section~\ref{ssec:sres}, this basic proof of concept investigation 
still requires further detailed study to verify the above idea.
%As far as we can tell with our present suite of numerical tools, 
%however, variable accretion and 
%the resulting intermittent UV feedback becomes even more prominent 
%at higher spatial resolution.

%--------------------------------------------------------------------%

As an extension of cases considered here, 
the formation of even more massive (or supermassive)
stars may be possible under certain favorable conditions;
the feasibility of this scenario will be examined in future studies.
Such studies will be necessary to determine
the maximum mass of primordial stars, 
a key quantity when considering the
origin of the supermassive black holes observed in the early universe.

{\acknowledgements 
The authors thank Hajime Susa, Eduard Vorobyov, Masayuki Umemura,
Shu-ichiro Inutsuka, and Ken Nomoto for fruitful discussions and comments. 
The numerical simulations were performed on the Cray XC30
at the Center for Computational Astrophysics, CfCA, 
of the National Astronomical Observatory of Japan.
Portions of this work were conducted at the Jet Propulsion Laboratory,
California Institute of Technology, operating under a contract with 
the National Aeronautics and Space Administration (NASA).
This work was financially supported 
by the Grants-in-Aid for Basic Research
by the Ministry of Education, Science and Culture of Japan 
(25800102, 15H00776: TH, 25287040: KO, 25287050: NY)
and by Grant-in-Aid for JSPS Fellows (SH).
RK acknowledges funding within the Emmy Noether Research Group on 
``Accretion Flows and Feedback in Realistic Models of Massive Star
Formation'' granted by the German Research Foundation (DFG) 
under grant no. KU 2849/3-1.
}

\clearpage
%=======================%
\bibliography{biblio}

\begin{thebibliography}{112}
\expandafter\ifx\csname natexlab\endcsname\relax\def\natexlab#1{#1}\fi

\bibitem[{{Abbott} {et~al.}(2016){Abbott}, {Abbott}, {Abbott}, {Abernathy},
  {Acernese}, {Ackley}, {Adams}, {Adams}, {Addesso}, {Adhikari}, \&
  et~al.}]{Abbott16}
{Abbott}, B.~P., {et~al.} 2016, Physical Review Letters, 116, 061102

\bibitem[{{Abel} {et~al.}(1997){Abel}, {Anninos}, {Zhang}, \&
  {Norman}}]{Abel97}
{Abel}, T., {Anninos}, P., {Zhang}, Y., \& {Norman}, M.~L. 1997, New Astronomy,
  2, 181

\bibitem[{{Abel} {et~al.}(2002){Abel}, {Bryan}, \& {Norman}}]{ABN02}
{Abel}, T., {Bryan}, G.~L., \& {Norman}, M.~L. 2002, Science, 295, 93

\bibitem[{{Aoki} {et~al.}(2014){Aoki}, {Tominaga}, {Beers}, {Honda}, \&
  {Lee}}]{Aoki14}
{Aoki}, W., {Tominaga}, N., {Beers}, T.~C., {Honda}, S., \& {Lee}, Y.~S. 2014,
  Science, 345, 912

\bibitem[{{Baruteau} {et~al.}(2011){Baruteau}, {Meru}, \&
  {Paardekooper}}]{Baruteau11}
{Baruteau}, C., {Meru}, F., \& {Paardekooper}, S.-J. 2011, \mnras, 416, 1971

\bibitem[{{Bate} {et~al.}(2010){Bate}, {Lodato}, \& {Pringle}}]{Bate10}
{Bate}, M.~R., {Lodato}, G., \& {Pringle}, J.~E. 2010, \mnras, 401, 1505

\bibitem[{{Becerra} {et~al.}(2015){Becerra}, {Greif}, {Springel}, \&
  {Hernquist}}]{Becerra15}
{Becerra}, F., {Greif}, T.~H., {Springel}, V., \& {Hernquist}, L.~E. 2015,
  \mnras, 446, 2380

\bibitem[{{Bromm} {et~al.}(2002){Bromm}, {Coppi}, \& {Larson}}]{Bromm02}
{Bromm}, V., {Coppi}, P.~S., \& {Larson}, R.~B. 2002, \apj, 564, 23

\bibitem[{{Bromm} \& {Loeb}(2003)}]{BL03}
{Bromm}, V., \& {Loeb}, A. 2003, \apj, 596, 34

\bibitem[{{Bromm} {et~al.}(2009){Bromm}, {Yoshida}, {Hernquist}, \&
  {McKee}}]{Bromm09}
{Bromm}, V., {Yoshida}, N., {Hernquist}, L., \& {McKee}, C.~F. 2009, \nat, 459,
  49

\bibitem[{{Caffau} {et~al.}(2011){Caffau}, {Bonifacio}, {Fran{\c c}ois},
  {Sbordone}, {Monaco}, {Spite}, {Spite}, {Ludwig}, {Cayrel}, {Zaggia},
  {Hammer}, {Randich}, {Molaro}, \& {Hill}}]{Caffau11}
{Caffau}, E., {et~al.} 2011, \nat, 477, 67

\bibitem[{{Clark} {et~al.}(2012){Clark}, {Glover}, \& {Klessen}}]{Clark12}
{Clark}, P.~C., {Glover}, S.~C.~O., \& {Klessen}, R.~S. 2012, \mnras, 420, 745

\bibitem[{{Clark} {et~al.}(2011){Clark}, {Glover}, {Smith}, {Greif}, {Klessen},
  \& {Bromm}}]{Cl11}
{Clark}, P.~C., {Glover}, S.~C.~O., {Smith}, R.~J., {Greif}, T.~H., {Klessen},
  R.~S., \& {Bromm}, V. 2011, Science, 331, 1040

\bibitem[{{Cooke} \& {Madau}(2014)}]{Cooke12}
{Cooke}, R.~J., \& {Madau}, P. 2014, \apj, 791, 116

\bibitem[{{DeSouza} \& {Basu}(2015)}]{DeSouza15}
{DeSouza}, A.~L., \& {Basu}, S. 2015, \mnras, 450, 295

\bibitem[{{Draine} \& {Bertoldi}(1996)}]{DB96}
{Draine}, B.~T., \& {Bertoldi}, F. 1996, \apj, 468, 269

\bibitem[{{Federrath} {et~al.}(2011){Federrath}, {Sur}, {Schleicher},
  {Banerjee}, \& {Klessen}}]{Federrath11}
{Federrath}, C., {Sur}, S., {Schleicher}, D.~R.~G., {Banerjee}, R., \&
  {Klessen}, R.~S. 2011, \apj, 731, 62

\bibitem[{{Fielding} {et~al.}(2015){Fielding}, {McKee}, {Socrates},
  {Cunningham}, \& {Klein}}]{Fielding15}
{Fielding}, D.~B., {McKee}, C.~F., {Socrates}, A., {Cunningham}, A.~J., \&
  {Klein}, R.~I. 2015, \mnras, 450, 3306

\bibitem[{{Forrey}(2013)}]{Forrey13}
{Forrey}, R.~C. 2013, \apjl, 773, L25

\bibitem[{{Frebel} {et~al.}(2015){Frebel}, {Chiti}, {Ji}, {Jacobson}, \&
  {Placco}}]{Frebel15}
{Frebel}, A., {Chiti}, A., {Ji}, A.~P., {Jacobson}, H.~R., \& {Placco}, V.~M.
  2015, \apjl, 810, L27

\bibitem[{{Galli} \& {Palla}(1998)}]{GP98}
{Galli}, D., \& {Palla}, F. 1998, \aap, 335, 403

\bibitem[{{Greif}(2014)}]{Greif14}
{Greif}, T.~H. 2014, \mnras, 444, 1566

\bibitem[{{Greif}(2015)}]{Greif15Rv}
---. 2015, Computational Astrophysics and Cosmology, 2, 3

\bibitem[{{Greif} {et~al.}(2012){Greif}, {Bromm}, {Clark}, {Glover}, {Smith},
  {Klessen}, {Yoshida}, \& {Springel}}]{Greif12}
{Greif}, T.~H., {Bromm}, V., {Clark}, P.~C., {Glover}, S.~C.~O., {Smith},
  R.~J., {Klessen}, R.~S., {Yoshida}, N., \& {Springel}, V. 2012, \mnras, 424,
  399

\bibitem[{{Greif} {et~al.}(2011){Greif}, {Springel}, {White}, {Glover},
  {Clark}, {Smith}, {Klessen}, \& {Bromm}}]{Greif11}
{Greif}, T.~H., {Springel}, V., {White}, S.~D.~M., {Glover}, S.~C.~O., {Clark},
  P.~C., {Smith}, R.~J., {Klessen}, R.~S., \& {Bromm}, V. 2011, \apj, 737, 75

\bibitem[{{Hartwig} {et~al.}(2015){Hartwig}, {Clark}, {Glover}, {Klessen}, \&
  {Sasaki}}]{Hartwig15}
{Hartwig}, T., {Clark}, P.~C., {Glover}, S.~C.~O., {Klessen}, R.~S., \&
  {Sasaki}, M. 2015, \apj, 799, 114

\bibitem[{{Hirano} {et~al.}(2015){Hirano}, {Hosokawa}, {Yoshida}, {Omukai}, \&
  {Yorke}}]{Hirano15}
{Hirano}, S., {Hosokawa}, T., {Yoshida}, N., {Omukai}, K., \& {Yorke}, H.~W.
  2015, \mnras, 448, 568

\bibitem[{{Hirano} {et~al.}(2014){Hirano}, {Hosokawa}, {Yoshida}, {Umeda},
  {Omukai}, {Chiaki}, \& {Yorke}}]{Hirano14}
{Hirano}, S., {Hosokawa}, T., {Yoshida}, N., {Umeda}, H., {Omukai}, K.,
  {Chiaki}, G., \& {Yorke}, H.~W. 2014, \apj, 781, 60

\bibitem[{{Hirano} \& {Yoshida}(2013)}]{Hirano13}
{Hirano}, S., \& {Yoshida}, N. 2013, \apj, 763, 52

\bibitem[{{Hollenbach} {et~al.}(1994){Hollenbach}, {Johnstone}, {Lizano}, \&
  {Shu}}]{HJLS94}
{Hollenbach}, D., {Johnstone}, D., {Lizano}, S., \& {Shu}, F. 1994, \apj, 428,
  654

\bibitem[{{Hosokawa} \& {Omukai}(2009)}]{HO09}
{Hosokawa}, T., \& {Omukai}, K. 2009, \apj, 691, 823

\bibitem[{{Hosokawa} {et~al.}(2012{\natexlab{a}}){Hosokawa}, {Omukai}, \&
  {Yorke}}]{HOY12}
{Hosokawa}, T., {Omukai}, K., \& {Yorke}, H.~W. 2012{\natexlab{a}}, \apj, 756,
  93

\bibitem[{{Hosokawa} {et~al.}(2011){Hosokawa}, {Omukai}, {Yoshida}, \&
  {Yorke}}]{HOYY11}
{Hosokawa}, T., {Omukai}, K., {Yoshida}, N., \& {Yorke}, H.~W. 2011, Science,
  334, 1250

\bibitem[{{Hosokawa} {et~al.}(2013){Hosokawa}, {Yorke}, {Inayoshi}, {Omukai},
  \& {Yoshida}}]{Hosokawa13}
{Hosokawa}, T., {Yorke}, H.~W., {Inayoshi}, K., {Omukai}, K., \& {Yoshida}, N.
  2013, \apj, 778, 178

\bibitem[{{Hosokawa} {et~al.}(2010){Hosokawa}, {Yorke}, \& {Omukai}}]{HYO10}
{Hosokawa}, T., {Yorke}, H.~W., \& {Omukai}, K. 2010, \apj, 721, 478

\bibitem[{{Hosokawa} {et~al.}(2012{\natexlab{b}}){Hosokawa}, {Yoshida},
  {Omukai}, \& {Yorke}}]{HYOY12}
{Hosokawa}, T., {Yoshida}, N., {Omukai}, K., \& {Yorke}, H.~W.
  2012{\natexlab{b}}, \apjl, 760, L37

\bibitem[{{Inayoshi} \& {Haiman}(2014)}]{IH14}
{Inayoshi}, K., \& {Haiman}, Z. 2014, \mnras, 445, 1549

\bibitem[{{Inayoshi} {et~al.}(2014){Inayoshi}, {Omukai}, \& {Tasker}}]{IOT14}
{Inayoshi}, K., {Omukai}, K., \& {Tasker}, E. 2014, \mnras, 445, L109

\bibitem[{{Ishigaki} {et~al.}(2014){Ishigaki}, {Tominaga}, {Kobayashi}, \&
  {Nomoto}}]{Ishigaki14}
{Ishigaki}, M.~N., {Tominaga}, N., {Kobayashi}, C., \& {Nomoto}, K. 2014,
  \apjl, 792, L32

\bibitem[{{John}(1988)}]{John88}
{John}, T.~L. 1988, \aap, 193, 189

\bibitem[{{Karlsson} {et~al.}(2008){Karlsson}, {Johnson}, \&
  {Bromm}}]{Karlsson08}
{Karlsson}, T., {Johnson}, J.~L., \& {Bromm}, V. 2008, \apj, 679, 6

\bibitem[{{Keller} {et~al.}(2014){Keller}, {Bessell}, {Frebel}, {Casey},
  {Asplund}, {Jacobson}, {Lind}, {Norris}, {Yong}, {Heger}, {Magic}, {da
  Costa}, {Schmidt}, \& {Tisserand}}]{Keller14}
{Keller}, S.~C., {et~al.} 2014, \nat, 506, 463

\bibitem[{{Kinugawa} {et~al.}(2014){Kinugawa}, {Inayoshi}, {Hotokezaka},
  {Nakauchi}, \& {Nakamura}}]{Kinugawa14}
{Kinugawa}, T., {Inayoshi}, K., {Hotokezaka}, K., {Nakauchi}, D., \&
  {Nakamura}, T. 2014, \mnras, 442, 2963

\bibitem[{{Kreckel} {et~al.}(2010){Kreckel}, {Bruhns}, {{\v C}{\'{\i}}{\v
  z}ek}, {Glover}, {Miller}, {Urbain}, \& {Savin}}]{Kreckel10}
{Kreckel}, H., {Bruhns}, H., {{\v C}{\'{\i}}{\v z}ek}, M., {Glover}, S.~C.~O.,
  {Miller}, K.~A., {Urbain}, X., \& {Savin}, D.~W. 2010, Science, 329, 69

\bibitem[{{Kuiper} {et~al.}(2010{\natexlab{a}}){Kuiper}, {Klahr}, {Beuther}, \&
  {Henning}}]{Kuiper10}
{Kuiper}, R., {Klahr}, H., {Beuther}, H., \& {Henning}, T. 2010{\natexlab{a}},
  \apj, 722, 1556

\bibitem[{{Kuiper} {et~al.}(2011){Kuiper}, {Klahr}, {Beuther}, \&
  {Henning}}]{Kuiper11}
---. 2011, \apj, 732, 20

\bibitem[{{Kuiper} {et~al.}(2010{\natexlab{b}}){Kuiper}, {Klahr}, {Dullemond},
  {Kley}, \& {Henning}}]{Kuiper10aa}
{Kuiper}, R., {Klahr}, H., {Dullemond}, C., {Kley}, W., \& {Henning}, T.
  2010{\natexlab{b}}, \aap, 511, A81

\bibitem[{{Kuiper} \& {Klessen}(2013)}]{KK13}
{Kuiper}, R., \& {Klessen}, R.~S. 2013, \aap, 555, A7

\bibitem[{{Kuiper} \& {Yorke}(2013)}]{KY13}
{Kuiper}, R., \& {Yorke}, H.~W. 2013, \apj, 772, 61

\bibitem[{{Kuiper} {et~al.}(2015){Kuiper}, {Yorke}, \& {Turner}}]{KYT15}
{Kuiper}, R., {Yorke}, H.~W., \& {Turner}, N.~J. 2015, \apj, 800, 86

\bibitem[{{Latif} {et~al.}(2013){Latif}, {Schleicher}, {Schmidt}, \&
  {Niemeyer}}]{Latif13}
{Latif}, M.~A., {Schleicher}, D.~R.~G., {Schmidt}, W., \& {Niemeyer}, J. 2013,
  \mnras, 433, 1607

\bibitem[{{Levermore} \& {Pomraning}(1981)}]{LP81}
{Levermore}, C.~D., \& {Pomraning}, G.~C. 1981, \apj, 248, 321

\bibitem[{{Machida} \& {Doi}(2013)}]{MD14}
{Machida}, M.~N., \& {Doi}, K. 2013, \mnras, 435, 3283

\bibitem[{{Machida} \& {Hosokawa}(2013)}]{Machida13}
{Machida}, M.~N., \& {Hosokawa}, T. 2013, \mnras, 431, 1719

\bibitem[{{Machida} {et~al.}(2011){Machida}, {Inutsuka}, \&
  {Matsumoto}}]{Machida11}
{Machida}, M.~N., {Inutsuka}, S.-i., \& {Matsumoto}, T. 2011, \apj, 729, 42

\bibitem[{{Martin} {et~al.}(1996){Martin}, {Schwarz}, \& {Mandy}}]{Martin96}
{Martin}, P.~G., {Schwarz}, D.~H., \& {Mandy}, M.~E. 1996, \apj, 461, 265

\bibitem[{{Matzner} \& {McKee}(1999)}]{Matzner99}
{Matzner}, C.~D., \& {McKee}, C.~F. 1999, \apj, 510, 379

\bibitem[{{McKee} \& {Tan}(2008)}]{MT08}
{McKee}, C.~F., \& {Tan}, J.~C. 2008, \apj, 681, 771

\bibitem[{{Meece} {et~al.}(2014){Meece}, {Smith}, \& {O'Shea}}]{Meece14}
{Meece}, G.~R., {Smith}, B.~D., \& {O'Shea}, B.~W. 2014, \apj, 783, 75

\bibitem[{{Mignone} {et~al.}(2007){Mignone}, {Bodo}, {Massaglia}, {Matsakos},
  {Tesileanu}, {Zanni}, \& {Ferrari}}]{Mignone07}
{Mignone}, A., {Bodo}, G., {Massaglia}, S., {Matsakos}, T., {Tesileanu}, O.,
  {Zanni}, C., \& {Ferrari}, A. 2007, \apjs, 170, 228

\bibitem[{{Millar}(1991)}]{Millar91}
{Millar}, T.~J. 1991, \aap, 242, 241

\bibitem[{{Oh} \& {Haiman}(2002)}]{OH02}
{Oh}, S.~P., \& {Haiman}, Z. 2002, \apj, 569, 558

\bibitem[{{Ohkubo} {et~al.}(2009){Ohkubo}, {Nomoto}, {Umeda}, {Yoshida}, \&
  {Tsuruta}}]{Ohkubo09}
{Ohkubo}, T., {Nomoto}, K., {Umeda}, H., {Yoshida}, N., \& {Tsuruta}, S. 2009,
  \apj, 706, 1184

\bibitem[{{Omukai}(2001)}]{Omukai01}
{Omukai}, K. 2001, \apj, 546, 635

\bibitem[{{Omukai} \& {Inutsuka}(2002)}]{OI02}
{Omukai}, K., \& {Inutsuka}, S.-i. 2002, \mnras, 332, 59

\bibitem[{{Omukai} \& {Nishi}(1998)}]{ON98}
{Omukai}, K., \& {Nishi}, R. 1998, \apj, 508, 141

\bibitem[{{Omukai} \& {Palla}(2003)}]{OP03}
{Omukai}, K., \& {Palla}, F. 2003, \apj, 589, 677

\bibitem[{{O'Shea} {et~al.}(2008){O'Shea}, {McKee}, {Heger}, \&
  {Abel}}]{OShea08b}
{O'Shea}, B.~W., {McKee}, C.~F., {Heger}, A., \& {Abel}, T. 2008, in American
  Institute of Physics Conference Series, Vol. 990, First Stars III, ed.
  {B.~W.~O'Shea \& A.~Heger}, D13

\bibitem[{{O'Shea} \& {Norman}(2008)}]{OShea08}
{O'Shea}, B.~W., \& {Norman}, M.~L. 2008, \apj, 673, 14

\bibitem[{{Osterbrock} \& {Ferland}(2006)}]{OF06}
{Osterbrock}, D.~E., \& {Ferland}, G.~J. 2006, {Astrophysics of gaseous nebulae
  and active galactic nuclei}, ed. {Osterbrock, D.~E.~\& Ferland, G.~J.}

\bibitem[{{Palla} {et~al.}(1983){Palla}, {Salpeter}, \& {Stahler}}]{PSS83}
{Palla}, F., {Salpeter}, E.~E., \& {Stahler}, S.~W. 1983, \apj, 271, 632

\bibitem[{{Palla} \& {Stahler}(1992)}]{PS92}
{Palla}, F., \& {Stahler}, S.~W. 1992, \apj, 392, 667

\bibitem[{{Papaloizou} \& {Terquem}(1995)}]{PT95}
{Papaloizou}, J.~C.~B., \& {Terquem}, C. 1995, \mnras, 274, 987

\bibitem[{{Peters} {et~al.}(2010){Peters}, {Klessen}, {Mac Low}, \&
  {Banerjee}}]{Peters10}
{Peters}, T., {Klessen}, R.~S., {Mac Low}, M.-M., \& {Banerjee}, R. 2010, \apj,
  725, 134

\bibitem[{{Regan} \& {Haehnelt}(2009)}]{Regan09}
{Regan}, J.~A., \& {Haehnelt}, M.~G. 2009, \mnras, 396, 343

\bibitem[{{Regan} {et~al.}(2014){Regan}, {Johansson}, \& {Haehnelt}}]{Regan14}
{Regan}, J.~A., {Johansson}, P.~H., \& {Haehnelt}, M.~G. 2014, \mnras, 439,
  1160

\bibitem[{{Richling} \& {Yorke}(1997)}]{RY97}
{Richling}, S., \& {Yorke}, H.~W. 1997, \aap, 327, 317

\bibitem[{{Sakurai} {et~al.}(2015){Sakurai}, {Hosokawa}, {Yoshida}, \&
  {Yorke}}]{Sakurai15}
{Sakurai}, Y., {Hosokawa}, T., {Yoshida}, N., \& {Yorke}, H.~W. 2015, \mnras,
  452, 755

\bibitem[{{Schleicher} {et~al.}(2010){Schleicher}, {Banerjee}, {Sur},
  {Arshakian}, {Klessen}, {Beck}, \& {Spaans}}]{Dominik10}
{Schleicher}, D.~R.~G., {Banerjee}, R., {Sur}, S., {Arshakian}, T.~G.,
  {Klessen}, R.~S., {Beck}, R., \& {Spaans}, M. 2010, \aap, 522, A115

\bibitem[{{Schleicher} {et~al.}(2013){Schleicher}, {Palla}, {Ferrara}, {Galli},
  \& {Latif}}]{Schleicher13}
{Schleicher}, D.~R.~G., {Palla}, F., {Ferrara}, A., {Galli}, D., \& {Latif}, M.
  2013, \aap, 558, A59

\bibitem[{{Schneider} {et~al.}(2012){Schneider}, {Omukai}, {Limongi},
  {Ferrara}, {Salvaterra}, {Chieffi}, \& {Bianchi}}]{Schneider12}
{Schneider}, R., {Omukai}, K., {Limongi}, M., {Ferrara}, A., {Salvaterra}, R.,
  {Chieffi}, A., \& {Bianchi}, S. 2012, \mnras, 423, L60

\bibitem[{{Shakura} \& {Sunyaev}(1973)}]{SS73}
{Shakura}, N.~I., \& {Sunyaev}, R.~A. 1973, \aap, 24, 337

\bibitem[{{Siess} {et~al.}(1997){Siess}, {Forestini}, \& {Bertout}}]{Siess97}
{Siess}, L., {Forestini}, M., \& {Bertout}, C. 1997, \aap, 326, 1001

\bibitem[{{Smith} {et~al.}(2012){Smith}, {Hosokawa}, {Omukai}, {Glover}, \&
  {Klessen}}]{Rowan12}
{Smith}, R.~J., {Hosokawa}, T., {Omukai}, K., {Glover}, S.~C.~O., \& {Klessen},
  R.~S. 2012, \mnras, 424, 457

\bibitem[{{Springel}(2005)}]{Springel05}
{Springel}, V. 2005, \mnras, 364, 1105

\bibitem[{{Stacy} \& {Bromm}(2013)}]{Stacy13}
{Stacy}, A., \& {Bromm}, V. 2013, \mnras, 433, 1094

\bibitem[{{Stacy} {et~al.}(2010){Stacy}, {Greif}, \& {Bromm}}]{Stacy10}
{Stacy}, A., {Greif}, T.~H., \& {Bromm}, V. 2010, \mnras, 403, 45

\bibitem[{{Stacy} {et~al.}(2012){Stacy}, {Greif}, \& {Bromm}}]{Stacy12}
---. 2012, \mnras, 2508

\bibitem[{{Stahler} {et~al.}(1986){Stahler}, {Palla}, \& {Salpeter}}]{SPS86}
{Stahler}, S.~W., {Palla}, F., \& {Salpeter}, E.~E. 1986, \apj, 302, 590

\bibitem[{{Stahler} {et~al.}(1980){Stahler}, {Shu}, \& {Taam}}]{SST80}
{Stahler}, S.~W., {Shu}, F.~H., \& {Taam}, R.~E. 1980, \apj, 241, 637

\bibitem[{{Susa}(2013)}]{Susa13}
{Susa}, H. 2013, \apj, 773, 185

\bibitem[{{Susa} {et~al.}(2014){Susa}, {Hasegawa}, \& {Tominaga}}]{Susa14}
{Susa}, H., {Hasegawa}, K., \& {Tominaga}, N. 2014, \apj, 792, 32

\bibitem[{{Tan} \& {Blackman}(2005)}]{TB05}
{Tan}, J.~C., \& {Blackman}, E.~G. 2005, \mnras, 362, 983

\bibitem[{{Tan} \& {McKee}(2004)}]{Tan04}
{Tan}, J.~C., \& {McKee}, C.~F. 2004, \apj, 603, 383

\bibitem[{{Tanaka} {et~al.}(2002){Tanaka}, {Takeuchi}, \& {Ward}}]{Tanaka02}
{Tanaka}, H., {Takeuchi}, T., \& {Ward}, W.~R. 2002, \apj, 565, 1257

\bibitem[{{Tanaka} {et~al.}(2013){Tanaka}, {Nakamoto}, \& {Omukai}}]{TKei13}
{Tanaka}, K.~E.~I., {Nakamoto}, T., \& {Omukai}, K. 2013, \apj, 773, 155

\bibitem[{{Tielens} \& {Hollenbach}(1985)}]{TH85}
{Tielens}, A.~G.~G.~M., \& {Hollenbach}, D. 1985, \apj, 291, 722

\bibitem[{{Truelove} {et~al.}(1997){Truelove}, {Klein}, {McKee}, {Holliman},
  {Howell}, \& {Greenough}}]{Truelove97}
{Truelove}, J.~K., {Klein}, R.~I., {McKee}, C.~F., {Holliman}, II, J.~H.,
  {Howell}, L.~H., \& {Greenough}, J.~A. 1997, \apjl, 489, L179

\bibitem[{{Turk} {et~al.}(2009){Turk}, {Abel}, \& {O'Shea}}]{Turk09}
{Turk}, M.~J., {Abel}, T., \& {O'Shea}, B. 2009, Science, 325, 601

\bibitem[{{Turk} {et~al.}(2012){Turk}, {Oishi}, {Abel}, \& {Bryan}}]{Turk12}
{Turk}, M.~J., {Oishi}, J.~S., {Abel}, T., \& {Bryan}, G.~L. 2012, \apj, 745,
  154

\bibitem[{{Vorobyov} \& {Basu}(2006)}]{VB06}
{Vorobyov}, E.~I., \& {Basu}, S. 2006, \apj, 650, 956

\bibitem[{{Vorobyov} \& {Basu}(2015)}]{VB15}
---. 2015, \apj, 805, 115

\bibitem[{{Vorobyov} {et~al.}(2013){Vorobyov}, {DeSouza}, \&
  {Basu}}]{Vorobyov13}
{Vorobyov}, E.~I., {DeSouza}, A.~L., \& {Basu}, S. 2013, \apj, 768, 131

\bibitem[{{Whalen} \& {Norman}(2008)}]{WN08}
{Whalen}, D., \& {Norman}, M.~L. 2008, \apj, 673, 664

\bibitem[{{Yoon} {et~al.}(2012){Yoon}, {Dierks}, \& {Langer}}]{YDL12}
{Yoon}, S.-C., {Dierks}, A., \& {Langer}, N. 2012, \aap, 542, A113

\bibitem[{{Yorke} \& {Bodenheimer}(2008)}]{YB08}
{Yorke}, H.~W., \& {Bodenheimer}, P. 2008, in Astronomical Society of the
  Pacific Conference Series, Vol. 387, Massive Star Formation: Observations
  Confront Theory, ed. H.~{Beuther}, H.~{Linz}, \& T.~{Henning}, 189

\bibitem[{{Yoshida} {et~al.}(2003){Yoshida}, {Abel}, {Hernquist}, \&
  {Sugiyama}}]{Y03}
{Yoshida}, N., {Abel}, T., {Hernquist}, L., \& {Sugiyama}, N. 2003, \apj, 592,
  645

\bibitem[{{Yoshida} {et~al.}(2007){Yoshida}, {Omukai}, \& {Hernquist}}]{Y07}
{Yoshida}, N., {Omukai}, K., \& {Hernquist}, L. 2007, \apjl, 667, L117

\bibitem[{{Yoshida} {et~al.}(2008){Yoshida}, {Omukai}, \& {Hernquist}}]{Y08}
---. 2008, Science, 321, 669

\bibitem[{{Yoshida} {et~al.}(2006){Yoshida}, {Omukai}, {Hernquist}, \&
  {Abel}}]{Y06}
{Yoshida}, N., {Omukai}, K., {Hernquist}, L., \& {Abel}, T. 2006, \apj, 652, 6

\bibitem[{{Zhu} {et~al.}(2012){Zhu}, {Hartmann}, {Nelson}, \& {Gammie}}]{Zhu12}
{Zhu}, Z., {Hartmann}, L., {Nelson}, R.~P., \& {Gammie}, C.~F. 2012, \apj, 746,
  110

\bibitem[{{Zinnecker} \& {Yorke}(2007)}]{ZY07}
{Zinnecker}, H., \& {Yorke}, H.~W. 2007, \araa, 45, 481

\end{thebibliography}
\bibliographystyle{apj}
%=======================%

%===================%
\appendix
%--------%

%%%%%%%%%%%%%%%%%%%%%%%%%%%%%%%%%%%%%%%%%%%%%%%%%%%
\section{Ideal Test Cases with a Homogeneous Cloud}
\label{sec:homog}
%%%%%%%%%%%%%%%%%%%%%%%%%%%%%%%%%%%%%%%%%%%%%%%%%%%

%------------------------------------------------------------------------------%
\begin{table*}
\label{tb:hmd}
\begin{center}
Table A1. Ideal Cases for Homogeneous Clouds Considered \\[3mm]
\begin{tabular}{lccc}
\hline
\hline
Cases & $N_R \cdot N_\theta \cdot N_\phi$ & $\Delta t$ ($10^4$~yr) 
& $M_* (M_\odot)$ \\
\hline
HM2D              & $64 \cdot 16 \cdot 1$    &   5 & 59.8 \\
HM2D-NF$^\dagger$ & $64 \cdot 16 \cdot 1$    &   5 &   \\
HM                & $64 \cdot 16 \cdot 64$   &   3 &   70.8 \\
HM-ssk$^\bullet$   & $64 \cdot 16 \cdot 64$   & 2.6 &  71.1 \\
HM-HR2$^\dagger$  & $128 \cdot 32 \cdot 128$ & 2.6  &  69.0 \\
\hline
\end{tabular}
\noindent
\end{center}
Col.\ 2: numbers of grid cells, Col.\ 3: time duration of simulations.\ 
Col.\ 4: final stellar masses.
The radial outer boundary is at $r_{\rm max} = 4 \times 10^4$~AU,
symmetry is assumed with respect to the equatorial plane ($\theta=0$), and the
initial cloud mass is $157~\msun$
%(for $0 < \theta < \pi$, i.e., twice of the mass contained in the
%hemispherical computational domain)
for all cases. \\
\
$^\dagger$: The suffixes have the same meanings as in Table.~1. \\
$^\bullet$: Smaller sink with $r_{\rm min} \simeq 10$~AU 
            ($r_{\rm min} = 30$~AU for the other cases).
\end{table*}
%---------------------------------------------------------------------------------%
%---------------------------------------------------------------------------------%
%%% Fig.A1 (Fig.26) %%%
\begin{figure}
  \begin{center}
\epsscale{0.5}
\plotone{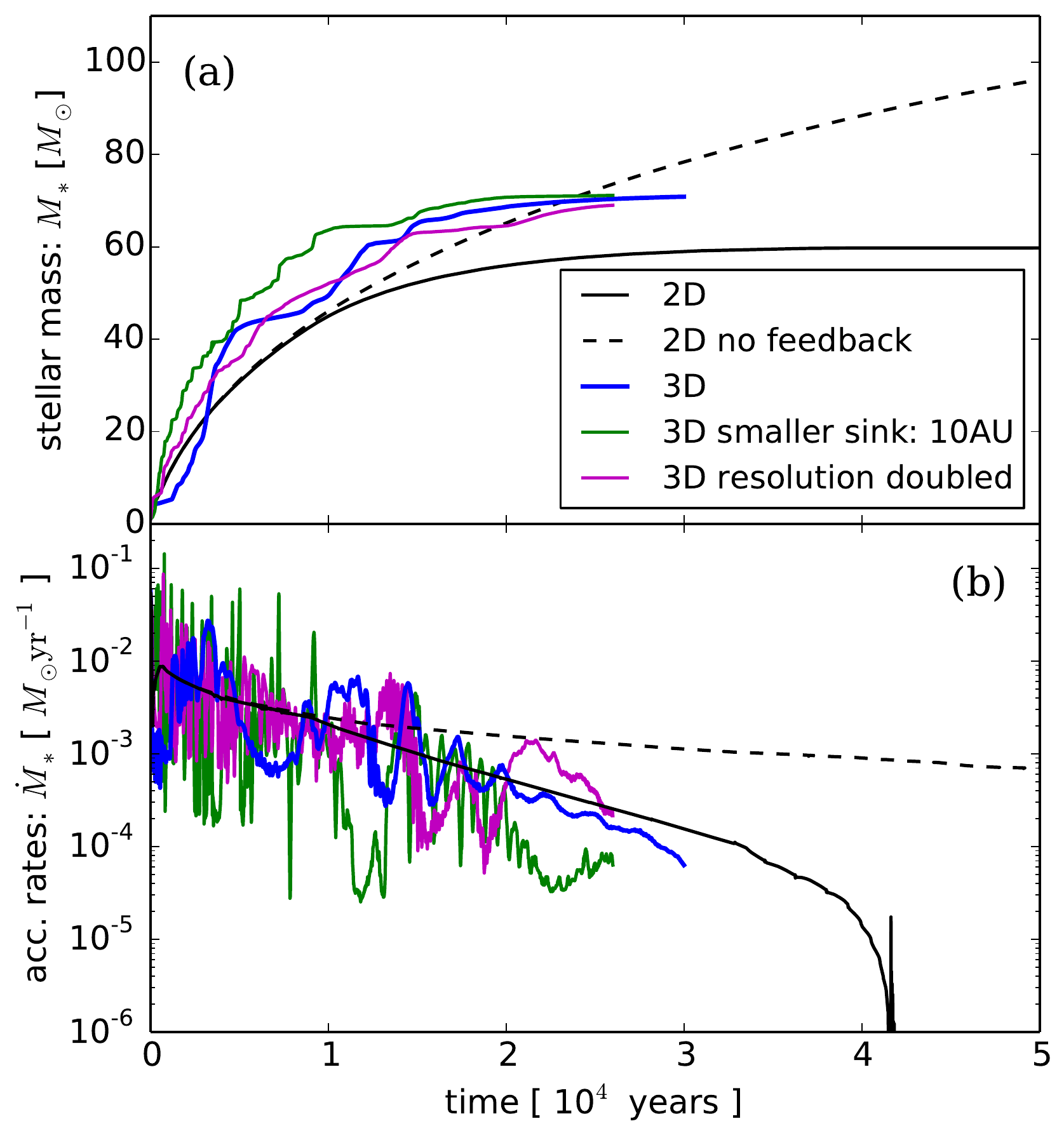}
\caption{Evolution of the stellar mass (panel a) and accretion rates
(panel b) for test cases starting with a homogeneous cloud
(Table~A1).
The black solid and dashed lines represent the 2D axisymmetric cases 
with (HM2D) and without (HM2D-NF) UV feedback.
The other lines represent 3D cases with different numerical settings: 
the fiducial case HM (blue), HM-ssk with a smaller sink (green), 
and HM-HR2 with doubled spatial resolution (magenta).   
The origin of time ($t = 0$) marks the epoch of stellar
birth after the runaway cloud collapse. 
}
\label{fig:xm_xmdot_t_ideal}
  \end{center}
\end{figure}
%---------------------------------------------------------------------------------%
%---------------------------------------------------------------------------------%
%%% Fig.A2 (Fig.27) %%%
\begin{figure}
  \begin{center}
\epsscale{0.5}
\plotone{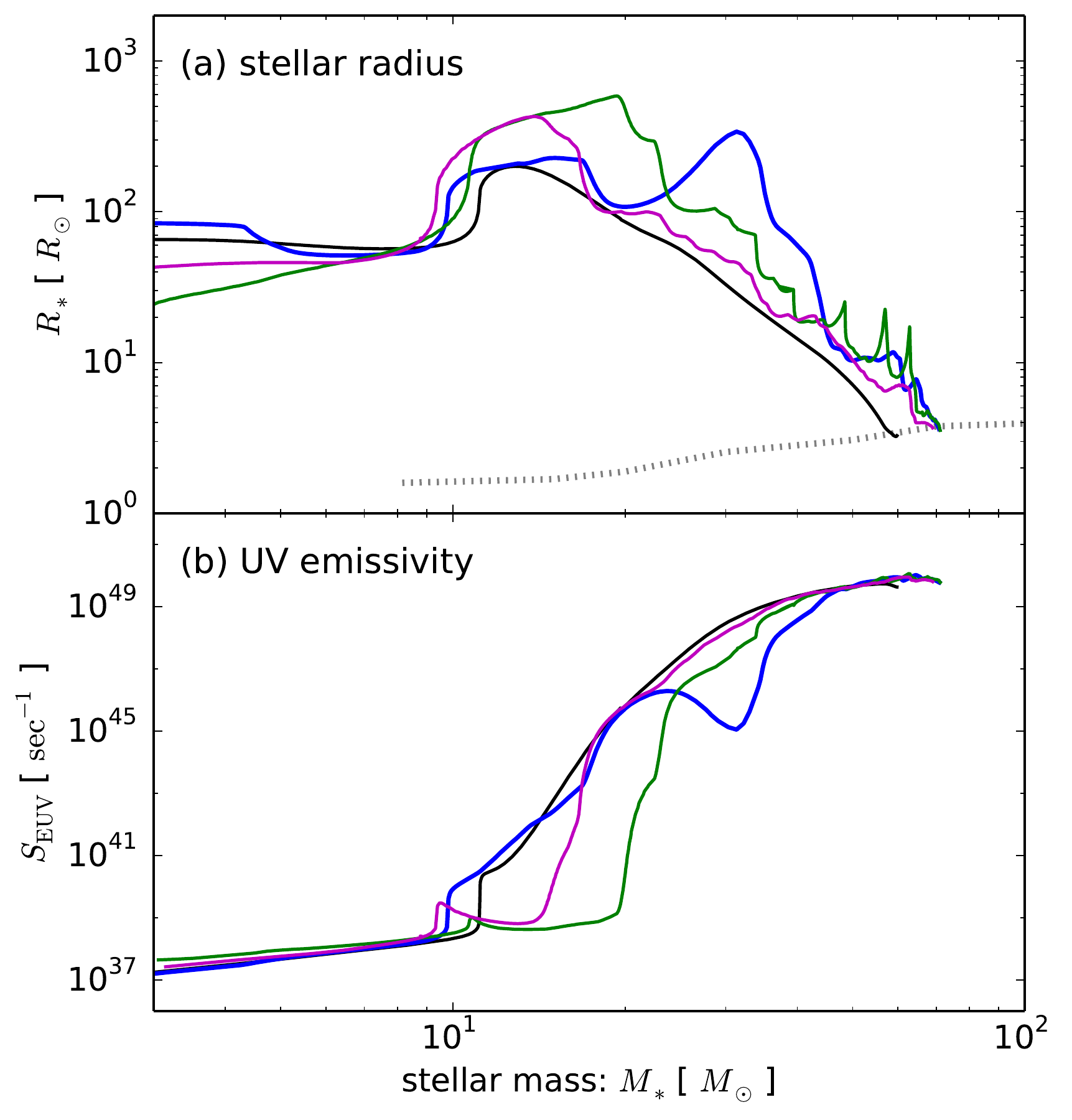}
\caption{Evolution of the stellar radius (panel a) and ionizing (EUV) photon
emissivity (panel b) for the test cases starting with a homogeneous cloud. 
The different lines represent the same cases as in
Figure~\ref{fig:xm_xmdot_t_ideal}.
In panel (a), the dotted line shows the mass-radius relationship for 
zero-age main-sequence stars.
}
\label{fig:xr_suv_xm_ideal}
  \end{center}
\end{figure}
%---------------------------------------------------------------------------------%

In addition to the cosmological cases studied above, 
we also perform simulations starting with idealized initial 
conditions: a zero-metallicity
homogeneous gas cloud with the initial density 
$\rho_0 = 3.5 \times 10^{-19}~{\rm cm}^{-3}$, temperature $T_0 = 200$~K, 
and chemical compositions $X_{\rm H_2} = 5 \times 10^{-3}$, 
$X_{\rm e} = X_{\rm HII} = 10^{-4}$, where $X_i$ is the number ratio
between the species $i$ and hydrogen nuclei.
We assume that the cloud is initially rigidly rotating at an angular velocity
$\Omega_0 = 4 \times 10^{-14}$~Hz, which is about $13~\%$ of the 
Kepler value for the enclosed gas mass $\sqrt{G M(<r)/r^3} = \sqrt{4 \pi G \rho_0/3}$,
independent of radius $r$.
Since the initial Jeans length $\simeq 3 \times 10^4$~AU is
smaller than the cloud's radial extension $4 \times 10^4$~AU, 
the cloud begins to gravitationally collapse in the well-known 
self-similar fashion \citep[e.g.,][]{ON98}.
For the current test cases, we use our modified version of
the mesh-based {\tt PLUTO} code from the beginning, unlike the 
cosmological cases for which the N-body/SPH code {\tt GADGET}-2 
had been used for the early collapse stage (see Section~\ref{sec:method}). 
The grid configuration is almost the same as described in Section~\ref{sec:method}, 
except that the current computational domain assumes
mirror symmetry with respect to the equator 
throughout the evolution.
We follow the collapse without a central sink cell 
until the central density reaches 
$\sim 10^{12}~\cmc$ and
the minimum Jeans length becomes poorly resolved.
We then introduce the sink cell to follow the
subsequent evolution in the protostellar accretion stage
using either 2D or 3D versions of our SE-RHD hybrid code. 

%-------------------------- accretion stage ----------------------------%

For all examined 3D cases, the evolution is similar to 
that obtained for the cosmologcal cases D and E (see Section~\ref{ssec:OMstar}); 
$M_* \simeq 60 - 70~\msun$ stars form after UV feedback
caused by an expanding bipolar \hii region efficiently halts mass accretion.
For the 2D cases we used a so-called $\alpha$-viscocity with $\alpha = 1.0$
to enable angular momentum transport under axial symmetry 
\citep[e.g.,][]{Kuiper11}.
Similarly, the evolution of stellar mass growth with and without UV feedback,
HM2D and HM2D-NF (Fig.~\ref{fig:xm_xmdot_t_ideal}), is comparable to that seen
in our previous 2D SE-RHD simulations in HS11, 
although a different hybrid code was used.
UV feedback begins to operate for $M_* \gtrsim 40~\msun$ as the
UV emissivity rises in the late KH contraction stage
(Fig.~\ref{fig:xr_suv_xm_ideal}).
The 3D cases show a comparable evolution, except for the
large fluctuations of accretion rates seen in 
Figure~\ref{fig:xm_xmdot_t_ideal}-(b).
This is due to the manner of mass and angular momentum transport in
self-gravitating circumstellar disks. 
We note that, for these idealized 3D cases, 
the assumed initial axial symmetry
is well preserved until the disk begins to grow during the accretion stage.
Thus, even with a much less disturbed structure of the accretion
envelope than for our cosmological cases, highly variable accretion
occurs in general.
Since, however, the mean accretion rate is relatively low for these test
cases, the protostar continues to contract to the ZAMS
without being brought to the supergiant stage by
accretion bursts.
This is still the same even with the doubled spatial resolution
(case HM-HR2), though the scatter of the accretion rates is
larger than the fiducial case HM, as analogously seen for cosmological
cases B and C (Section~\ref{ssec:sres}).

%------------------------ effect of small sink -------------------------%

For the above cases, the stellar radius is always much smaller 
than the default sink size $30$~AU.
We thus also examine the effects of reducing the sink size to 
10~AU (case HM-ssk).
Figure~\ref{fig:xm_xmdot_t_ideal} shows that the resulting 
accretion history is more variable with the smaller sink.
This trend is also consistent with our findings for cosmological case B,
for which a larger sink resulted in less accretion variability
(Fig.~\ref{fig:hn1p2_B}).
Although the stellar mass growth is even more rapid 
than for the other 3D cases for the first $10^4$ years
(Fig.~\ref{fig:xm_xmdot_t_ideal}-a), UV feedback limits
the mass accretion once a bipolar \hii region begins to expand.
Since the 10~AU sink is still larger than the stellar radius, 
we still need further studies dedicated to resolving the flow connecting 
the disk inner edge and stellar surface.
It is also possible that tight binaries with separations much smaller 
than $\sim$10~AU might form as seen in the Galaxy 
(Section~\ref{ssec:multi}), whose formation mechanism is poorly understood.

%===================%

\end{document}